\documentclass[journal]{IEEEtran}

\usepackage{amsmath} 
\usepackage{amssymb}  
\usepackage{amsfonts} 

\usepackage[utf8]{inputenc}

\usepackage[pdftex]{graphicx}
\usepackage[usenames, dvipsnames]{color}
\usepackage{color,soul}
\setstcolor{red}

\usepackage{subcaption}

\usepackage{multirow}
\usepackage{makecell}
\usepackage[linesnumbered,ruled]{algorithm2e}

\usepackage{array}
\usepackage{url}

\usepackage{nomencl}

\usepackage{amsthm}
\usepackage{tikz}

\theoremstyle{definition}

\theoremstyle{remark}

\usepackage{makecell}

\usepackage{lipsum}
\usepackage{multicol}

\usepackage{bm}

\usepackage{relsize}
\usepackage{cleveref}
\usepackage{epsfig}
\usepackage{graphicx}
\usepackage{array}
\usepackage{booktabs}
\usepackage{xcolor}
\usepackage{cite}
\usepackage{multirow}

\usepackage{bm}
\usepackage{epstopdf}
\usepackage{enumerate}

\DeclareMathOperator*{\argmax}{arg\,max}
\DeclareMathOperator*{\argmin}{arg\,min}

\title{Game-Theoretic Modeling of Multi-Vehicle Interactions at Uncontrolled Intersections}

\author{Nan Li, Yu Yao, Ilya Kolmanovsky, Ella Atkins, and Anouck Girard
\thanks{*This research has been supported by the National Science Foundation Award Number CNS 1544844.}
\thanks{Nan Li, Ilya Kolmanovsky, and Anouck Girard are with the Department of Aerospace Engineering,
        University of Michigan, Ann Arbor, MI 48109, USA
        {\tt\small \{nanli,ilya,anouck\}@umich.edu}. 
        Yu Yao, and Ella Atkins are with the Robotics Institute,
        University of Michigan, Ann Arbor, MI 48109, USA {\tt\small \{brianyao,ematkins\}@umich.edu}.}
}

\usepackage{mathrsfs}

\begin{document}

\maketitle

\begin{abstract}
Motivated by the need to develop simulation tools for verification and validation of autonomous driving systems operating in traffic consisting of both autonomous and human-driven vehicles, we propose a framework for modeling vehicle interactions at uncontrolled intersections. The proposed interaction modeling approach is based on game theory with multiple concurrent leader-follower pairs, and accounts for common traffic rules. We parameterize the intersection layouts and geometries to model uncontrolled intersections with various configurations, and apply the proposed approach to model the interactive behavior of vehicles at these intersections. Based on simulation results in various traffic scenarios, we show that the model exhibits reasonable behavior expected in traffic, including the capability of reproducing scenarios extracted from real-world traffic data and reasonable performance in resolving traffic conflicts. The model is further validated based on the level-of-service traffic quality rating system and demonstrates manageable computational complexity compared to traditional multi-player game-theoretic models.
\end{abstract}

\IEEEpeerreviewmaketitle

\section{Introduction}\label{sec:intro}

To provide safer, cleaner, and more efficient transportation is the promise of autonomous driving technologies~\cite{anderson2014autonomous}. Thanks to the serious efforts that have been made in both academia and industry to pursue this goal, advances in perception, decision-making/planning, control theory, and computing systems have made fully autonomous driving possible \cite{buehler2009darpa}. However, before autonomous vehicles can be deployed in mass production, their control systems need to be tested and validated in terms of guaranteeing safety and performance when operating in various traffic environments, which remains a challenging problem \cite{Kalra16safe}. On the one hand, simulation tools can be used for quick and safe virtual tests of these systems and to reduce the time and cost of road tests. On the other hand, the reliability of virtual tests depends on the fidelity of the simulations in terms of modeling traffic scenarios.

In the near to medium term, autonomous vehicles will operate in traffic scenarios together with human-driven vehicles, where interactions between autonomous vehicles and human-driven vehicles will constantly occur. Among different traffic scenarios, the interactive behavior of vehicles at intersections may be particularly complex. An autonomous driving system must account for these interactions to be able to operate safely at an intersection.

By types of traffic control, intersections can be classified as signal-controlled, ``stop'' or ``yield'' sign-controlled, and uncontrolled \cite{UIIG}. Uncontrolled intersections are intersections without traffic signals or signs, and are common in both urban and rural settings over the world \cite{bjorklund2005driver,liu2014analyzing,patil2016microscopic}. According to the U.S. National Highway Traffic Safety Administration's fatality analysis report, more than one fourth of fatal crashes in the U.S. occur at or are related to intersections, and about $50\%$ of these occur at uncontrolled intersections \cite{NHTSA}.

At an uncontrolled intersection, due to the lack of guidance from traffic signals or signs, drivers/automations need to decide whether, when, and how to enter and pass through the intersection on their own; in this case, accounting for the interactions among vehicles is particularly important. Failures in accounting for these interactions may cause deadlocks if driving overly conservatively -- the vehicles may get stuck and never pass through the intersection, or may cause collisions if driving overly aggressively.

Advanced strategies that have been proposed for handling interactive traffic at intersections include cooperative driving, where vehicles cooperate with each other and with road infrastructure to resolve traffic conflicts. They may cooperate through vehicle-to-vehicle negotiations \cite{carlino2013auction,de2013autonomous,ahmane2013modeling}, or through coordination by a centralized traffic ``manager'' in the approach called ``autonomous intersection management'' \cite{dresner2008multiagent,wu2012cooperative,lee2012development}. Although strategies based on cooperative driving have been shown to be capable of improving intersection traffic safety and efficiency, they rely on dense penetration of vehicle-to-vehicle and/or vehicle-to-infrastructure communications as well as autonomous driving systems, which will likely not be the case in the near to medium term.

Alternative strategies have been focused on individual control of the autonomous ego vehicle. To account for the interactions among vehicles, approaches based on, e.g., online verification using reachability analysis \cite{althoff2009model,althoff2014online}, receding-horizon optimization \cite{liniger2015optimization,schwarting2017safe}, learning \cite{you2019advanced}, and game theory \cite{mandiau2008behaviour,sadigh2016planning,bahram2016game,yu2018human,dreves2018generalized}, may be used. Although these approaches establish theoretical foundations of creating autonomous vehicles that are capable of handling interactive traffic at uncontrolled intersections, they must be calibrated and validated to achieve control systems that provide promised safety and performance.

Simulation tools used for virtual tests of these control systems are supposed to be capable of representing the interactive behavior of vehicles with reasonable fidelity, which motivates the development of approaches to modeling vehicle interactions. In this paper, we propose a novel game-theoretic vehicle interaction modeling approach for uncontrolled intersections.

Game theory is in general a suitable tool for modeling strategic interaction between rational decision-makers \cite{myerson2013game}, and has been exploited for modeling driver/vehicle interactions at intersections by several researchers.

In \cite{mandiau2008behaviour}, the vehicle-to-vehicle interactions at an intersection are modeled based on normal-form games -- the vehicles select actions between ``Stop'' and  ``Go'' based on their payoff matrices. The performance of the approach in \cite{mandiau2008behaviour} is limited by the limited number of action choices (i.e., two) and the fact that the dynamic behavior of the vehicles is not explicitly taken into account when the payoff matrices are designed. For instance, as the number of interacting vehicles increases to $6$, almost half of the simulation runs following the approach of \cite{mandiau2008behaviour} lead to deadlocks. In \cite{sadigh2016planning}, the interactions between a human-driven vehicle and an autonomous vehicle are modeled based on a two-player game formulation, where vehicle dynamics are explicitly accounted for. The results of a two-vehicle traffic scenario on an one-lane four-way intersection where both vehicles are going straight to cross the intersection are reported. Extensions of this approach to more vehicles have not been reported and may not be straightforward due to both theoretical limitations and computational challenges.

In our previous work \cite{oyler2016game,li2016hierarchical,li2018game_2}, a game-theoretic framework for modeling vehicle-to-vehicle interactions in multi-vehicle highway traffic scenarios has been proposed. The framework is based on the application of level-$\mathcal{K}$ game theory \cite{nagel1995unraveling,stahl1995players} and explicitly takes into account the dynamic behavior of the vehicles. The vehicle driving policies are determined using reinforcement learning. Once the policies have been obtained offline, highway traffic scenarios with a possibly large number of interacting vehicles can be modeled with minimum online computational effort. Such a level-$\mathcal{K}$ game-theoretic framework has also been extended to model the interactions between two vehicles at an uncontrolled two-lane four-way intersection in \cite{li2018game}. However, generalizations to more complex intersection traffic scenarios, e.g., with more than $2$ interacting vehicles and at intersections of various configurations, have not been addressed.

The contributions of the present paper are: 1) We propose a novel framework based on a formulation of dynamic/sequential leader-follower games with multiple concurrent leader-follower pairs and receding-horizon optimization for modeling the interactive behavior of vehicles at uncontrolled intersections. The framework explicitly accounts for the dynamic behavior of the vehicles, decision-making delays, and common traffic rules. It is generalizable to traffic scenarios with more than $2$ interacting vehicles (results of up to $10$ vehicles are reported) and to intersections of various configurations. 2) We describe an intersection model that parameterizes the intersection layouts and geometries so that uncontrolled intersections with a wide range of configurations can be modeled using a finite set of parameters. 3) We apply our interaction modeling approach to the intersection model to simulate the interactive behavior of vehicles in various uncontrolled intersection traffic scenarios (with various numbers of interacting vehicles, intersection layouts and geometries, etc). 4) Based on simulation results and statistical evaluations, we show that the model exhibits reasonable behavior expected in traffic -- it can reproduce scenarios extracted from real-world traffic data and has reasonable performance in resolving traffic conflicts in complex intersection traffic scenarios. Furthermore, the model demonstrates a manageable increase in computational complexity as the number of interacting vehicles increases. 5) We also describe a generalized version of the interactive decision-making model of vehicles proposed in \cite{li2018game} based on level-$\mathcal{K}$ games, simulate the interactions between the model proposed in this paper and this alternative model, and demonstrate that the model proposed in this paper is capable of resolving conflicts with different drivers.

This paper is organized as follows: In Section~\ref{sec:lf_game}, we present our game-theoretic approach to model interactive decision-making of vehicles at uncontrolled intersections. In Section~\ref{sec:intersection_path}, we describe our intersection model with parameterized layouts and geometries, to which our vehicle interaction modeling approach is applied. In Section~\ref{sec:kinematics_rewards}, we introduce the kinematics model to represent vehicles' dynamic behavior at uncontrolled intersections and the reward function design to represent drivers' decision-making objectives. In Section~\ref{sec:additions}, we incorporate several additional considerations in our model to improve the fidelity of our model in imitating the decision-making processes of human drivers. In Section~\ref{sec:levelK}, we consider a previously proposed interactive decision-making model of vehicles based on level-$\mathcal{K}$ game theory. In Section~\ref{sec:simulations}, we run multiple simulation case studies to comprehensively illustrate and evaluate our proposed framework for modeling vehicle interactions at uncontrolled intersections. The paper is summarized and concluded in Section~\ref{sec:sum}.

\section{Vehicle interaction modeling based on leader-follower games}\label{sec:lf_game}

In this section, we introduce our game-theoretic approach to model interactive decision-making of vehicles at uncontrolled intersections. We first describe the logic for leader-follower role assignment to vehicles at uncontrolled intersections in Section~\ref{sec:leader_follower}, which is the foundation for formulating our leader-follower games. We then describe our vehicle interactive decision-making model based on leader-follower games in two-vehicle interaction settings in Section~\ref{sec:game_2}, and generalize it to multi-vehicle interactions based on our proposed ``pairwise leader-follower games'' in Section~\ref{sec:game_n}.

\subsection{Leader-follower role assignment to vehicles at uncontrolled intersections}\label{sec:leader_follower}

Human drivers can usually resolve traffic conflicts at uncontrolled intersections by following the ``right-of-way'' rules \cite{NHTSA_2}. The right-of-way rules help the drivers decide who proceeds first at an intersection. Motivated by the right-of-way rules, we assign a leader-follower relationship to each pair of vehicles (denoted by $(i,j)$) at an intersection based on the following logic:
\begin{enumerate}[(1)]
\item If vehicles $i,j$ have both entered the intersection, the vehicle with a strictly smaller signed distance to the exit of the intersection is the leader.

\item If at most one of vehicles $i,j$ has entered the intersection, the vehicle with a strictly smaller signed distance to the entrance of the intersection is the leader.

\item If no leader-follower relationship has been assigned based on (1) or (2), then the vehicle on the right is the leader when the two vehicles are coming from adjacent road arms.

\item If no leader-follower relationship has been assigned based on (1), (2), or (3), then the vehicle going straight is the leader when the other vehicle is making a turn.
\end{enumerate}

We note that if a vehicle has entered (resp. exited) the intersection, then its signed distance to the entrance (resp. exit) of the intersection is the negative of the corresponding distance. The entrance and exit points of an intersection (see Fig.~\ref{fig:vehicle_kinematics}) are defined in Section~\ref{sec:intersection_path} for arbitrary intersection layouts and geometries.

If vehicle $i$ is the leader of the pair $(i,j)$, we write $i \prec j$; if $i$ is not the leader (i.e., either $j$ is the leader or no leader-follower relationship has been assigned based on (1)-(4)), we write $i \succeq j$.

We note that the relations $\prec$ and $\succeq$ are not (pre)orders, as they do not have the transitivity property: if $i \prec j$ and $j \prec k$ (resp. $i \succeq j$ and $j \succeq k$), then $i \prec k$ (resp. $i \succeq k$). This can be seen by considering the traffic scenario where four vehicles $i$, $j$, $k$, and $l$ coming from different road arms arrive at the entrances of a four-way intersection at the same time. Then, based on the above role assignment logic we have $i \prec j$, $j \prec k$, $k \prec l$, and $l \prec i$. Indeed, this scenario and similar scenarios where such a cyclic pattern occurs are challenging scenarios for both human drivers and autonomous vehicles -- they may lead to deadlocks, i.e., no one decides to enter the intersection or everyone gets stuck in the middle of the intersection.

The leader-follower role assignment is presented formally as an algorithm in Section~\ref{sec:kinematics_rewards}, which incorporates the above logic with vehicle kinematics, intersection layouts, perception imperfections, etc.

\subsection{Leader-follower game to model two-vehicle interactions}\label{sec:game_2}

Once a leader-follower relationship has been assigned to a pair of vehicles $(i,j)$, we use a leader-follower game (also referred to as a Stackelberg game) to model their interactive decision-making. We choose to use the Stackelberg model because it incorporates the asymmetric roles of the two players and grants one player advantages over the other \cite{basar1999dynamic}, which can be used to account for common traffic rules such as that a car arriving earlier to the intersection typically has the right of way over a car arriving later to the intersection.

Let $\gamma_{l}$ (resp. $\gamma_{f}$) denote an action of the leader (resp. follower), taking values in an action set $\Gamma_{l}$ ($\Gamma_{f}$). Either player makes decisions on its action choices to maximize a reward function, denoted by $\mathbb{R}_{l}({\bf s},\gamma_{l},\gamma_{f})$ for the leader and by $\mathbb{R}_{f}({\bf s},\gamma_{l},\gamma_{f})$ for the follower, where ${\bf s} \in {\bf S}$ denotes the present state in which the two players are making their decisions. When modeling the interactions of two vehicles, ${\bf s}$ contains the states of these two vehicles, i.e., ${\bf s} = (s_i,s_j)$, where $s_i$ (resp. $s_j$) denotes the state of vehicle $i$ (resp. vehicle $j$) (its detailed definition depends on the vehicle kinematics model to be used, which is introduced in Section~\ref{sec:kinematics}). The dependence of either player's reward on both players' states and actions reflects the interactive nature of such a decision-making process.

Following the concept of Stackelberg equilibrium \cite{basar1999dynamic}, one could model the leader-follower decision-making process as follows:
\begin{align}\label{equ:leader_follower_2_1}
\mathbb{Q}_{l}'({\bf s},\gamma_{l}) &:= \min_{\gamma_{f} \in \Gamma_{f}^{*}{\mkern-3mu'}({\bf s},\gamma_{l})} \mathbb{R}_{l}({\bf s},\gamma_{l},\gamma_{f}), \nonumber \\[-1pt]
\Gamma_{f}^{*}{\mkern-2mu'}({\bf s},\gamma_{l}) &:= \big\{\gamma_{f}' \in \Gamma_{f} : \mathbb{R}_{f}({\bf s},\gamma_{l},\gamma_{f}') \ge \mathbb{R}_{f}({\bf s},\gamma_{l},\gamma_{f}), \nonumber \\
& \quad\quad\quad\quad\quad\quad\;\, \forall \gamma_{f} \in \Gamma_{f} \big\},  \nonumber \\[2pt]
\gamma_{l}^* &\in \argmax_{\gamma_{l} \in \Gamma_{l}}\, \mathbb{Q}_{l}'({\bf s},\gamma_{l}), \nonumber \\
\gamma_{f}^* &\in \argmax_{\gamma_{f} \in \Gamma_{f}}\, \mathbb{R}_{f}({\bf s},\gamma^*_{l},\gamma_{f}).
\end{align}

The actions of the leader and the follower are interdependent. In particular, the leader has the so-called ``first mover advantage'': the leader controls the follower's set of rational actions $\Gamma_{f}^{*}{\mkern-2mu'}({\bf s},\gamma_{l})$ through the leader's own action choice $\gamma_{l}$. In such a game formulation, it is assumed that the leader is aware that the follower is capable of observing the leader's action $\gamma_{l}$ before selecting its own action $\gamma_{f}$.

However, in the setting of drivers making decisions in traffic, each driver responds to the actions of other drivers with a reaction delay. More specifically, each driver can only observe the actions of other drivers that are applied at time step $t$ and take them into account in his/her own decision-making at the next time step $t+1$. From the follower's standpoint, since it cannot instantly observe and respond to the instant action of the leader, to secure its possible rewards against the uncertain action choices of the leader, we assume that it applies a ``maximin'' strategy, i.e.,
\begin{align}\label{equ:follower_2}
\mathbb{Q}_{f}({\bf s},\gamma_{f}) &:= \min_{\gamma_{l} \in \Gamma_{l}} \mathbb{R}_{f}({\bf s},\gamma_{l},\gamma_{f}),  \nonumber \\
\gamma_{f}^* &\in \argmax_{\gamma_{f} \in \Gamma_{f}}\, \mathbb{Q}_{f}({\bf s},\gamma_{f}).
\end{align}
We assume that the leader is aware that the follower is using such a maximin strategy to secure its rewards. Taking this awareness into account, the leader makes rational decisions based on:
\begin{align}\label{equ:leader_2}
\mathbb{Q}_{l}({\bf s},\gamma_{l}) &:= \min_{\gamma_{f} \in \Gamma_{f}^*({\bf s})} \mathbb{R}_{l}({\bf s},\gamma_{l},\gamma_{f}), \nonumber \\
\Gamma_{f}^*({\bf s}) &:= \big\{\gamma_{f}' \in \Gamma_{f} : \mathbb{Q}_{f}({\bf s},\gamma_{f}') \ge \mathbb{Q}_{f}({\bf s},\gamma_{f}), \forall \gamma_{f} \in \Gamma_{f} \big\},  \nonumber \\[4pt]
\gamma_{l}^* &\in \argmax_{\gamma_{l} \in \Gamma_{l}}\, \mathbb{Q}_{l}({\bf s},\gamma_{l}).
\end{align}

We now make assumptions on the uniqueness of maximizers as follows:
\begin{align}\label{equ:uniqueness}
&\,\, \forall\, ({\bf s},\gamma_{f}) \in {\bf S} \times \Gamma_{f},\, \exists !\, \gamma_{l}' \in \Gamma_{l} \text{ such that } \nonumber \\
&\,\, \mathbb{R}_{l}({\bf s},\gamma_{l}',\gamma_{f}) \ge \mathbb{R}_{l}({\bf s},\gamma_{l},\gamma_{f}),\, \forall\, \gamma_{l} \in \Gamma_{l}; \nonumber \\[4pt]
&\,\, \forall\, {\bf s} \in {\bf S},\, \exists !\, \gamma_{f}' \in \Gamma_{f} \text{ such that } \nonumber \\
& \min_{\gamma_{l} \in \Gamma_{l}} \mathbb{R}_{f}({\bf s},\gamma_{l},\gamma_{f}') \ge \min_{\gamma_{l} \in \Gamma_{l}} \mathbb{R}_{f}({\bf s},\gamma_{l},\gamma_{f}),\, \forall\, \gamma_{f} \in \Gamma_{f}.
\end{align}
Assumption \eqref{equ:uniqueness} means that at each traffic state ${\bf s}$, for either player ($l$ or $f$), there is one action that is strictly better than the others to use. Although not strictly required in the leader-follower decision-making process described by \eqref{equ:follower_2} and \eqref{equ:leader_2}, assumption \eqref{equ:uniqueness} can simplify the mathematical expression of \eqref{equ:follower_2}-\eqref{equ:leader_2}, i.e.,
\begin{align}\label{equ:leader_follower_2_2}
\mathbb{Q}_{l}({\bf s},\gamma_{l}) &= \mathbb{R}_{l}({\bf s},\gamma_{l},\gamma_{f}^*), \nonumber \\[2pt]
\mathbb{Q}_{f}({\bf s},\gamma_{f}) &= \min_{\gamma_{l} \in \Gamma_{l}} \mathbb{R}_{f}({\bf s},\gamma_{l},\gamma_{f}), \nonumber \\
\gamma_{l}^* &= \argmax_{\gamma_{l} \in \Gamma_{l}}\, \mathbb{Q}_{l}({\bf s},\gamma_{l}), \nonumber \\
\gamma_{f}^* &= \argmax_{\gamma_{f} \in \Gamma_{f}}\, \mathbb{Q}_{f}({\bf s},\gamma_{f}).
\end{align}
We also note that based on our reward function design that is introduced in Section~\ref{sec:rewards}, assumption \eqref{equ:uniqueness} holds.

Note that in the game formulation \eqref{equ:follower_2} and \eqref{equ:leader_2} (and in \eqref{equ:leader_follower_2_2}), the leader has been given the ``first mover advantage'' as the follower applies a maximin strategy, a conservative strategy assuming worst-case scenarios, whereas the leader is able to select comparatively more aggressive actions taking advantage of its awareness of the follower's maximin strategy. An alternative formulation is to let the leader apply a conservative maximin strategy, which corresponds to an assumption that the leader knows that it cannot control the follower's set of rational actions $\Gamma_{f}^*({\bf s})$ through its instant action choice $\gamma_{l}$ since $\gamma_{l}$ is not instantly observable to the follower. Then, \eqref{equ:leader_follower_2_1} becomes
\begin{align}\label{equ:leader_follower_2_3}
\mathbb{Q}_{l}'({\bf s},\gamma_{l}) &= \min_{\gamma_{f} \in \Gamma_{f}} \mathbb{R}_{l}({\bf s},\gamma_{l},\gamma_{f}), \nonumber \\
\gamma_{l}^* &= \argmax_{\gamma_{l} \in \Gamma_{l}}\, \mathbb{Q}_{l}'({\bf s},\gamma_{l}), \nonumber \\
\gamma_{f}^* &= \argmax_{\gamma_{f} \in \Gamma_{f}}\, \mathbb{R}_{f}({\bf s},\gamma^*_{l},\gamma_{f}).
\end{align}
It is clear that \eqref{equ:leader_follower_2_3} is equal to \eqref{equ:leader_follower_2_2} up to a switch of the roles ``leader'' and ``follower.'' Based on the role assignment criterion introduced in Section~\ref{sec:leader_follower}, we choose to use formulation \eqref{equ:leader_follower_2_2} as it agrees with the common traffic rule that a leader, e.g., a car arriving earlier to the intersection, typically has the right of way over a follower, e.g., a car arriving later to the intersection.

\subsection{Pairwise leader-follower game to model multi-vehicle interactions}\label{sec:game_n}

This section discusses a computationally scalable generalization of the vehicle interaction modeling framework based on leader-follower games proposed in Section~\ref{sec:game_2} to intersection traffic scenarios with $n$ interacting vehicles, where $n \ge 2$.

Although $2$-player leader-follower games may be generalized to $n$-player games through considering a multi-level decision-making hierarchy, e.g., player~$k$ being the leader of players~$k+1,\cdots,n$ and being the follower of players~$1,\cdots,k-1$ for every $k \in \{1,\cdots,n\}$, or allowing a level to accommodate multiple players, e.g., players~$2,\cdots,n$ being the followers of player~$1$ and applying Nash equilibrium-based strategies among themselves, such generalizations require exponentially increased computational efforts to solve for solutions as the number of players increases. For instance, a Stackelberg equilibrium solution can be difficult to compute when $n > 3$ \cite{yoo2018predictive}.

Therefore, to handle intersection traffic scenarios with a possibly large number of traffic participants, we propose an alternative generalization approach. Our approach relies on pairwise leader-follower relationships defined for all vehicle pairs at the intersection, and each vehicle's decision-making accounts for all the pairwise leader-follower relationships related to itself. In particular, vehicle $i$ makes decisions on its action choices according to:
\begin{align}\label{equ:leader_follower_n_1}
\underline{\mathbb{Q}}_{i}({\bf s}_{\text{traffic}},\gamma_{i}) &:= \min_{j \in \{1,\cdots,n\},\, j \neq i} \mathbb{Q}_{i,j}({\bf s}_{i,j},\gamma_{i}), \nonumber \\
\mathbb{Q}_{i,j}({\bf s}_{i,j},\gamma_{i}) &:= \begin{cases} \mathbb{Q}_{l}({\bf s}_{i,j},\gamma_{i}) & \text{if } i \prec j, \nonumber \\
\mathbb{Q}_{f}({\bf s}_{i,j},\gamma_{i}) & \text{if } i \succeq j, \end{cases} \\[2pt]
\gamma_{i}^* &\in \argmax_{\gamma_i \in \Gamma_i}\, \underline{\mathbb{Q}}_{i}({\bf s}_{\text{traffic}},\gamma_{i}),
\end{align}
where $\mathbb{Q}_{l}({\bf s}_{i,j},\gamma_{i})$ (resp. $\mathbb{Q}_{f}({\bf s}_{i,j},\gamma_{i})$) is defined in \eqref{equ:leader_follower_2_2} with player $i$ being the leader $l$ (resp. the follower $f$); the traffic state ${\bf s}_{\text{traffic}}$ contains the states of all interacting vehicles at the intersection, i.e., ${\bf s}_{\text{traffic}} = (s_1,\cdots,s_n)$; and ${\bf s}_{i,j} = (s_i,s_j)$ represents the state of the vehicle pair $(i,j)$.

The decision-making model \eqref{equ:leader_follower_n_1} can be interpreted as follows: If $i$ is the follower of $j$, the secured reward of action $\gamma_{i}$ is the least reward $i$ may get due to the uncertain action choice of $j$; if $i$ is the leader of $j$, $i$ is aware that the most aggressive action that $j$ can choose is subject to $j$'s maximin principle between their pairwise interactions, and thus $i$ predicts the reward of action $\gamma_{i}$ by assuming $j$ to apply its maximin action of their pair. On top of this, to account for its interactions with all other players, $i$ maximizes the minimum of its secured/predicted rewards over all pairwise interactions.

We note that when $i$ is the leader of $j$, its reward prediction may be inaccurate as the actually applied action of $j$ is not only subject to $j$'s maximin principle between their pairwise interactions but also subject to $j$'s interactions with the other players. However, in the setting of vehicle interactions at uncontrolled intersections where the central question is ``who goes first,'' the above strategy for the leader, i.e., predicting action rewards by assuming the follower to apply the maximin action of their pair, is reasonable -- if the follower's maximin action of their pair is not to go first, the follower will likely not choose to go first when it maximizes the minimum of its secured/predicted rewards over all pairwise interactions. With this strategy, the pairwise leader is able to select comparatively more aggressive actions than the pairwise follower. And as a result, if there is an overall leader, i.e., the leader in every pairwise leader-follower relationship related to itself, it can take comparatively most aggressive actions than all other players, e.g., to go first. The effectiveness of decision-making model \eqref{equ:leader_follower_n_1} in resolving traffic conflicts at uncontrolled intersections, i.e., driving every vehicle safely through the intersection without causing collisions or deadlocks, is illustrated through multiple simulation case studies in Section~\ref{sec:simulations}. Furthermore, decision-making model \eqref{equ:leader_follower_n_1} decouples the $n$-player interactions into pairwise interactions, and thus significantly decreases the computational complexity in solving for solutions. It also agrees with intuition -- when driving in traffic, a driver may focus more on the interactions between each neighbouring driver and him/herself than on the interactions among the other drivers.

We note also that our strategy based on leader-follower games is not equivalent to a rule-based strategy where ``who goes first'' is determined by specified rules or logic, e.g., a strategy where a follower always waits until a leader passes through the intersection before the follower itself enters the intersection. Our leader-follower game based strategy allows a follower to enter the intersection even when a leader is still in the intersection, for instance, in situations where the follower's action choices have minor conflicts with the leader's action choices. Whether or not the follower's action choices have conflicts with the leader's action choices and how these conflicts influence their interactive decision-making are represented and automatically handled by the decision-making process \eqref{equ:leader_follower_n_1}.

\section{Parameterized intersection and vehicle path modeling} \label{sec:intersection_path}

Simulation tools used for verification and validation of autonomous vehicles are supposed to cover a sufficiently rich set of traffic scenarios. For instance, intersections in real-world road networks can have different layouts (e.g., number of road arms) and geometries (e.g., angles between road arms and lane width). To model traffic scenarios at intersections of various layouts and geometries, in this section, we first describe an intersection model that parameterizes the intersection layouts and geometries and then present an approach to model the paths of vehicles at the intersections. Although there has been a rich literature on path planning for vehicles \cite{schwarting2018planning}, our vehicle path model is simple but sufficient for our purpose. Moreover, both the intersection and vehicle path models described in this section are designed in such a way that they are convenient for the application of our vehicle interaction model described in Section~\ref{sec:lf_game}.

\subsection{Parameterized intersection modeling}\label{sec:intersection}

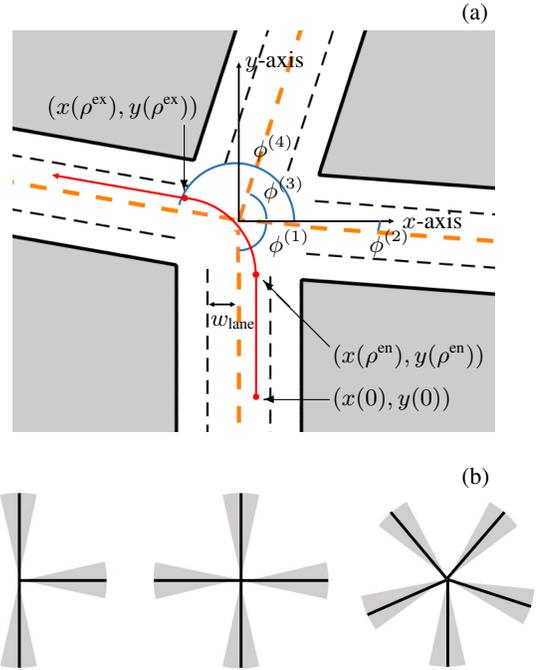
\begin{figure}[h!]
\begin{center}
\begin{picture}(195.0, 246.0)
            \put(  5,  90){\epsfig{file=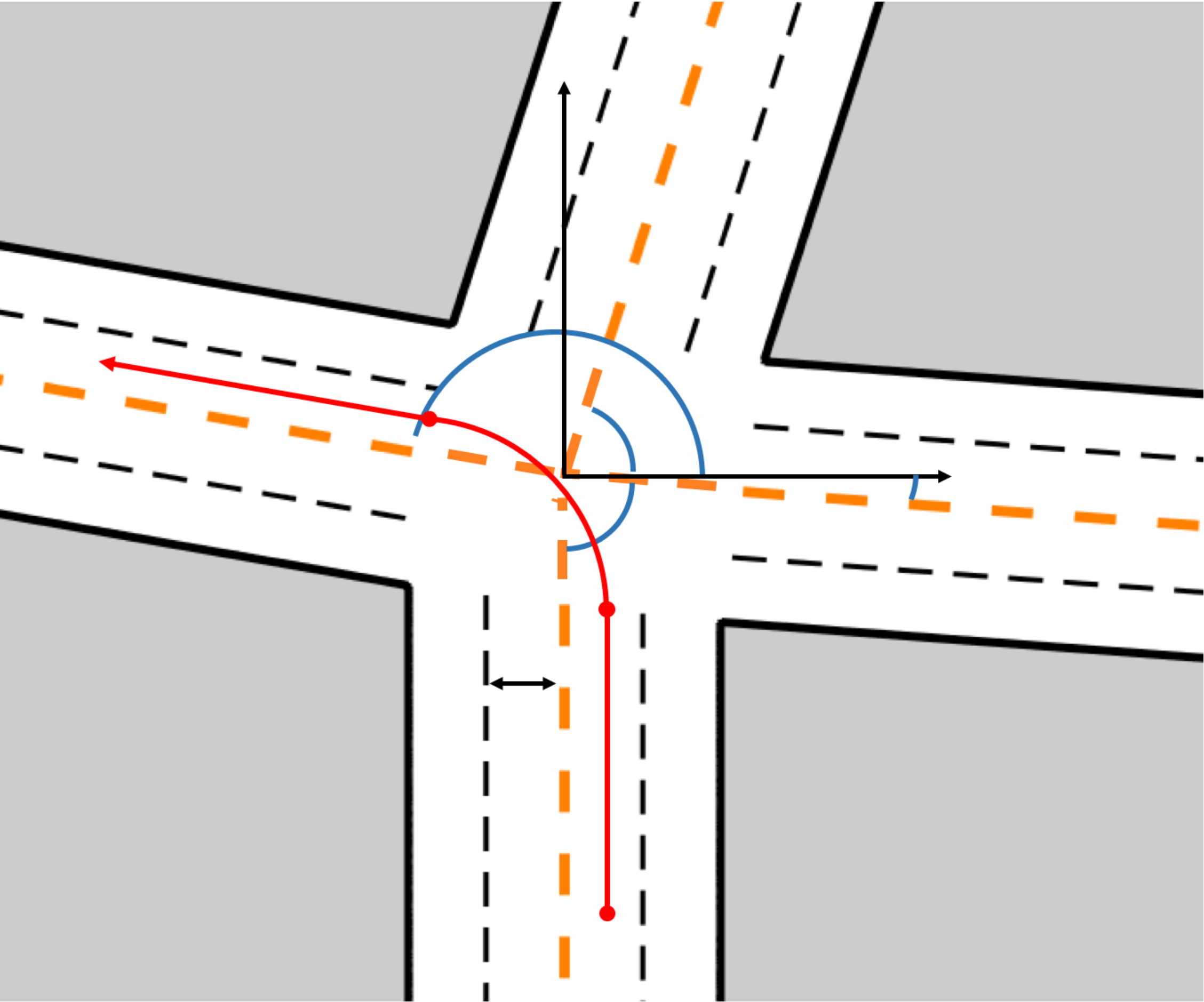,height=2.1in}} 
            \put(  0,  0){\epsfig{file=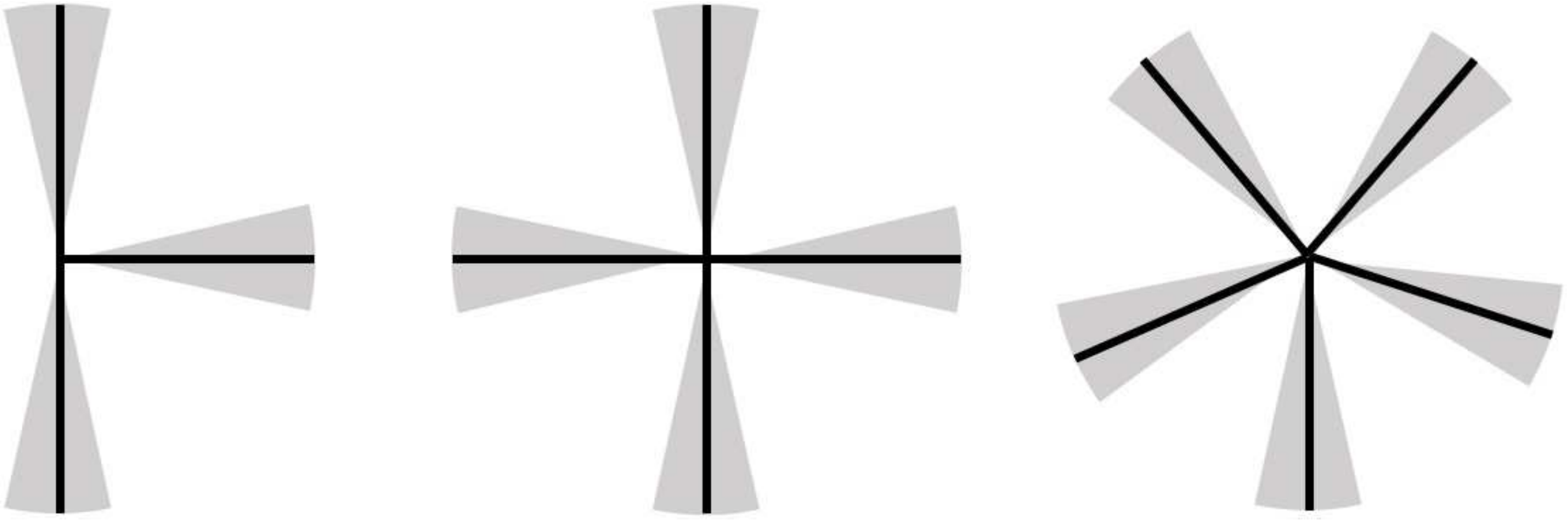,height=0.93in}} 
            \small
            \put(  152.5,  167){$x$-axis}
            \put(  93,  228){$y$-axis}
            \put(  102,  158){$\phi^{(1)}$}  
            \put(  140,  158){$\phi^{(2)}$}  
            \put(  100,  178){$\phi^{(3)}$}  
            \put(  96,  194){$\phi^{(4)}$}  
            \put(  126,  100){$(x(0),y(0))$}  
            \put(  126,  116){$(x(\rho^{\text{en}}), y(\rho^{\text{en}}))$}  
            \put(  18,  210){$(x(\rho^{\text{ex}}), y(\rho^{\text{ex}}))$}  
            \put(  80,  130){$w_{\text{lane}}$}
            \put(  125, 102){\vector(-1,0){25}}
            \put(  125, 122){\vector(-1,1){25}}
            \put(  70, 206){\vector(0,-1){25}} 
            \put(  175,  246){(a)}
            \put(  175,  70){(b)}
            \normalsize
\end{picture}
\end{center}
      \caption{Intersection geometry and topology. (a) A four-way intersection, where the orange dashed lines are the road centerlines, the black dashed lines are the lane markings that separate lanes of traffic moving in the same directions, the black solid lines are the road boundaries, and the shaded polygons are off-road regions. (b) The topologies of three-way, four-way, and five-way intersections, where the black sticks indicate road arms in their nominal directions and the shaded areas indicate the admissible directional variations for the road arms.}
     \label{fig:intersection}
\end{figure}

We characterize the layout and geometry of an intersection using a set of parameters, i.e.,
\begin{equation}\label{equ:inter_para}
    \big(N, \{M_{\text{f}}^{(m)}\}_{m=1}^N, \{M_{\text{b}}^{(m)}\}_{m=1}^N, \{\phi^{(m)}\}_{m=1}^N, w_{\text{lane}} \big),
\end{equation}
where $N$ is the number of road arms of the intersection, $M_{\text{f}}^{(m)}\in\{0,1,2,\cdots\}$ and $M_{\text{b}}^{(m)}\in\{0,1,2,\cdots\}$ are, respectively, the numbers of forward and backward lanes\footnote{A ``forward'' lane (resp. a ``backward'' lane) is a lane for traffic ``entering the intersection'' (resp. ``moving away from the intersection'').} of the $m$th arm, $\phi^{(m)}$ is the counter-clockwise angle of the $m$th arm with respect to the $x$-axis, and $w_{\text{lane}}$ is the lane width\footnote{We assume that all of the lanes have the same width although in principle they do not have to.} (see Fig.~\ref{fig:intersection}(a)). We note that $M_{\text{f}}^{(m)} = 0$ (or $M_{\text{b}}^{(m)} = 0$) represents one-way road, and $M_{\text{f}}^{(m)}=0$ and $M_{\text{b}}^{(m)}=0$ should not happen at the same time. We assume that the road centerlines\footnote{The road centerlines are the lane markings that separate lanes of traffic moving in the opposite directions.} of all the road arms intersect at the same point, which is referred to as the intersection center with coordinates $(x_\text{o},y_\text{o}) = (0,0)$.

In this paper, we consider three-way, four-way, and five-way intersections (see Fig.~\ref{fig:intersection}(b)), i.e., $N \in \{3,4,5\}$, as they are most common in real-world road networks.

Given a set of parameters \eqref{equ:inter_para}, the lane markings and road boundaries of the $m$th arm can be expressed according to
\begin{equation}\label{equ:lane_func}
        x\sin(\phi^{(m)}) - y\cos(\phi^{(m)}) + \frac{k w_{\text{lane}}}{2} = 0,
\end{equation}
where $k \in \{-2M_{\text{b}}^{(m)}, \cdots, 2M_{\text{f}}^{(m)}\}$. When $k = 2M_{\text{f}}^{(m)}$ (resp. $k = -2M_{\text{b}}^{(m)}$), \eqref{equ:lane_func} represents the right-hand-side road boundary when looking in the forward direction (resp. in the backward direction); when $k \in 2\{M_{\text{f}}^{(m)}-1, \cdots, 1\}$ (resp. $k \in 2\{-M_{\text{b}}^{(m)}+1, \cdots, -1\}$), \eqref{equ:lane_func} represents a lane marking that separates two lanes of traffic moving in the forward direction (resp. in the backward direction); when $k = 0$, \eqref{equ:lane_func} represents the road centerline; and when $k \in 2\{M_{\text{f}}^{(m)}, \cdots, 1\}-1$ (resp. $k \in 2\{-M_{\text{b}}^{(m)}, \cdots, -1\}+1$), \eqref{equ:lane_func} represents the center of a forward lane (resp. backward lane).

On the basis of \eqref{equ:lane_func}, we assign an ``entrance point'' to each forward lane, $(x(\rho^{\text{en}}), y(\rho^{\text{en}}))$, which indicates entering the intersection, as follows: We first locate the $N$ intersection corners\footnote{An intersection corner is the intersection point of two adjacent road boundaries.}. The line segment connecting each pair of adjacent corners is referred to as the ``entrance line'' of the corresponding road arm. The entrance point of each forward lane is determined as the intersection point of its center and the entrance line of the road arm it belongs to.

On the other hand, we also assign ``exit points'' indicating exiting the intersection, $(x(\rho^{\text{ex}}), y(\rho^{\text{ex}}))$, which are used in determining the leader-follower relationships between vehicles (see Section~\ref{sec:leader_follower}). In particular, the determination of an exit point is coupled with our model for vehicle path, which is described in the next section.

We note that the parameterized intersection model described above corresponds to the right-hand traffic \cite{kincaid1986rule}. To model intersections in the context of left-hand traffic requires corresponding and straightforward modifications.

\subsection{Vehicle path modeling}\label{sec:path}

We assume that vehicles can plan their paths according to their origin lanes and target lanes\footnote{The origin lane and the target lane of a vehicle are, respectively, the lane where it is driving before entering the intersection and the lane that it is going to after exiting the intersection.} before entering the intersection and follow these pre-planned paths to pass through the intersection. Such an assumption is often adopted in the literature \cite{chen2016cooperative}. When there are conflicts between vehicles, they can adjust their speeds along the paths according to their interactions with each other. In this section, we describe our vehicle path model.

At first, we specify the origin lane and target lane of each vehicle to be modeled. When specifying the origin lane and target lane, some constraints representing common traffic rules can be enforced, including: 1) given an origin road arm, if the target road arm corresponds to a left turn, then the origin lane (resp. the target lane) must be the leftmost forward lane of the origin road arm (resp. the leftmost backward lane of the target road arm); 2) given an origin road arm, if the target road arm corresponds to a right turn, then the origin lane (resp. the target lane) must be the rightmost forward lane of the origin road arm (resp. the rightmost backward lane of the target road arm); and 3) when going straight, if the origin lane is the $\eta$th forward lane from the left of the origin road arm, then the target lane must be the $\eta'$th backward lane from the left of the target road arm $m$, where $\eta' = \min(\eta,M_{\text{b}}^{(m)})$.

In particular, given the intersection layout and geometry, we determine whether a vehicle is ``making a left turn,'' ``going straight,'' or ``making a right turn'' based on the angle between its origin road arm and target road arm. When the clockwise angle from its origin road arm to its target road arm is in the interval $(0,\frac{3\pi}{4}]$, then it is ``making a left turn''; when the angle is in the interval $(\frac{3\pi}{4},\frac{5\pi}{4})$, then it is ``going straight''; it is ``making a right turn'' otherwise. We also note that U-turns are not considered in this paper, i.e., the origin lane and the target lane of a vehicle must belong to two different road arms.

Once the origin lane and target lane of a vehicle is specified, we assign the vehicle an ``initial point,'' $(x^{\text{ini}},y^{\text{ini}})$, located in the center of its origin lane and a ``terminal point,'' $(x^{\text{term}},y^{\text{term}})$, located in the center of its target lane. After that, we model the vehicle's path, $\mathcal{P}$, as a curve composed of three segments. The first segment is a line segment with the two end points being $(x^{\text{ini}},y^{\text{ini}})$ and the intersection entrance point $(x(\rho^{\text{en}}), y(\rho^{\text{en}}))$ of the origin lane (see Section~\ref{sec:intersection}). Similarly, the third segment is a line segment with $(x^{\text{term}},y^{\text{term}})$ as one of its end points and extending in the direction of the target lane. The second segment is an arc that connects the first segment and the third segment, tangential to the first segment at $(x(\rho^{\text{en}}), y(\rho^{\text{en}}))$ and tangential to the third segment, where the point of tangency is defined as the intersection exit point, $(x(\rho^{\text{ex}}), y(\rho^{\text{ex}}))$, associated with path $\mathcal{P}$\footnote{So $(x(\rho^{\text{ex}}), y(\rho^{\text{ex}}))$ is the other end point of the third segment.}. Thus, $\mathcal{P}$ is a smooth curve. For any point on the curve, $(x,y) \in \mathcal{P}$, we define $\rho$ as the length of the curve piece from $(x^{\text{ini}},y^{\text{ini}})$ to $(x,y)$. Note that the curve $\mathcal{P}$ can be expressed as an injective function of $\rho$ since $\mathcal{P}$ does not intersect itself, i.e., any point on the curve can be determined by a unique $\rho$. In particular, we have $(x(0),y(0)) = (x^{\text{ini}},y^{\text{ini}})$, i.e., the initial point is the location of the vehicle when its traveled distance along the path is zero. Also, we let $\rho^{\text{en}}$ (resp. $\rho^{\text{ex}}$) denote the value of $\rho$ corresponding to the point on $\mathcal{P}$ with coordinates $(x(\rho^{\text{en}}), y(\rho^{\text{en}}))$ (resp. $(x(\rho^{\text{ex}}), y(\rho^{\text{ex}}))$)\footnote{So it is reasonable to name the intersection entrance (resp. exit) point as $(x(\rho^{\text{en}}), y(\rho^{\text{en}}))$ (resp. $(x(\rho^{\text{ex}}), y(\rho^{\text{ex}}))$) in the first place.}, i.e., the vehicle enters (resp. exits) the intersection when its traveled distance along the path is $\rho^{\text{en}}$ (resp. $\rho^{\text{ex}}$).

To facilitate reproducing the results of this paper, the functions used to generate $\mathcal{P}$ given the coordinates of $(x^{\text{ini}},y^{\text{ini}})$, $(x^{\text{term}},y^{\text{term}})$, and $(x(\rho^{\text{en}}), y(\rho^{\text{en}}))$ are explicitly provided in Appendix~A.

Based on the above descriptions, the path of a vehicle can be determined given its origin lane and target lane and represented as
\begin{align}\label{equ:center_coordinate}
    \mathcal{P}:&\,\,\, \mathbf{R} \to \mathbf{R}^2, \quad\quad \rho \mapsto \begin{bmatrix} x(\rho) \\ y(\rho) \end{bmatrix},
\end{align}
where $\rho$ represents the distance the vehicle has traveled along the path. In particular, we define the signed distance of the vehicle to the entrance (resp. exit) of the intersection as $\Delta \rho^{\text{en}} = \rho^{\text{en}} - \rho$ (resp. $\Delta \rho^{\text{ex}} = \rho^{\text{ex}} - \rho$) so that $\Delta \rho^{\text{en}}(t) < 0$ (resp. $\Delta \rho^{\text{ex}}(t) < 0$) represents that the vehicle has entered (resp. exited) the intersection.

\section{Vehicle kinematics and rewards at uncontrolled intersections}\label{sec:kinematics_rewards}

In this section, we introduce the model to represent vehicles' dynamic behavior and the reward function to represent drivers' objectives at uncontrolled intersections.

\subsection{Vehicle kinematics}\label{sec:kinematics}

We represent a vehicle using a rectangle bounding the vehicle's geometric contour projected onto the ground. This rectangle is referred to as the ``collision zone'' ($c$-zone) as an overlap of two vehicles' $c$-zones indicates a danger of collision. To fully characterize the $c$-zone of a vehicle, we need a $5$-tuple, $(x,y,\theta,l_c,w_c)$, where $(x,y)$ are the coordinates of its geometric center, $\theta$ is the vehicle's heading angle (the counter-clockwise angle of the vehicle's heading direction with respect to the $x$-axis), and $l_c$ (resp. $w_c$) is the length (resp. width) of the rectangle.

On the basis of the vehicle path model \eqref{equ:center_coordinate} and the assumption that the vehicle can follow its pre-planned path $\mathcal{P}$ perfectly, $(x,y)$ can be written as functions of $\rho$, i.e., $(x(\rho),y(\rho))$, and the vehicle's heading angle $\theta(\rho)$ can be computed using the path geometry as follows:
\begin{equation}\label{equ:heading_angle_1}
    \theta(\rho) = \lim_{h \to 0^+} \text{arctan2} \Big(y(\rho + h) -y(\rho),x(\rho + h) - x(\rho)\Big),
\end{equation}
which, using the fact that $\mathcal{P}$ is smooth, can be written as
\begin{equation}\label{equ:heading_angle_2}
    \theta(\rho) = \text{arctan2} \Big(\frac{\text{d}y}{\text{d}\rho},\frac{\text{d}x}{\text{d}\rho}\Big).
\end{equation}

Based on \eqref{equ:center_coordinate}-\eqref{equ:heading_angle_2} and the assumption above, the dynamic behavior of a vehicle can be fully characterized by the dynamics of $\rho$ as follows:
\begin{align}\label{equ:kinematics}
    \rho(t+1) &= \rho(t) + v(t)\, \Delta t, \nonumber \\
    v(t+1) &= v(t) + a(t)\, \Delta t,
\end{align}
where $t$ denotes the discrete time instant, $v(t) \in [v_{\min},v_{\max}]$ and $a(t)$ denote, respectively, the vehicle's speed and acceleration at $t$, and $\Delta t$ is the sampling period.

We collect all relevant variables and define the state of a vehicle as an $8$-tuple, i.e.,
\begin{align}\label{equ:state}
    s(t) =&\, \big(\mathcal{P},\rho(t),v(t),x(\rho(t)),y(\rho(t)),\theta(\rho(t)), \nonumber \\
    &\,\,\,\, \Delta \rho^{\text{en}}(t),\Delta \rho^{\text{ex}}(t)\big).
\end{align}

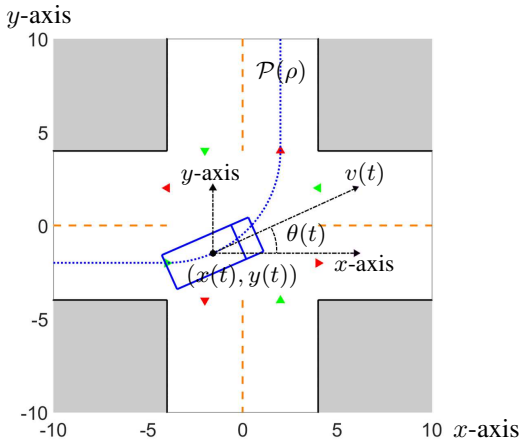
\begin{figure}[h!]
\begin{center}
\begin{picture}(188.0, 163.0)
\put(  0,  -10){\epsfig{file=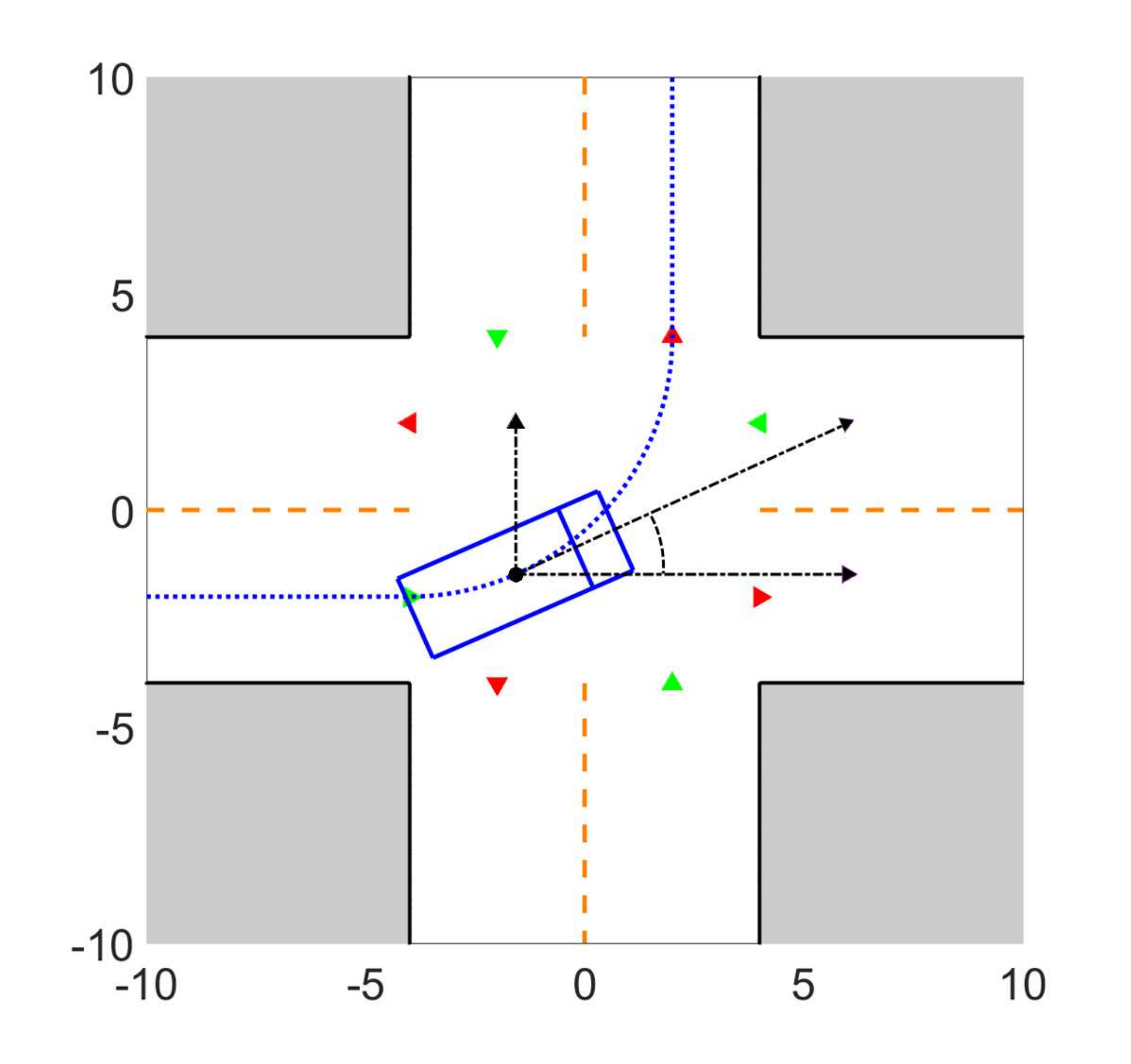,height=2.4in}}  
\put(  175,  0){$x$-axis}  
\put(  6,  157){$y$-axis}  
\small
\put(  130,  63){$x$-axis}  
\put(  72,  97.5){$y$-axis}  
\put(  134,  97.5){$v(t)$}  
\put(  112,  74.5){$\theta(t)$}  
\put(  74,  59){$(x(t),y(t))$}  
\put(  100.5,  136){$\mathcal{P}(\rho)$}  
\normalsize
\end{picture}
\end{center}
      \caption{Vehicle kinematics model.}
      \label{fig:vehicle_kinematics}
\end{figure}

The vehicle kinematics at a typical two-lane four-way intersection are illustrated in Fig.~\ref{fig:vehicle_kinematics}. The blue rectangle represents the vehicle's $c$-zone where the end with double lines is the vehicle's front end. The blue dotted curve represents the pre-planned path $\mathcal{P}$. The states $x(t)$, $y(t)$ and $\theta(t)$ can be computed using the traveled distance along the path $\rho(t)$ and the path geometry. The green triangles represent the intersection entrance points $(x(\rho^{\text{en}}), y(\rho^{\text{en}}))$ and the red triangles represent the intersection exit points $(x(\rho^{\text{ex}}), y(\rho^{\text{ex}}))$.

To adjust speed along the path, we assume that a vehicle has a finite number of acceleration levels to choose from at each time step, i.e.,
\begin{equation}\label{equ:accelerations}
    a(t) \in A = \big\{a^1, \cdots, a^{\mathcal{M}}\big\}, \quad \forall\, t.
\end{equation}

On the basis of the model representing vehicles' dynamic behavior at uncontrolled intersections described above, we now present the leader-follower role assignment logic in Section~\ref{sec:leader_follower} formally as Algorithm~\ref{alg:Role}.

\begin{algorithm}
    \caption{Leader-follower role assignment}
    \label{alg:Role}
    \SetKwInOut{Input}{Input}
    \SetKwInOut{Output}{Output}
    \Input{an ordered pair of vehicles $(i,j)$ and their states $\big(s_i(t),s_j(t)\big)$ }
    \Output{whether $i$ is the leader of $j$ }
    \lIf{($\Delta \rho_i^{\text{en}}(t) \le 0$ and $\Delta \rho_j^{\text{en}}(t) \le 0$) and $\Delta \rho_i^{\text{ex}}(t) < \Delta \rho_j^{\text{ex}}(t) - \delta$}{$i \prec j$}
    \lElseIf{($\Delta \rho_i^{\text{en}}(t) > 0$ or $\Delta \rho_j^{\text{en}}(t) > 0$) and $\Delta \rho_i^{\text{en}}(t) < \Delta \rho_j^{\text{en}}(t) - \delta$}{$i \prec j$}
    \lElseIf{$i$ and $j$ are coming from adjacent ways and $i$'s way is on the right of $j$'s way}{$i \prec j$}
    \lElseIf{$i$ is going straight and $j$ is making a turn}{$i \prec j$}
    {\bf else} $i \succeq j$.
\end{algorithm}

In Algorithm~\ref{alg:Role}, $\delta \ge 0$ is a threshold for differentiating the distances, accounting for the fact that drivers can only estimate the distances with limited accuracy. In particular, we assume that drivers cannot recognize which distance is smaller when $|\Delta \rho_i^{\text{en}}(t) - \Delta \rho_j^{\text{en}}(t)| \le \delta$ (resp. $|\Delta \rho_i^{\text{ex}}(t) - \Delta \rho_j^{\text{ex}}(t)| \le \delta$). On the basis of Algorithm~\ref{alg:Role}, at most one of outcomes $i \prec j$ or $j \prec i$ can take place. It may happen that $i \succeq j$ and $j \succeq i$. In such a case, both vehicles will view themselves as followers and thus make conservative decisions. In line~4, ``going straight'' and ``making a turn'' need to be differentiated, which has been described for arbitrary intersection layouts and geometries in Section~\ref{sec:path}.

\subsection{Reward function}\label{sec:rewards}

Basic goals of a driver at an intersection include: 1) to maintain safety, e.g., to not have a collision with another vehicle, 2) to keep a reasonable distance from other vehicles to improve safety and comfort, and 3) to pass through the intersection and get to his/her target under traffic rules and in a timely manner.

We assume that common traffic rules, such as that a left turn can only be made when the vehicle is entering the intersection from a left-turn lane (usually the leftmost forward lane), and speed limits, have been incorporated in path planning (see Section~\ref{sec:path}) and in speed bounds $v(t) \in [v_{\min},v_{\max}]$. Then, the other goals can be represented using a reward function as follows:
\begin{equation}\label{equ:reward}
    \mathbb{R}(t) = \sum_{\tau = 1}^{\mathcal{N}} \lambda^{\tau-1} R(\tau|t),
\end{equation}
where $R(\tau|t)$ is a predicted stage reward at time instant $t+\tau$ with the prediction made at the current time instant $t$, $\mathcal{N}$ is the prediction horizon, and $\lambda \in [0,1]$ is a factor discounting future rewards. The stage reward is defined as a linear combination of three terms, each of which represents a goal introduced above, i.e.,
\begin{equation}\label{equ:stage_reward}
    R(\tau|t) = w_1 \hat{c}(\tau|t) + w_2 \hat{s}(\tau|t) + w_3 \hat{v}(\tau|t),
\end{equation}
where $w_i > 0$, $i \in \{1,2,3\}$, are weighting factors, and the terms $\hat{c}(\tau|t)$, $\hat{s}(\tau|t)$, and $\hat{v}(\tau|t)$ are further explained below.

On the basis of our decision-making model \eqref{equ:leader_follower_n_1}, the ego vehicle making decisions considers its interactions with each of the other vehicles separately. Let vehicle $i$ denote the ego vehicle and vehicle $j$ denote the vehicle in the pairwise interaction with vehicle $i$.

\noindent $\boldsymbol{\cdot}$ Collision avoidance, $\hat{c}\,$:
\begin{equation*}
\hat{c}(\tau|t) = - \big(1 + S_c(\tau|t) + \hat{w} |v_i(\tau|t) v_j(\tau|t)| \big)\, \mathbb{I}\big(S_c(\tau|t)>0\big),
\end{equation*}
where $S_c(\tau|t) \ge 0$ is the predicted area of the intersection of vehicle $i$ and $j$'s $c$-zones, $v_i(\tau|t)$ and $v_j(\tau|t)$ are, respectively, vehicle $i$ and $j$'s predicted speeds, $\hat{w}>0$ is a tunable parameter, and $\mathbb{I}(\cdot)$ is an indicator function taking $1$ if $(\cdot)$ holds and $0$ otherwise. The $c$-zone of a vehicle is defined at the beginning of Section~\ref{sec:kinematics}, which is a rectangle characterized by the 5-tuple $(x,y,\theta,l_c,w_c)$. Thus, $S_c(\tau|t)$ is determined by the predicted states of vehicle $i$ and $j$, i.e., $S_c(\tau|t) = S_c\big(s_i(\tau|t),s_j(\tau|t)\big)$.

The term $\hat{c}$ is designed in the above form so that if $S_c(\tau|t) = 0$, i.e., there is no danger of collision, then $\hat{c}(\tau|t) = 0$; if $S_c(\tau|t) > 0$, i.e., there is a danger of collision, then the penalty depends on the $c$-zone intersection area $S_c$ and the vehicle speeds $v_i$ and $v_j$. In particular, larger penalties are imposed for larger $S_c$ values and for larger absolute values of $v_i$ and $v_j$ as they imply more severe collisions; the parameter $\hat{w}>0$ adjusts the relative contribution of intersection area versus speeds; and the addition of $1$ ensures a minimum penalty for collisions.

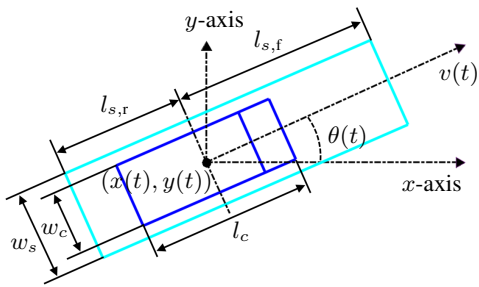
\begin{figure}[h!]
\begin{center}
\begin{picture}(186.0, 105.0)
\put(  0,  -5){\epsfig{file=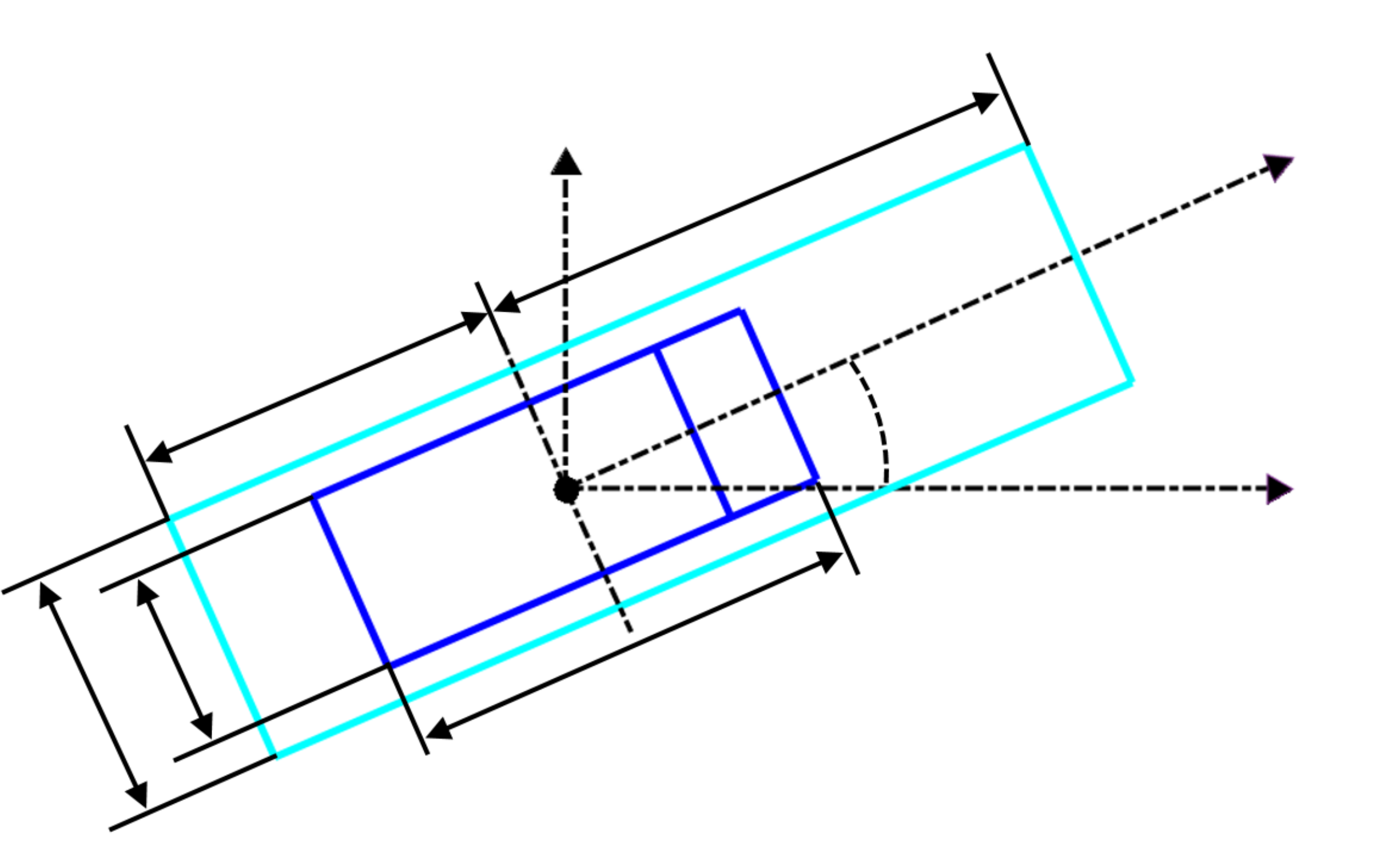,height=1.6in}}  
\small
\put(  150,  34){$x$-axis}  
\put(  68,  96){$y$-axis}  
\put(  164,  76){$v(t)$}  
\put(  122,  52){$\theta(t)$}  
\put(  36,  35){$(x(t),y(t))$}  
\put(  86,  16){$l_c$}  
\put(  13.5,  17.5){$w_c$}  
\put(  94,  88.5){$l_{s,\text{f}}$}  
\put(  36,  64){$l_{s,\text{r}}$}  
\put(  2,  12){$w_s$}  
\normalsize
\end{picture}
\end{center}
      \caption{The $c$-zone (dark blue rectangle) and $s$-zone (light blue rectangle) of a vehicle.}
      \label{fig:vehicle_zones}
\end{figure}

\noindent $\boldsymbol{\cdot}$ Separation, $\hat{s}$:
\begin{equation*}
\hat{s}(\tau|t) = - \big(1 + S_s(\tau|t) + \hat{w} |v_i(\tau|t) v_j(\tau|t)| \big)\, \mathbb{I}\big(S_s(\tau|t)>0\big),
\end{equation*}
where $S_s(\tau|t) \ge 0$ is the predicted area of the intersection of vehicle $i$ and $j$'s ``separation zones'' ($s$-zones). The $s$-zone of a vehicle is defined as a rectangle that shares the same longitudinal line of symmetry with the vehicle's $c$-zone and over-bounds the $c$-zone with a safety margin (see Fig.~\ref{fig:vehicle_zones}). It can be fully characterized by a 6-tuple $(x,y,\theta,l_{s,\text{f}},l_{s,\text{r}},w_s)$, where $l_{s,\text{f}},l_{s,\text{r}} \ge l_c/2$ and $w_s \ge w_c$. In particular, when vehicle $i$ is the leader (resp. the follower) in its pairwise interaction with vehicle $j$, vehicle $i$ assumes that both vehicles have their $s$-zones of the same size, denoted by $(l_{s,\text{f}}^l,l_{s,\text{r}}^l,w_s^l)$ (resp. $(l_{s,\text{f}}^f,l_{s,\text{r}}^f,w_s^f)$). We let $l_{s,\text{f}}^l \le l_{s,\text{f}}^f$, $l_{s,\text{r}}^l \le l_{s,\text{r}}^f$, and $w_s^l \le w_s^f$ to further encourage the leader to choose comparatively more aggressive actions than the follower.

\noindent $\boldsymbol{\cdot}$ Velocity, $\hat{v}$: As the vehicle is assumed to follow its pre-planned path, its status of approaching its target can be characterized by its speed along the path. In particular, we set the term $\hat{v}(\tau|t) = v_i(\tau|t)$.

Note that the predicted stage reward \eqref{equ:stage_reward} depends on the predicted states of the two vehicles $\big(s_i(\tau|t),s_j(\tau|t)\big)$. On the basis of the vehicle kinematics model \eqref{equ:center_coordinate}-\eqref{equ:state}, $s_\xi(\tau|t)$, $\xi \in \{i,j\}$, is uniquely determined by the current state of vehicle $\xi$, $s_\xi(t)$, and the predicted acceleration sequence of vehicle $\xi$, $\{a_\xi(k|t)\}_{k=0}^{\tau-1}$, i.e., $s_\xi(\tau|t) = s_\xi \big(s_\xi(t),\{a_\xi(k|t)\}_{k=0}^{\tau-1}\big)$. We define an action choice of vehicle $\xi$, $\xi \in \{i,j\}$, at the current time instant $t$ as $\gamma_\xi(t) = \{a_\xi(\tau|t)\}_{\tau=0}^{\mathcal{N}-1}$, taking values in the action set $\Gamma = A^\mathcal{N}$. Therefore, the cumulative reward \eqref{equ:reward} is a function of ${\bf s}_{i,j}(t)$, $\gamma_i(t)$, and $\gamma_j(t)$, in consistent with the expressions in our decision-making model \eqref{equ:leader_follower_n_1}.

The closed-loop operation of each vehicle is based on receding-horizon optimization, i.e., once an acceleration sequence $\gamma_i(t) = \{a_i(\tau|t)\}_{\tau=0}^{\mathcal{N}-1}$ is determined, the ego vehicle $i$ applies the first acceleration value $a_i(0|t)$ for one time step, i.e., $a_i(t) = a_i(0|t)$, then repeats the decision-making process \eqref{equ:leader_follower_n_1} at the next time instant.

\section{Additional modeling considerations}\label{sec:additions}

To model the interactive behavior of human-driven vehicles with higher fidelity, we incorporate several additional considerations in our model. They are discussed in this section.

\subsection{Courteous driving}

A vehicle is not supposed to interrupt other vehicles' nominal drives. More specifically, a vehicle is not supposed to intentionally choose an action that would cause a collision when other vehicles maintain their speeds. We account for this by adjusting the action set for each vehicle, i.e., modify the decision-making model \eqref{equ:leader_follower_n_1} according to
\begin{equation}\label{equ:leader_follower_n_2}
\gamma_{i}^*(t) \in \argmax_{\gamma_i \in \Gamma_i(t)}\, \underline{\mathbb{Q}}_{i}({\bf s}_{\text{traffic}}(t),\gamma_{i}),
\end{equation}
where $\Gamma_i(t) \subset \Gamma = A^\mathcal{N}$ is defined as
\begin{align}\label{equ:courteous}
\Gamma_i(t) &:= A_i(t) \times A^{\mathcal{N}-1}, \nonumber \\
A_i(t) &:= \Big\{a_i(t) \in A \,\big|\, a_i(t) \text{ satisfies either } \forall j \neq i, a_j(t) = 0 \nonumber \\
&\, \implies S_c\big({\bf s}_{i,j}(t+1)\big) = S_c\big({\bf s}_{i,j}(t),a_i(t),a_j(t)\big) = 0, \nonumber \\
&\,\,\, \text{ or } a_i(t) = \min \{a \in A\} \Big\}.
\end{align}

\subsection{Limited perception ranges}\label{sec:ranges}

Human drivers have limited ranges of visual perception. To account for this, we assume that a driver only considers his/her interactions with the other vehicles that are in a certain vicinity of his/her own. In particular, we further modify the decision-making model \eqref{equ:leader_follower_n_1} according to
\begin{equation}\label{equ:leader_follower_n_3}
\underline{\mathbb{Q}}_{i}({\bf s}_{\text{traffic}}(t),\gamma_{i}) = \min_{j \in \Omega_i(t)} \mathbb{Q}_{i,j}({\bf s}_{i,j}(t),\gamma_{i}),
\end{equation}
where $\Omega_i(t) \subset \{1,\cdots,n\}$ is defined as
\begin{align}
\Omega_i(t) &:= \Big\{j \in \{1,\cdots,n\} \,\big|\, j \neq i \text{ and} \nonumber \\
&\, \sqrt{\big(x_j(t)-x_i(t)\big)^2+\big(y_j(t)-y_i(t)\big)^2} \le \omega_i \Big\},
\end{align}
with $\omega_i > 0$ representing vehicle $i$'s maximum perception distance.

We also note that such a modification decreases the computational complexity of our model to a greater extent by reducing the number of interacting vehicle pairs $(i,j)$ in the decision-making process \eqref{equ:leader_follower_n_1} of each vehicle, and as a result, further improves the scalability of our modeling framework.

\subsection{Breakage of deadlocks via exploratory actions}\label{sec:randomness}

As discussed at the end of Section~\ref{sec:leader_follower}, in some scenarios cyclic patterns, such as $i \prec j$, $j \prec k$, $k \prec l$, and $l \prec i$, may occur and lead to deadlocks -- no one decides to enter the intersection or everyone gets stuck in the middle of the intersection.

Such cyclic patterns also exist in real-world traffic. However, human drivers can usually break a deadlock. When a deadlock occurs, we usually observe that one or more human drivers will probe the possibility of going first and such probes can often help the drivers reach an agreement on their orders of passing through the intersection. On the basis of such an observation, we propose a strategy to break deadlocks via random exploratory actions, which is presented as Algorithm~\ref{alg:Break_D}.

\begin{algorithm}
    \caption{Breakage of deadlocks via exploratory actions}
    \label{alg:Break_D}
    \SetKwInOut{Input}{Input}
    \SetKwInOut{Output}{Output}
    \Input{the states of all vehicles ${\bf s}_{\text{traffic}}(t) = \big(s_1(t),\cdots,s_n(t)\big)$ and the acceleration choices of all vehicles $\big(a_1(t),\cdots,a_n(t)\big)$ obtained based on \eqref{equ:leader_follower_n_1}}
    \Output{the modified acceleration choices of all vehicles $\big(a_1(t),\cdots,a_n(t)\big)$}
    $\Omega_{\text{conflict}} = \text{Null}$; \\
    \For{$i = 1,\cdots,n$}{
        \lIf{$i$ is the first vehicle coming from its origin lane that has not exited the intersection ($\Delta \rho_i^{\text{ex}}(t) > 0$)}{add $i$ to $\Omega_{\text{conflict}}$}}
    \If{$v_i(t) = 0$ and $a_i(t) = 0,\, \forall i \in \Omega_{\rm{conflict}}$}{
        \For{$i \in \Omega_{\rm{conflict}}$}{\If{$\{a \in A_i(t) \,|\, a > 0\} \neq \emptyset$}{reset $a_i(t)$ based on
            $
            a_i(t) =
                \begin{cases}
                \min \{a \in A_i(t) \,|\, a > 0\}, & \text{with prob.} = p_i, \\
                0, &\!\!\!\!\!\!\!\!\!\! \text{with prob.} = 1-p_i.
                \end{cases}
            $ \\
            }}}
\end{algorithm}

In Algorithm~\ref{alg:Break_D}, lines~2-4 are used to identify the vehicles that are in conflict. For example, vehicle $i$ that has exited the intersection is not a vehicle in conflict. As another example, if there is a vehicle $j$ that is entering/has entered the intersection from the same lane as $i$, drives in front of $i$\footnote{There shall be no ambiguity in ``in front of'' here since $i$ and $j$ are entering/have entered the intersection from the same lane.}, and has not exited the intersection, then vehicle $i$ is not a vehicle in conflict. Line~5 is used to identify the occurrence of a deadlock, i.e., all of the vehicles in conflict have stopped and no one decides to move according to decisions of \eqref{equ:leader_follower_n_1}. Then, lines~6-10 assign the vehicles in conflict probabilities of making slight movements to probe the possibility of going first.

The effectiveness of Algorithm~\ref{alg:Break_D} in terms of breaking deadlocks is illustrated through the simulation case studies in Section~\ref{sec:simulations}.

\section{An alternative vehicle interactive decision-making model}\label{sec:levelK}

Human drivers can usually resolve traffic conflicts even when they are interacting with other drivers whose driving styles are a priori unknown. A model representing driver decision-making is supposed to have reasonable robustness against uncertainties in the behavior of interacting drivers.

To illustrate the robustness of decision-making using the proposed model \eqref{equ:leader_follower_n_1} based on pairwise leader-follower games, we also consider another interactive decision-making model, which is based on level-$\mathcal{K}$ game theory \cite{nagel1995unraveling,stahl1995players} and has been developed in \cite{li2018game}. We simulate the interactions between model \eqref{equ:leader_follower_n_1} and this alternative model in the simulation case studies in Section~\ref{sec:simulations} and compare them.

The level-$\mathcal{K}$ model is premised on the idea that each strategic agent in a non-cooperative multi-agent setting (each vehicle at an uncontrolled intersection, in our setting) makes decisions through a finite number of reasoning steps (called ``levels''), and different agents may have different reasoning levels. The reasoning process starts from level-$0$. Level-$0$ typically represents an agent's decisions of minimal rationality, e.g., instinctive decisions, to pursue its goals without strategically accounting for its interactions with the other agents. On the contrary, level-$\mathcal{K}$, $\mathcal{K} \ge 1$, represents an agent's decisions optimally responding to the level-($\mathcal{K}-1$) decisions of the other agents. In particular, once the level-$0$ decisions, either as time series or as a function of a vehicle's own state and its interacting vehicles' states (denoted respectively by $s_i$ and $\mathbf{s}_{-i} = (s_j)_{j \neq i}$ for vehicle $i$), are formulated, the corresponding level-$\mathcal{K}$, $\mathcal{K} \ge 1$, decisions can be computed through solving the optimization problems
\begin{equation}\label{equ:levelK}
    \gamma_i^{\mathcal{K}} \in \argmax_{\gamma_i \in \Gamma_i}\, \mathbb{R}_i(s_i,\mathbf{s}_{-i},\gamma_i,{\bm \gamma}_{-i}^{\mathcal{K}-1})
\end{equation}
sequentially for $\mathcal{K} = 1,2,\cdots$, where ${\bm \gamma}_{-i}^{\mathcal{K}-1} = (\gamma_j^{\mathcal{K}-1})_{j \neq i}$ denotes the level-($\mathcal{K}-1$) decisions of the vehicles interacting with vehicle $i$, and $\gamma_j^0 = \gamma_j^0(s_j,\mathbf{s}_{-j})$ is the formulated level-$0$ decision of vehicle $j$.

To make this level-$\mathcal{K}$ reasoning and decision-making process more explicit, we present it for the case of $2$-agent interactions formally as Algorithm~\ref{alg:levelK_2agents}. Algorithm~\ref{alg:levelK_2agents} can be generalized to the case of $n$-agent interactions straightforwardly.

\begin{algorithm}
    \caption{Level-$\mathcal{K}$ decision-making process in $2$-agent interactions}
    \label{alg:levelK_2agents}
    \SetKwInOut{Input}{Input}
    \SetKwInOut{Output}{Output}
    \Input{the states of the ego agent $s_1$ and the other agent $s_2$, the level of the ego agent $\mathcal{K}$, a level-$0$ decision rule}
    \Output{the level-$\mathcal{K}$ decision of the ego agent}
    $k = 1$; \\
    \eIf{$\text{mod}(\mathcal{K},2)=0$}{\While{$k < \mathcal{K}$}{
        $\gamma_2^{k} \in \argmax_{\gamma_2 \in \Gamma_2} \mathbb{R}_2(s_1,s_2,\gamma_1^{k-1},\gamma_2)$; \\
        $\gamma_1^{k+1} \in \argmax_{\gamma_1 \in \Gamma_1} \mathbb{R}_1(s_1,s_2,\gamma_1,\gamma_2^{k})$;
        $k \leftarrow k+2$;
    }}{\While{$k < \mathcal{K}-1$}{
        $\gamma_1^{k} \in \argmax_{\gamma_1 \in \Gamma_1} \mathbb{R}_1(s_1,s_2,\gamma_1,\gamma_2^{k-1})$; \\
        $\gamma_2^{k+1} \in \argmax_{\gamma_2 \in \Gamma_2} \mathbb{R}_2(s_1,s_2,\gamma_1^{k},\gamma_2)$;
        $k \leftarrow k+2$;
    }
    $\gamma_1^{\mathcal{K}} \in \argmax_{\gamma_1 \in \Gamma_1} \mathbb{R}_1(s_1,s_2,\gamma_1,\gamma_2^{\mathcal{K}-1})$;
    }
    output $\gamma_1^{\mathcal{K}}$.
\end{algorithm}

In this paper, the state $s_i(t)$, the action $\gamma_i(t) = \{a_i(\tau|t)\}_{\tau=0}^{\mathcal{N}-1}$, the kinematics model relating the predicted states $s_i(\tau|t)$ to the action $\gamma_i(t)$, and the reward function $\mathbb{R}_i(t)$ of vehicle $i$ are defined in the same way as in the leader-follower game based decision-making model \eqref{equ:leader_follower_n_1} (see Section~\ref{sec:kinematics_rewards}). In particular, since there are no leader-follower relationships defined in the scheme of level-$\mathcal{K}$ models, when computing its rewards, a level-$\mathcal{K}$ vehicle assumes that the $s$-zones of all vehicles (including itself and all other vehicles in interaction) are of the same size, denoted by $(l_{s,\text{f}}^\mathcal{K},l_{s,\text{r}}^\mathcal{K},w_s^\mathcal{K})$.

In this paper, a level-$0$ decision of vehicle $i$ is to maximize the reward $\mathbb{R}_i(t)$ while treating the other vehicles as stationary obstacles, i.e., $v_j(\tau|t) = 0$ for all $j \neq i$ and $\tau = 0,\cdots, \mathcal{N}-1$. In this setting, a level-$0$ vehicle may represent an aggressive driver in real-world traffic who assumes that the other drivers will yield the right of way.

Note that a level-$\mathcal{K}$ agent optimally responds to level-($\mathcal{K}-1$) interacting agents. When the levels of interacting agents are not known a priori, online estimation and adaptation can be incorporated in decision-making. In particular, we consider the following decision-making model \cite{li2018game}:
\begin{align}\label{equ:model_D}
    & \gamma_i^{\mathfrak{K}} \in \argmax_{\gamma_i \in \Gamma_i} \\
    &\! \sum_{\substack{k_j \in \{0,\cdots,k_{\max}\}, \\ j \neq i }}\!\! \bigg[ \Big(\prod_{j \neq i} \mathbb{P}(\mathcal{K}_j = k_j) \Big) \mathbb{R}_i \Big(s_i,\mathbf{s}_{-i},\gamma_i,(\gamma_j^{k_j})_{j \neq i}\Big) \bigg], \nonumber
\end{align}
where $\mathbb{P}(\mathcal{K}_j = k_j)$ represents vehicle $i$'s belief in that vehicle $j$ can be modeled as level-$k_j$, which is updated after each time step $t$ according to the following algorithm: For each $j \neq i$, if there exist $k_j, k_j' \in \{0,\cdots,k_{\max}\}$ such that $a_j^{k_j}(0|t) \neq a_j^{k_j'}(0|t)$, then
\begin{align}\label{equ:model_estimation}
    & \tilde{k}_j \in \argmin_{k_j \in \{0,\cdots,k_{\max}\}} \big|a_j^{k_j}(0|t) - a_j^{\text{actual}}(t)\big|, \nonumber \\
    & \mathbb{P}(\mathcal{K}_j = \tilde{k}_j) \leftarrow \mathbb{P}(\mathcal{K}_j = \tilde{k}_j) + \Delta \mathbb{P}, \nonumber \\
    & \mathbb{P}(\mathcal{K}_j = k_j) \leftarrow \mathbb{P}(\mathcal{K}_j = k_j) \bigg/ \Big(\sum_{k_j' = 0}^{k_{\max}} \mathbb{P}(\mathcal{K}_j = k_j')\Big), \nonumber \\[-2pt]
    & \quad\quad\quad\quad\quad\quad\;\; \forall\, k_j = 0,\cdots,k_{\max};
\end{align}
otherwise, $\mathbb{P}(\mathcal{K}_j = k_j) \leftarrow \mathbb{P}(\mathcal{K}_j = k_j)$ for all $k_j = 0,\cdots,k_{\max}$, where $a_j^{k_j}(0|t)$ is vehicle $j$'s predicted acceleration for time step $t$ that is predicted by vehicle $i$ at the time instant $t$ and using the level-$k_j$ model, and $a_j^{\text{actual}}(t)$ is vehicle $j$'s actual acceleration over time step $t$ that is observed by vehicle $i$ after the time step $t$ is over.

The model estimation algorithm \eqref{equ:model_estimation} is to increase the belief in the level-$\tilde{k}_j$ model whose prediction matches the actual behavior of vehicle $j$ best. It is triggered only when there is at least one predicted value $a_j^{k_j}(0|t)$ of some level-$k_j$ model different from others; otherwise, when the predictions of all models are the same, the ego vehicle has no information to improve its beliefs \cite{li2018game,tian2018adaptive}. The decision-making process \eqref{equ:model_D} selects actions to maximize the weighted sum of rewards corresponding to all possible level combinations of the interacting vehicles where the weights are the ego vehicle's beliefs in each level combination, i.e., to maximize the reward expectation.

Since the decision-making model \eqref{equ:model_D} is developed upon the set of decision-making models \eqref{equ:levelK} induced from level-$\mathcal{K}$ game theory, and it adapts itself to uncertain interacting agents using online model estimation \eqref{equ:model_estimation}, we refer to it as an ``adaptive level-$\mathcal{K}$'' decision-making model or model $\mathfrak{K}$.

\section{Simulation case studies}\label{sec:simulations}

In this section, we illustrate our game-theoretic framework to model vehicle interactions at uncontrolled intersections through multiple simulation case studies. The values of parameters used in all of the reported simulation results are collected in Table~\ref{tab:para} in Appendix~B.

\subsection{Case study 1: Reproducing real-world traffic scenarios}

\begin{figure}[h!]
\begin{center}
\begin{picture}(204.0, 310)
\put(  0,  226){\epsfig{file=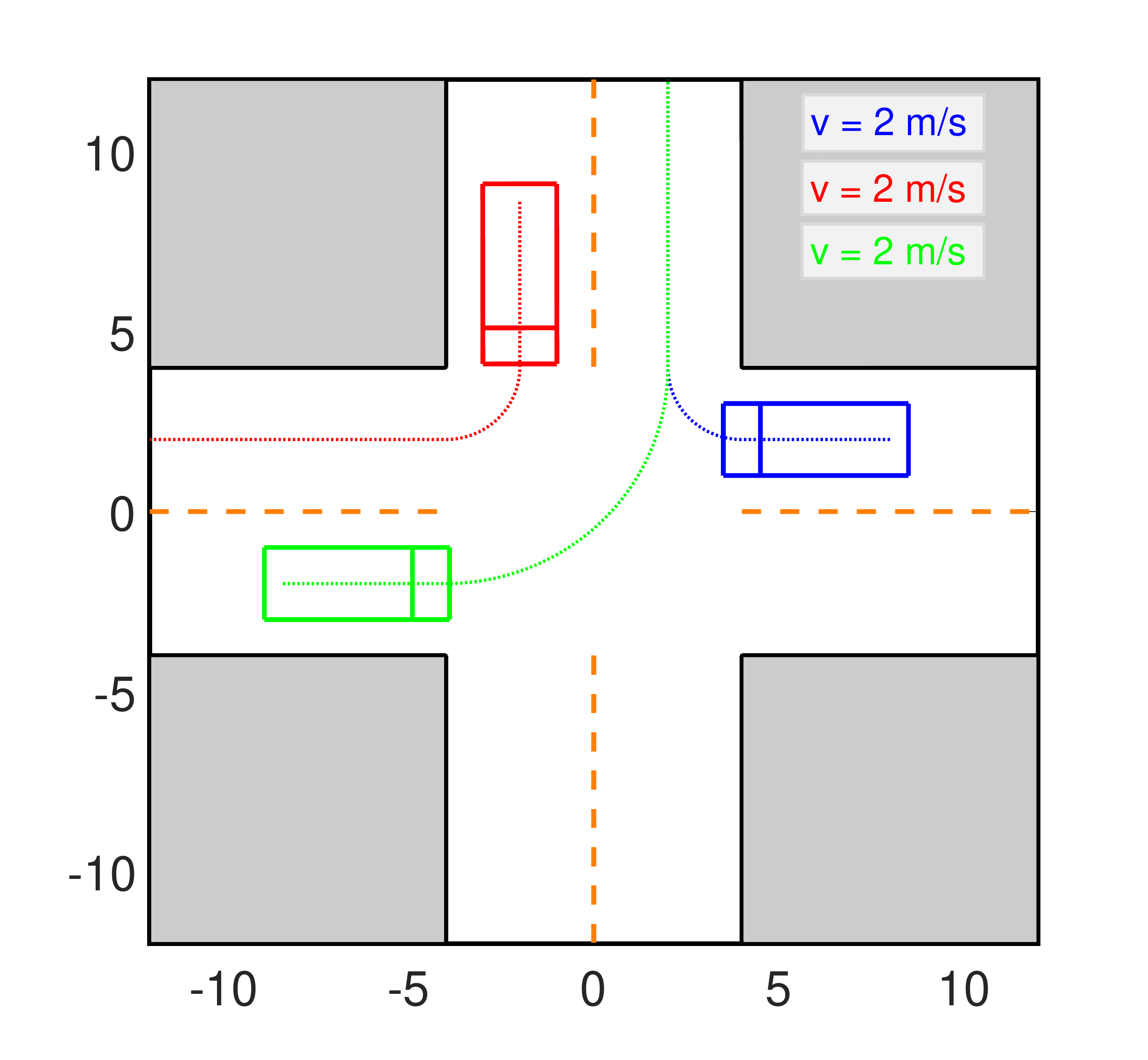,height=1.1in}}  
\put(  98,  236){\epsfig{file=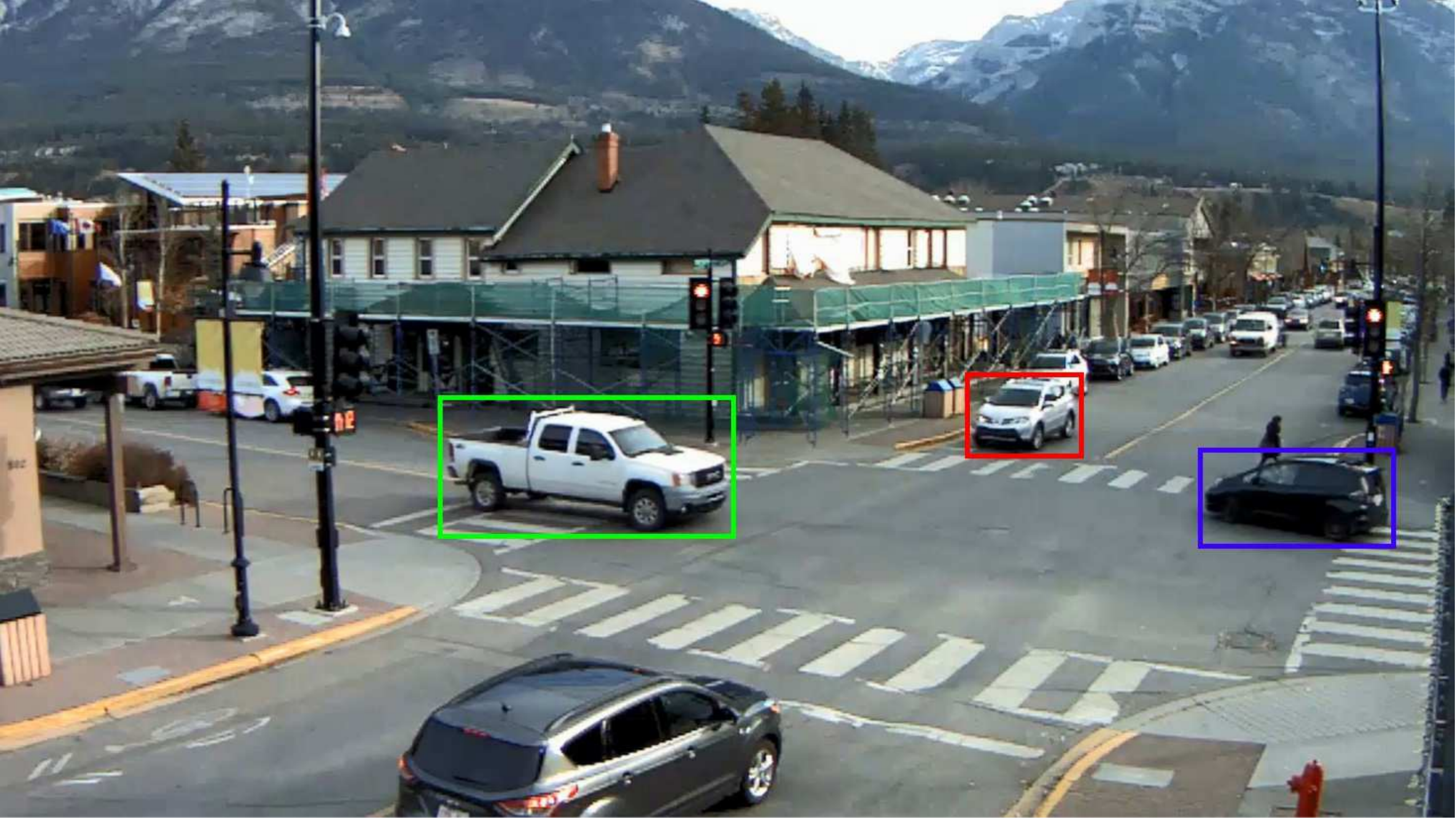,height=0.85in}}  
\put(  0,  149){\epsfig{file=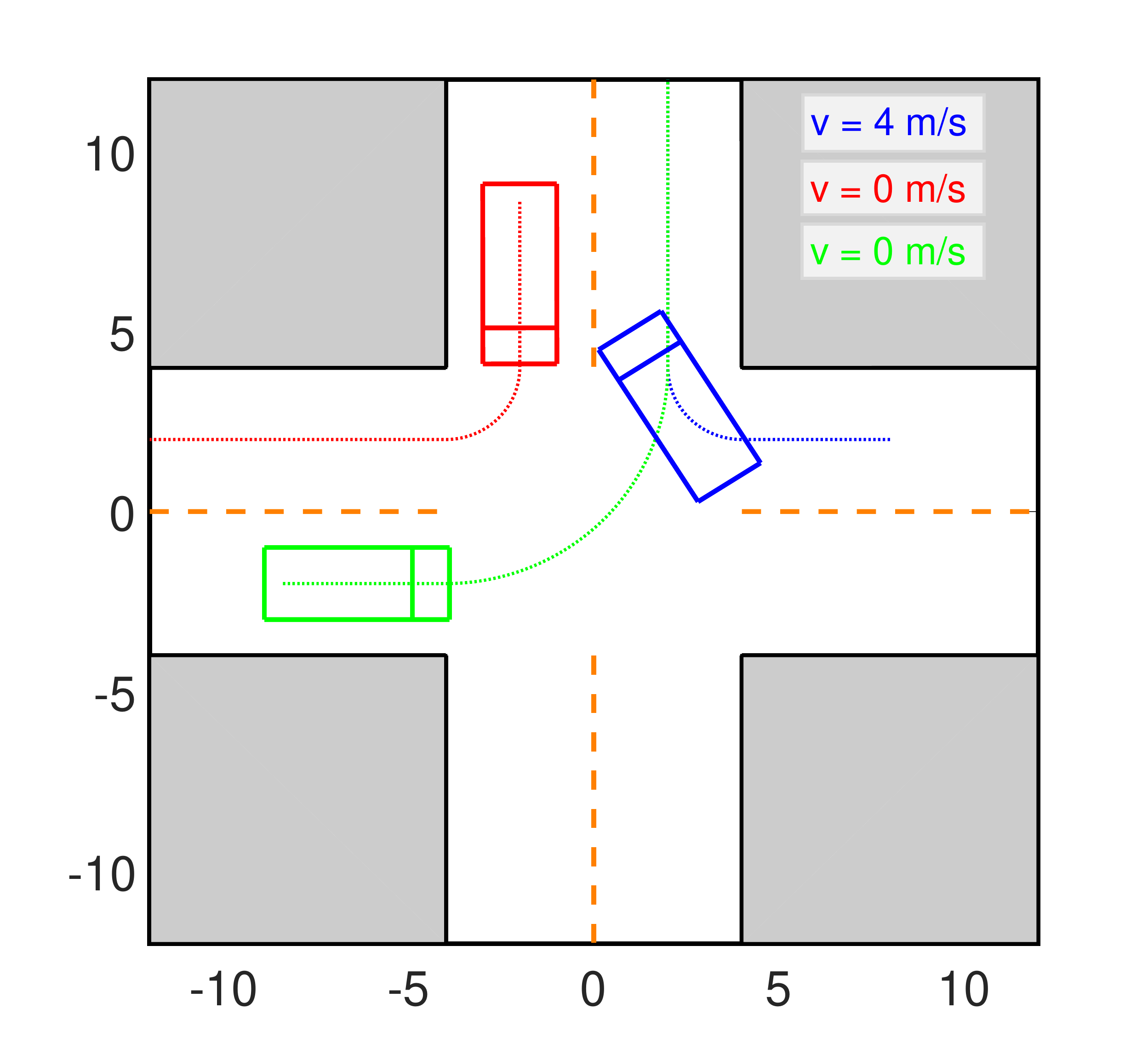,height=1.1in}}  
\put(  98,  159){\epsfig{file=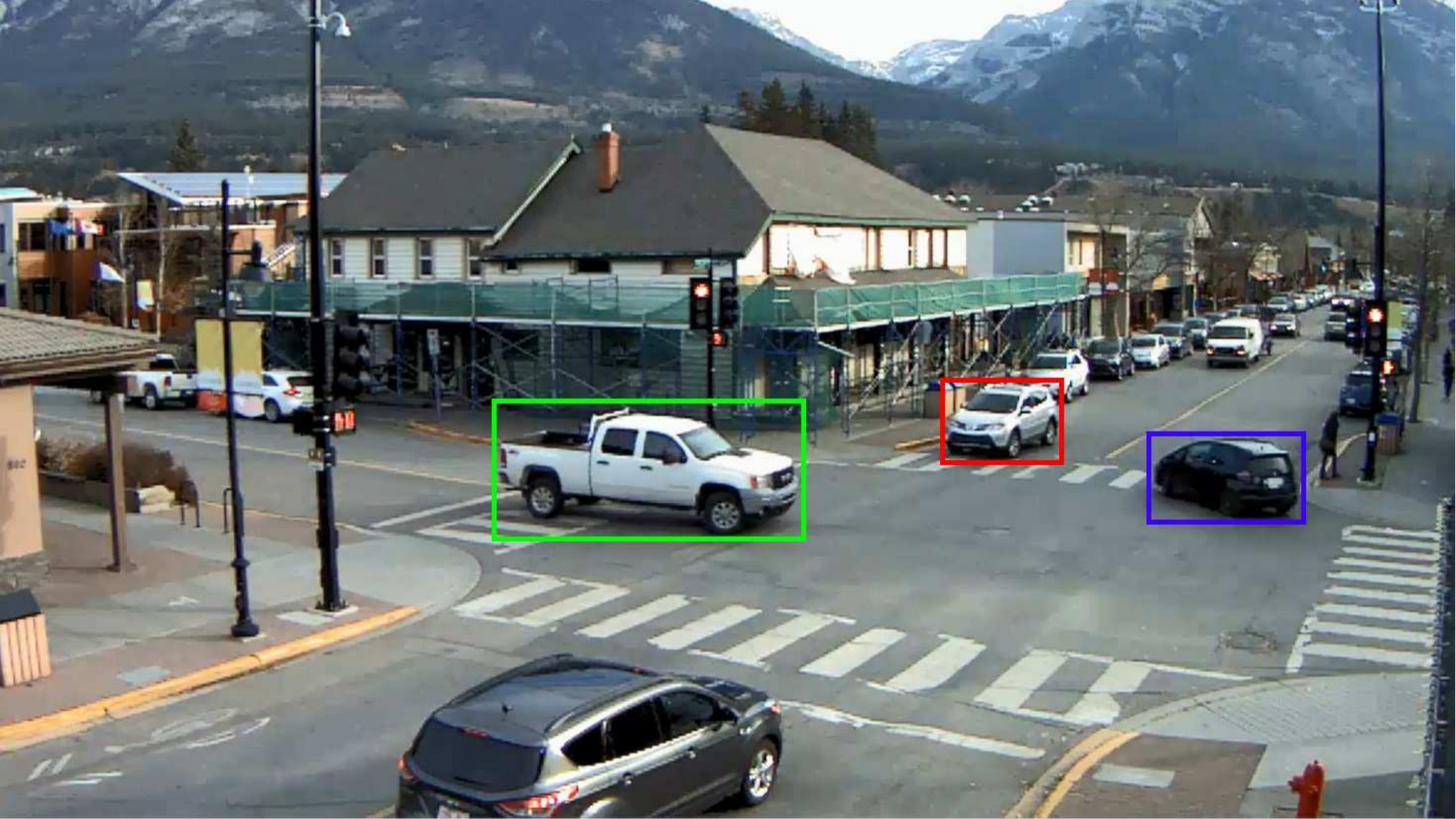,height=0.85in}}  
\put(  0,  72){\epsfig{file=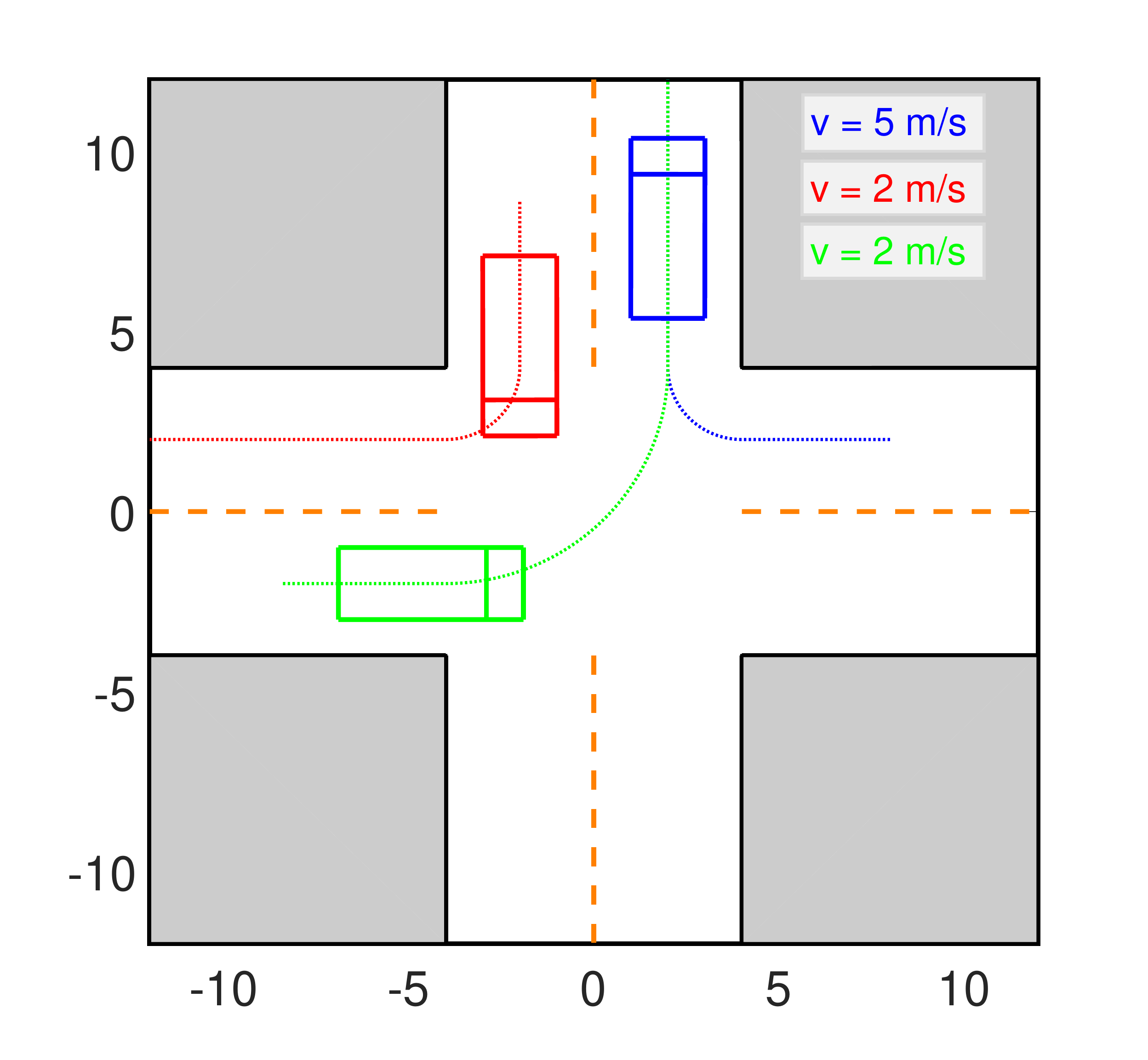,height=1.1in}}  
\put(  98,  82){\epsfig{file=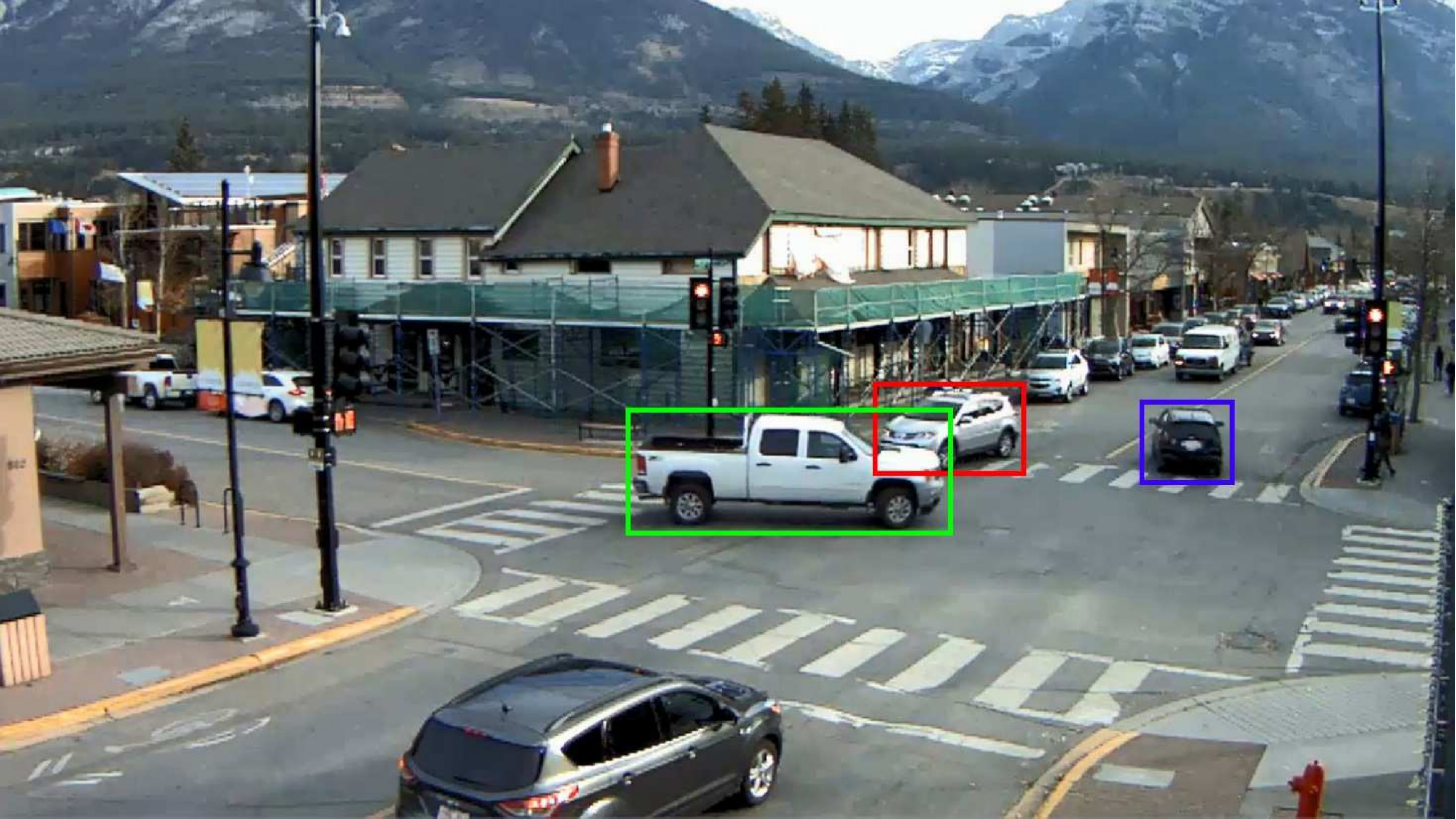,height=0.85in}}  
\put(  0,  -5){\epsfig{file=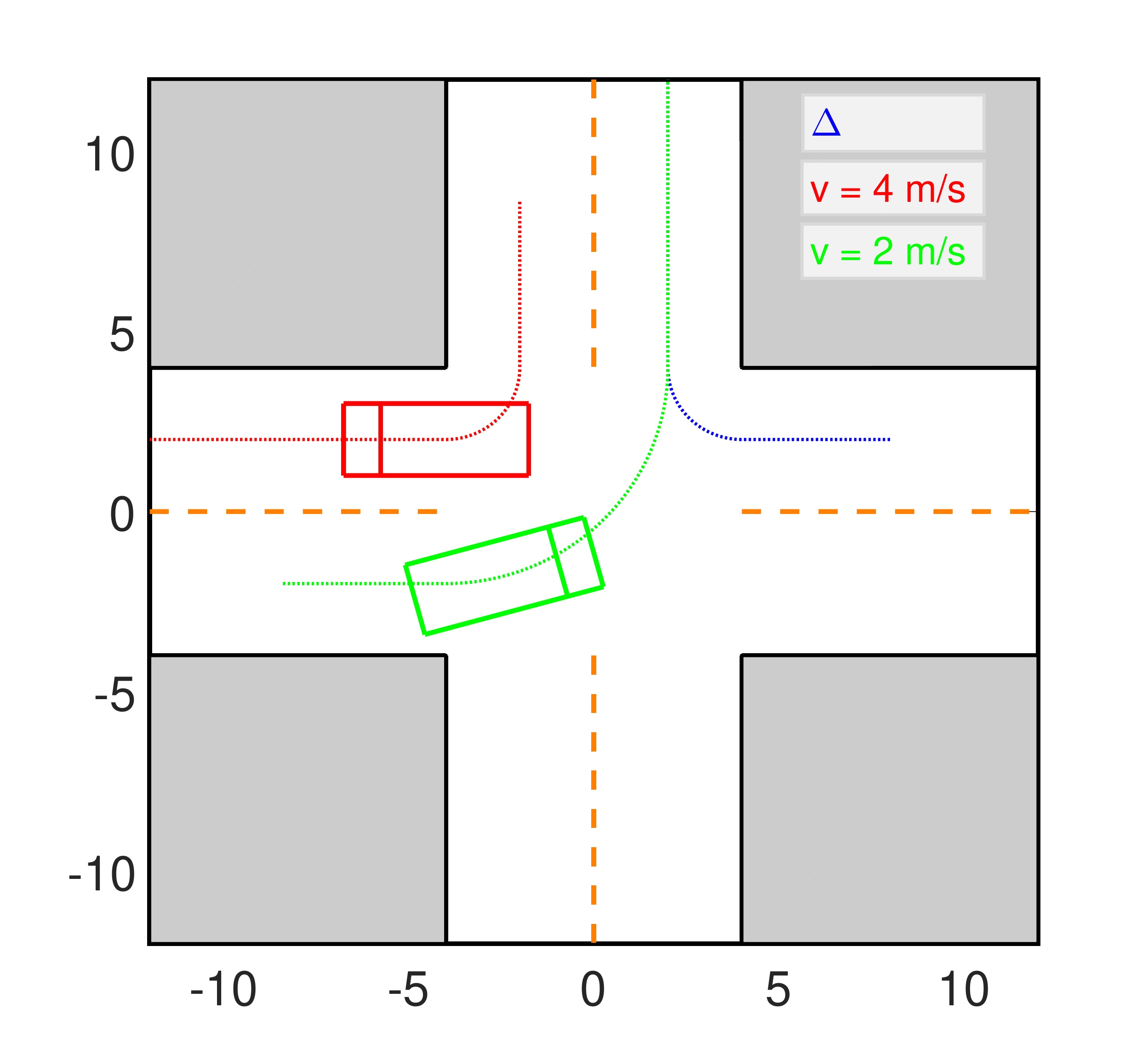,height=1.1in}}  
\put(  98,  5){\epsfig{file=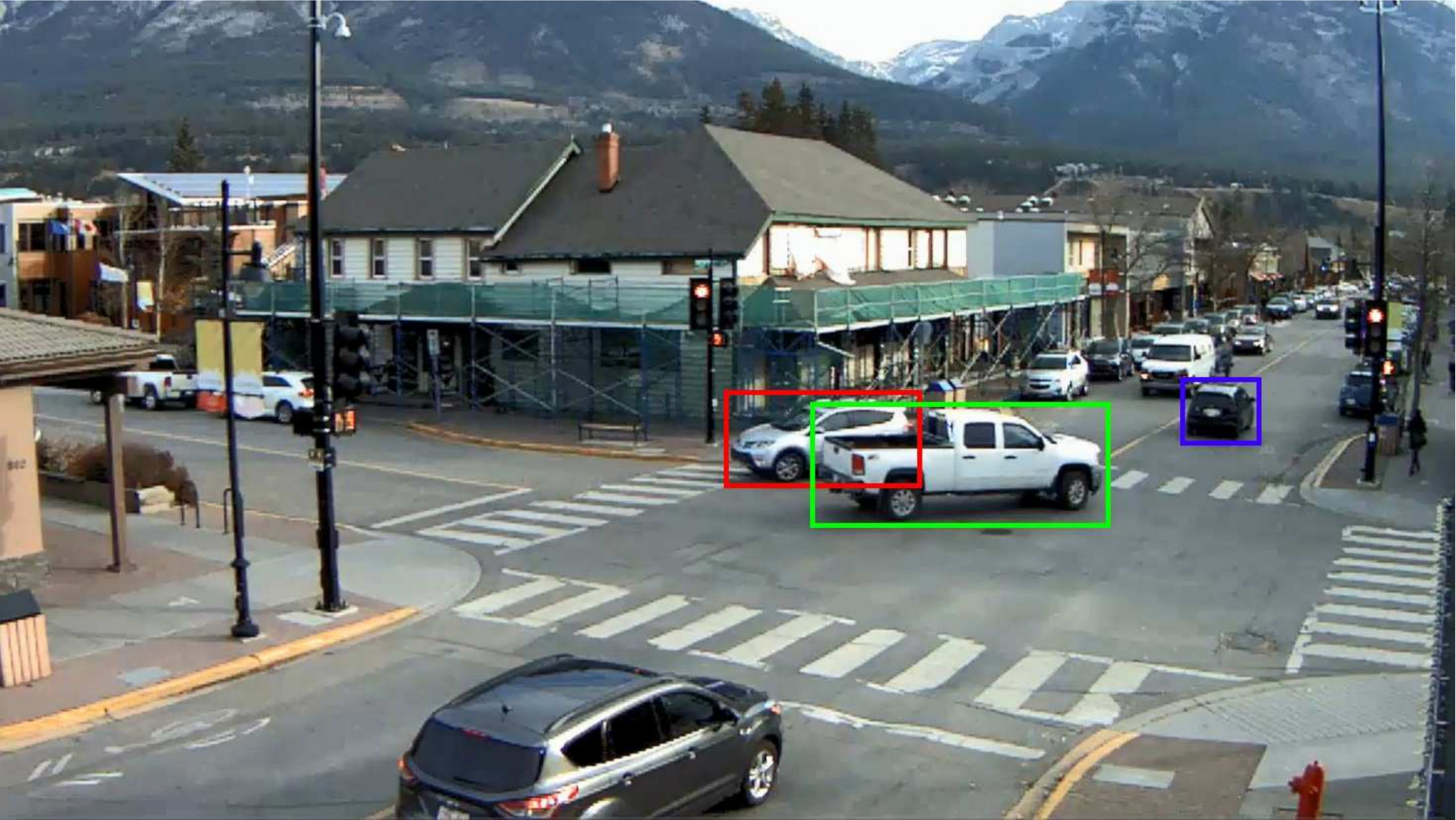,height=0.85in}}  
\small
\put(195, 301){(a)}
\put(195, 224){(b)}
\put(195, 147){(c)}
\put(195, 70){(d)}
\normalsize
\end{picture}
\end{center}
      \caption{Real-world traffic scenario with 3 interacting vehicles.}
      \label{fig:real_1}
\end{figure}
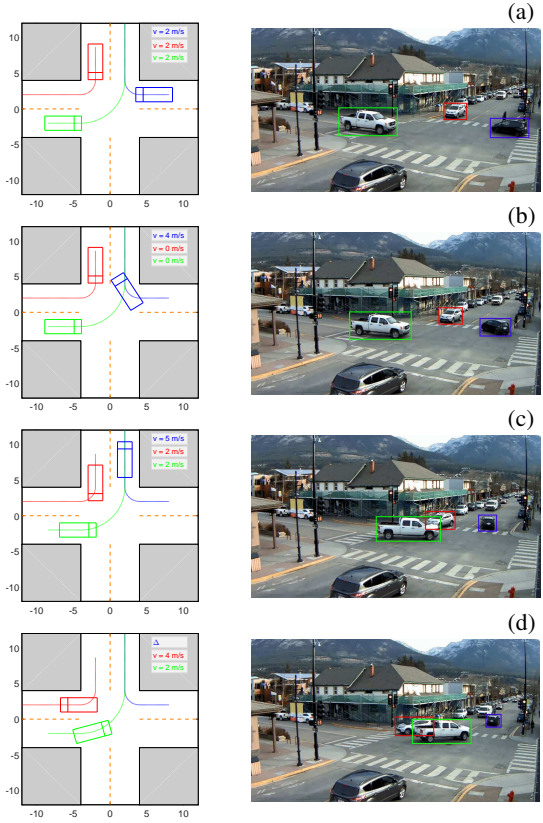

\begin{figure}[h!]
\begin{center}
\begin{picture}(204.0, 235)
\put(  0,  149){\epsfig{file=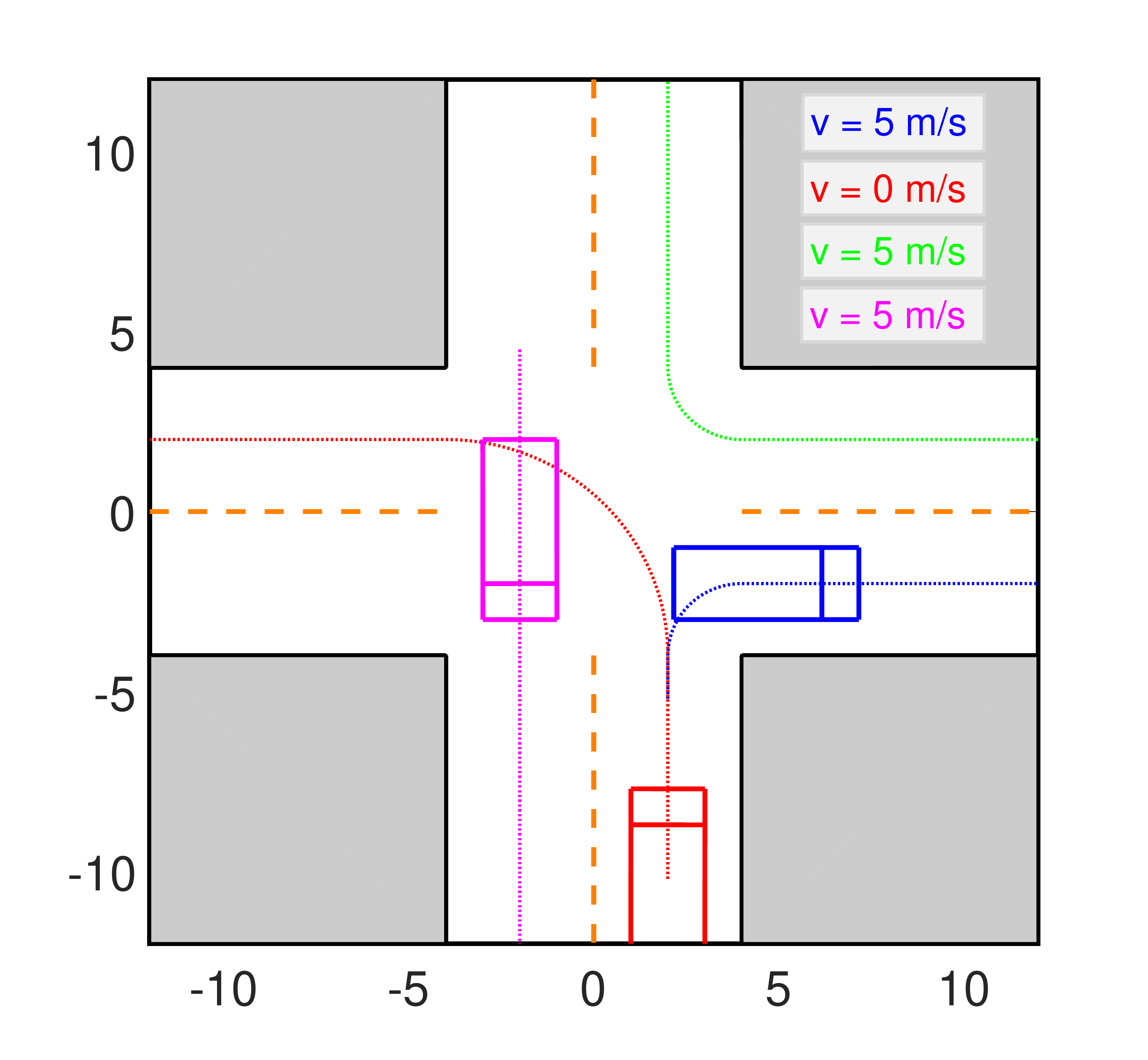,height=1.1in}}  
\put(  98,  159){\epsfig{file=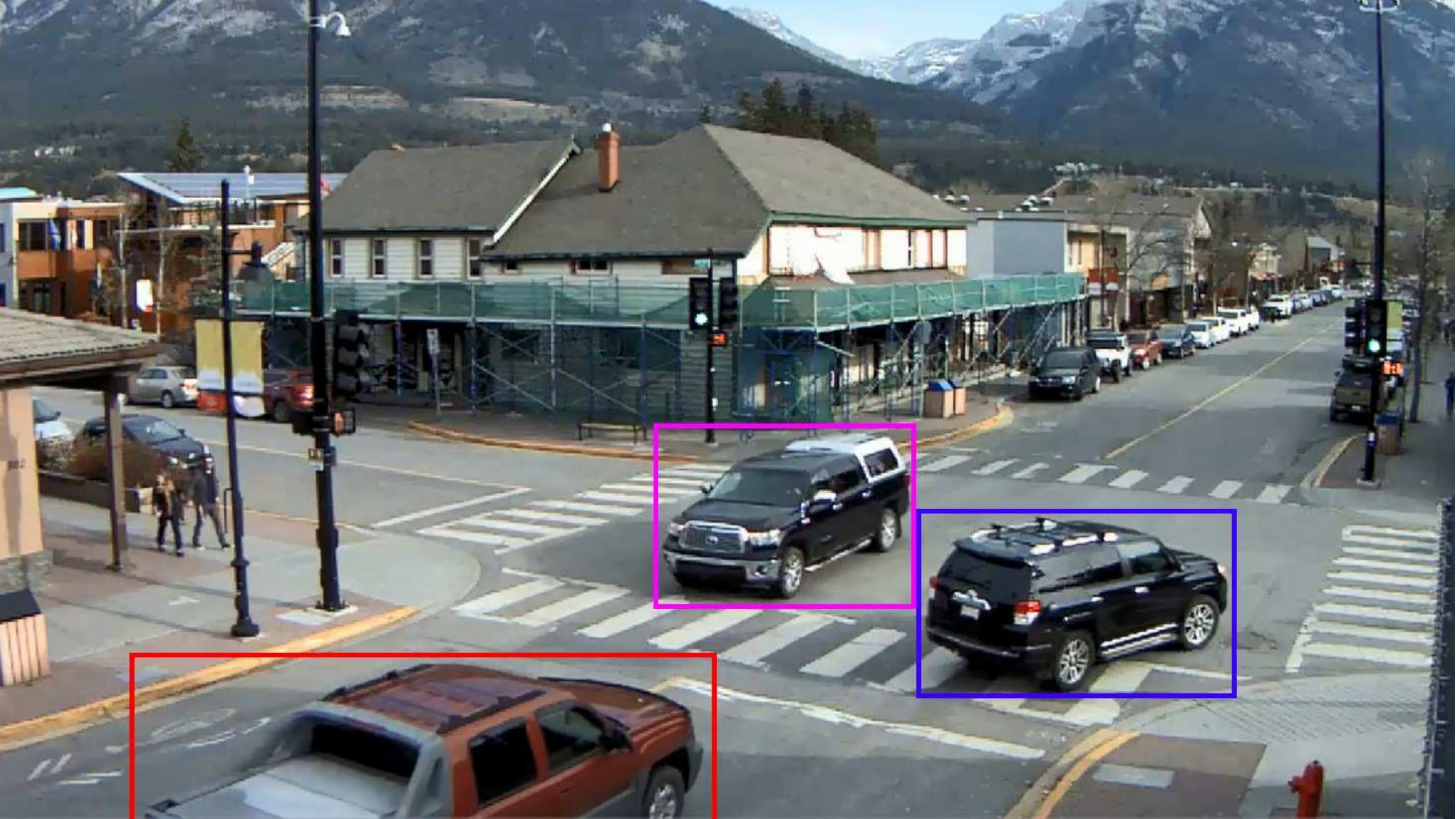,height=0.85in}}  
\put(  0,  72){\epsfig{file=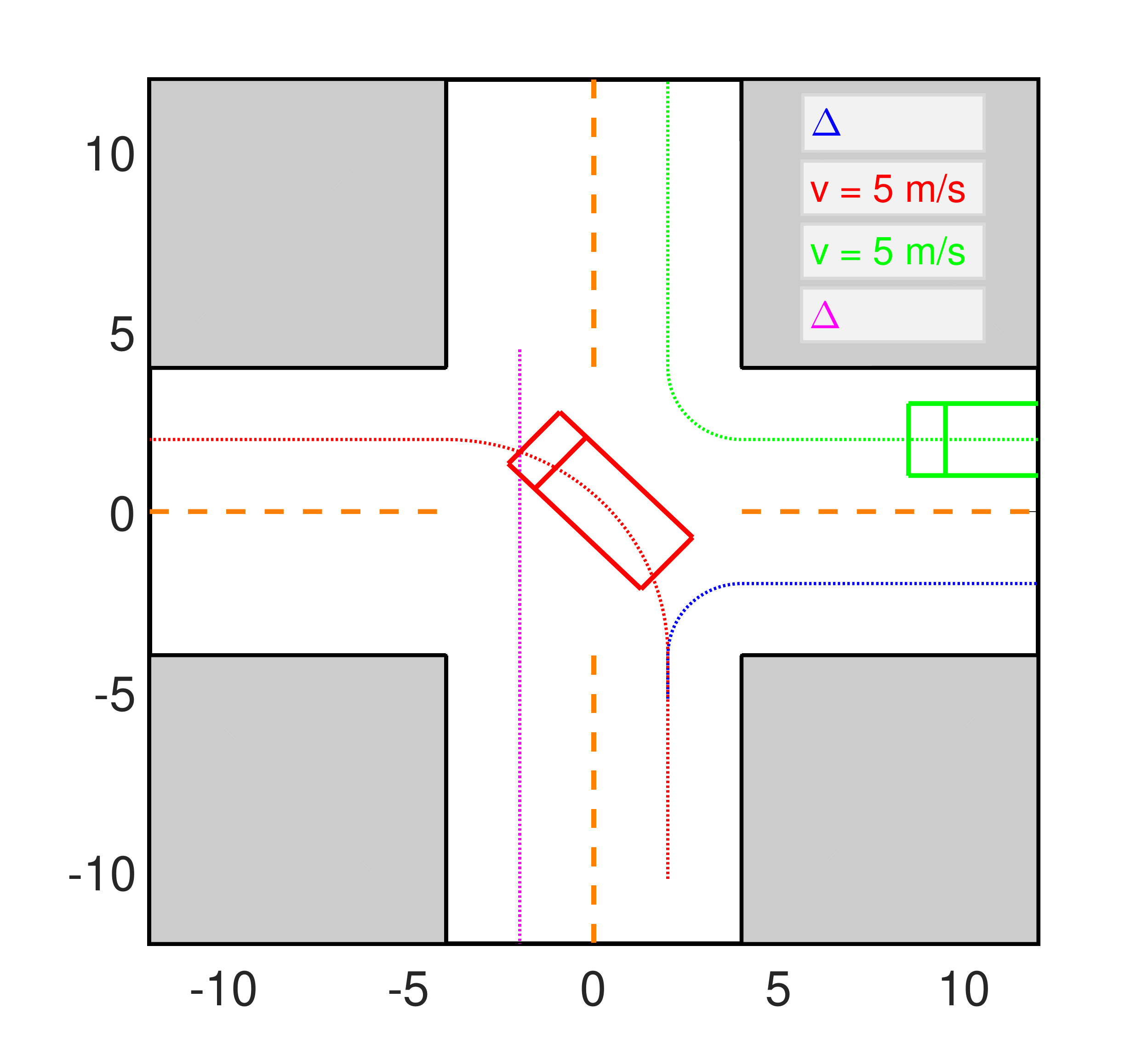,height=1.1in}}  
\put(  98,  82){\epsfig{file=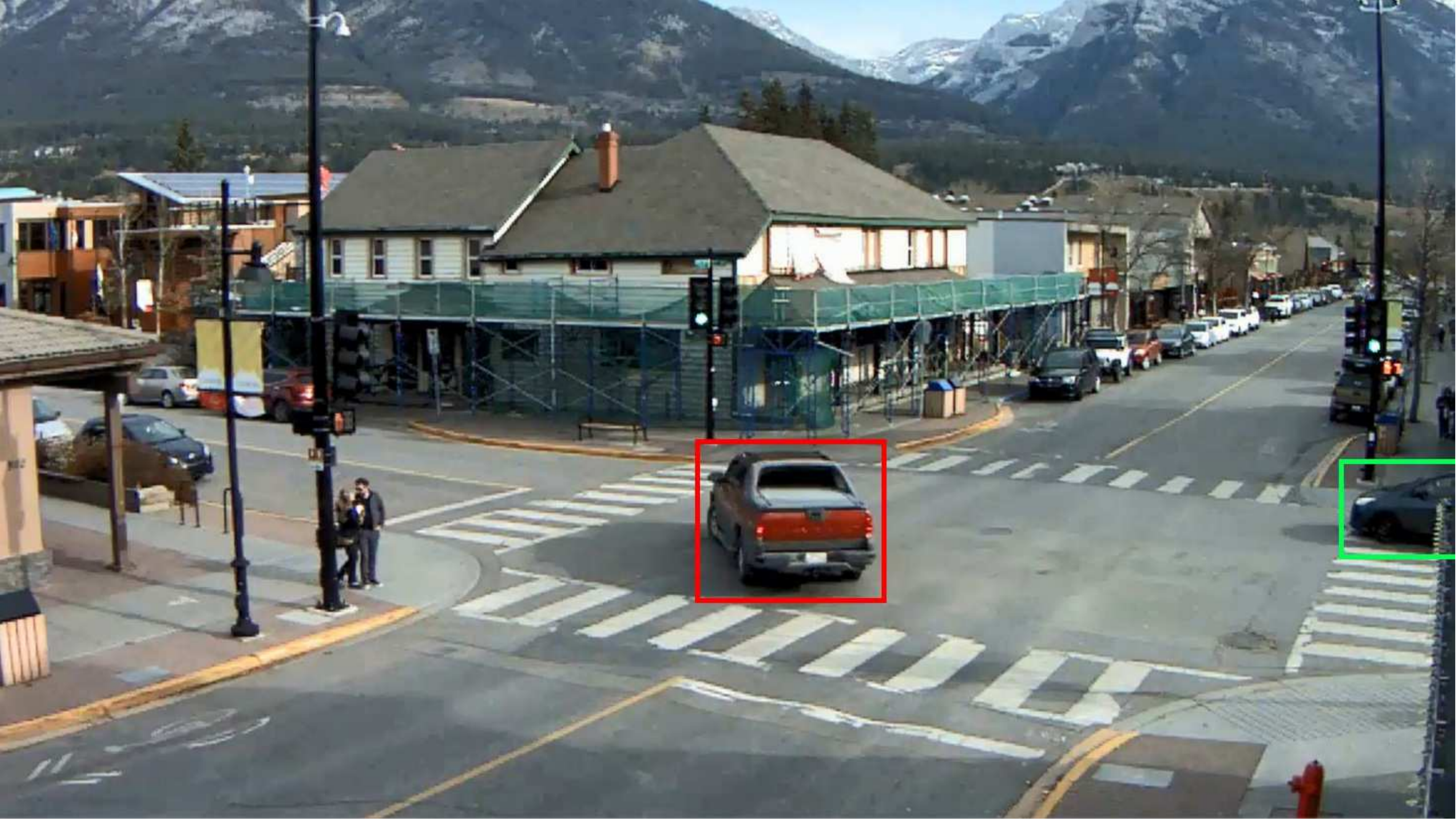,height=0.85in}}  
\put(  0,  -5){\epsfig{file=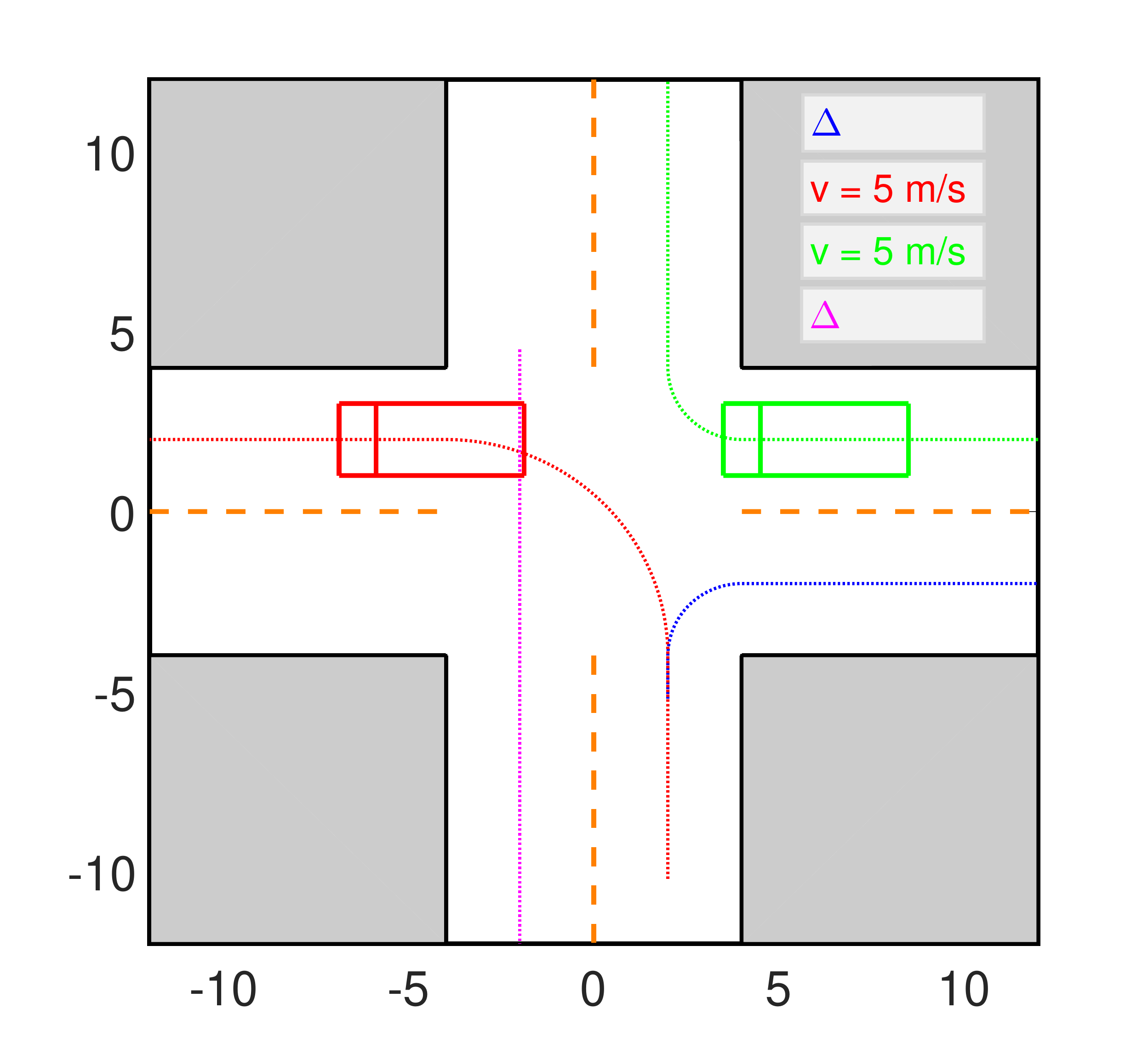,height=1.1in}}  
\put(  98,  5){\epsfig{file=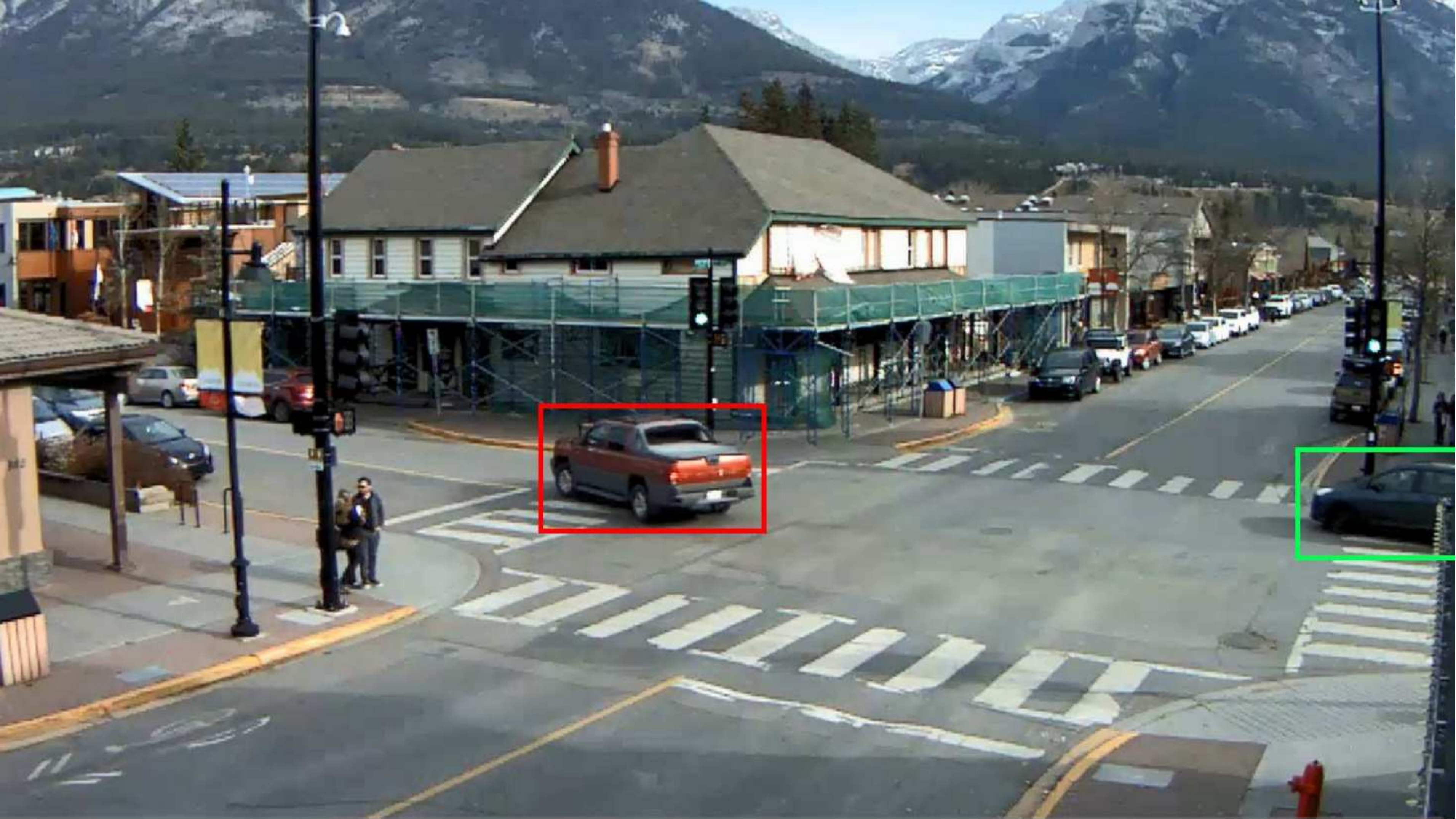,height=0.85in}}  
\small
\put(195, 224){(a)}
\put(195, 147){(b)}
\put(195, 70){(c)}
\normalsize
\end{picture}
\end{center}
      \caption{Real-world traffic scenario with 4 interacting vehicles.}
      \label{fig:real_2}
\end{figure}
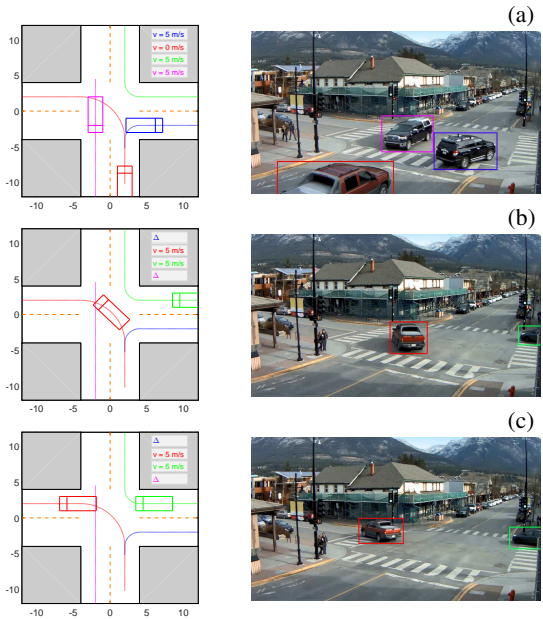

We first show that our proposed vehicle interaction model can reproduce real-world traffic scenarios. The scenarios we consider are extracted from the video dataset used in \cite{ren2018learning}. We note that although the traffic data are collected at a signalized intersection in Canmore, Alberta, we consider only the vehicles that are legally allowed to enter the intersection (i.e., under a green light or making a right turn), the behavior of which can be modeled similarly to that at uncontrolled intersections. In particular, we initialize the vehicles in our simulation based on the positions and velocities of the corresponding vehicles in the video, and compare the evolution of the scenario simulated using our model to that provided by the video. The results of a scenario involving 3 interacting vehicles are shown in Fig.~\ref{fig:real_1} and those of a scenario involving 4 interacting vehicles are shown in Fig.~\ref{fig:real_2}. It can be observed that our model reproduces the scenarios with satisfactory accuracy.

\subsection{Case study 2: Completely symmetric}

As discussed at the end of Section~\ref{sec:leader_follower} and at the beginning of Section~\ref{sec:randomness}, one type of challenging scenarios at uncontrolled intersections for both human drivers and autonomous vehicles are scenarios where no one has a determinable role of the leader. Among these scenarios, the ones where all the vehicles arrive at the entrances of a geometrically symmetrical intersection at the same time with the same speed may be particularly challenging. In this section, we show the simulation results of two such ``completely symmetric'' cases.

\begin{figure}[h!]
\begin{center}
\begin{picture}(234.0, 312.0)
\put(  0,  206){\epsfig{file=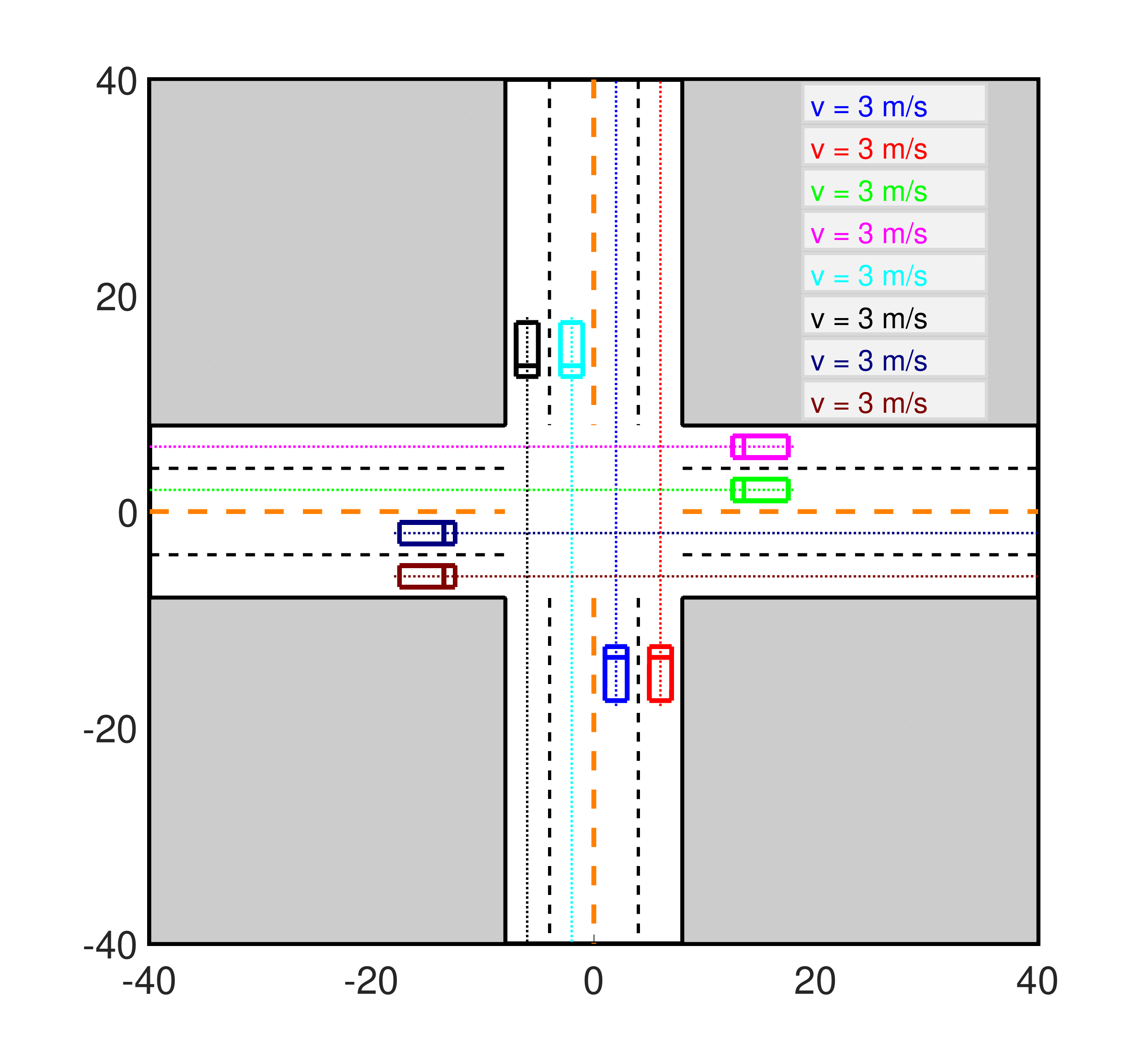,height=1.52in}}  
\put(  112,  206){\epsfig{file=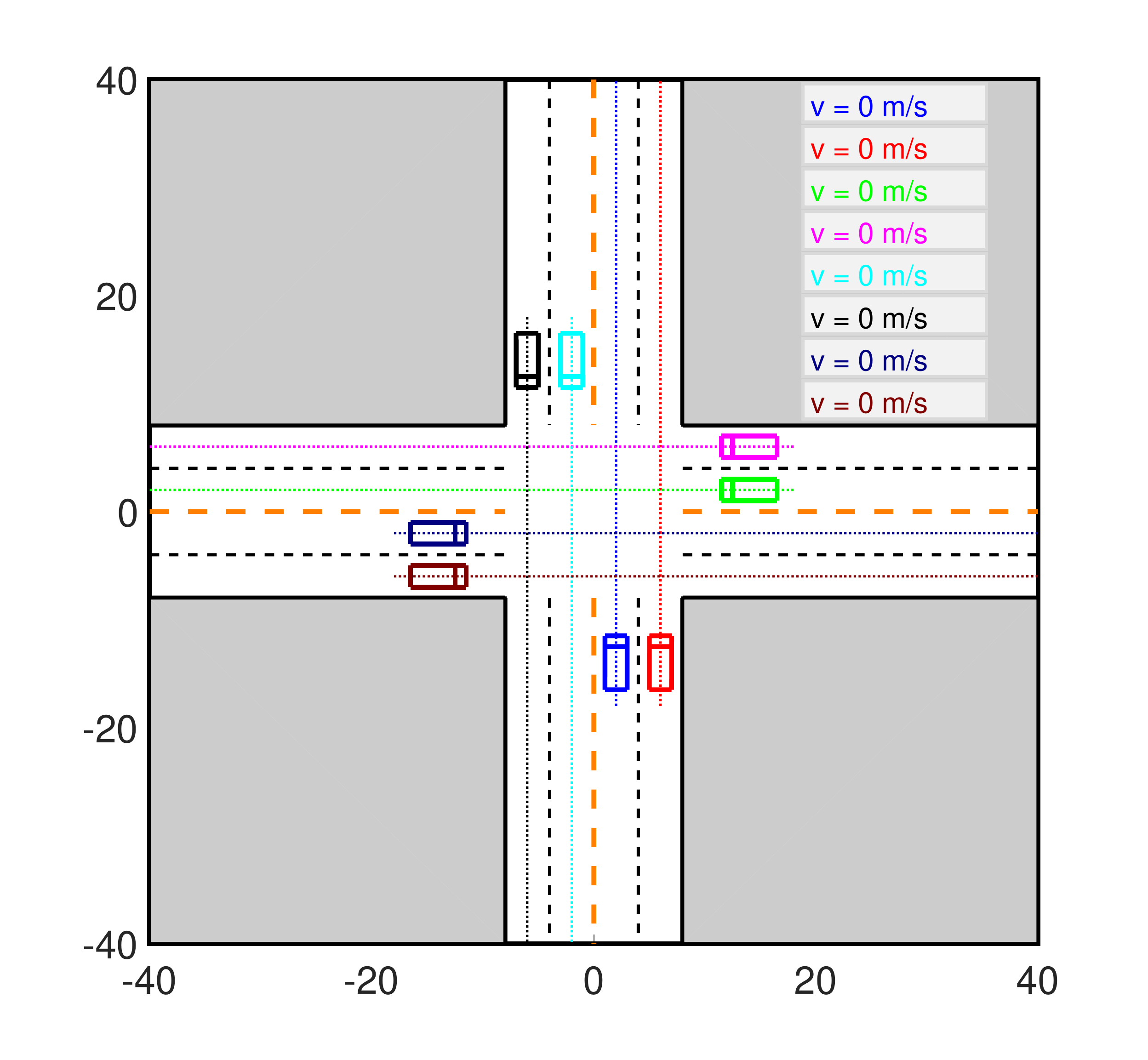,height=1.52in}}  
\put(  0,  100){\epsfig{file=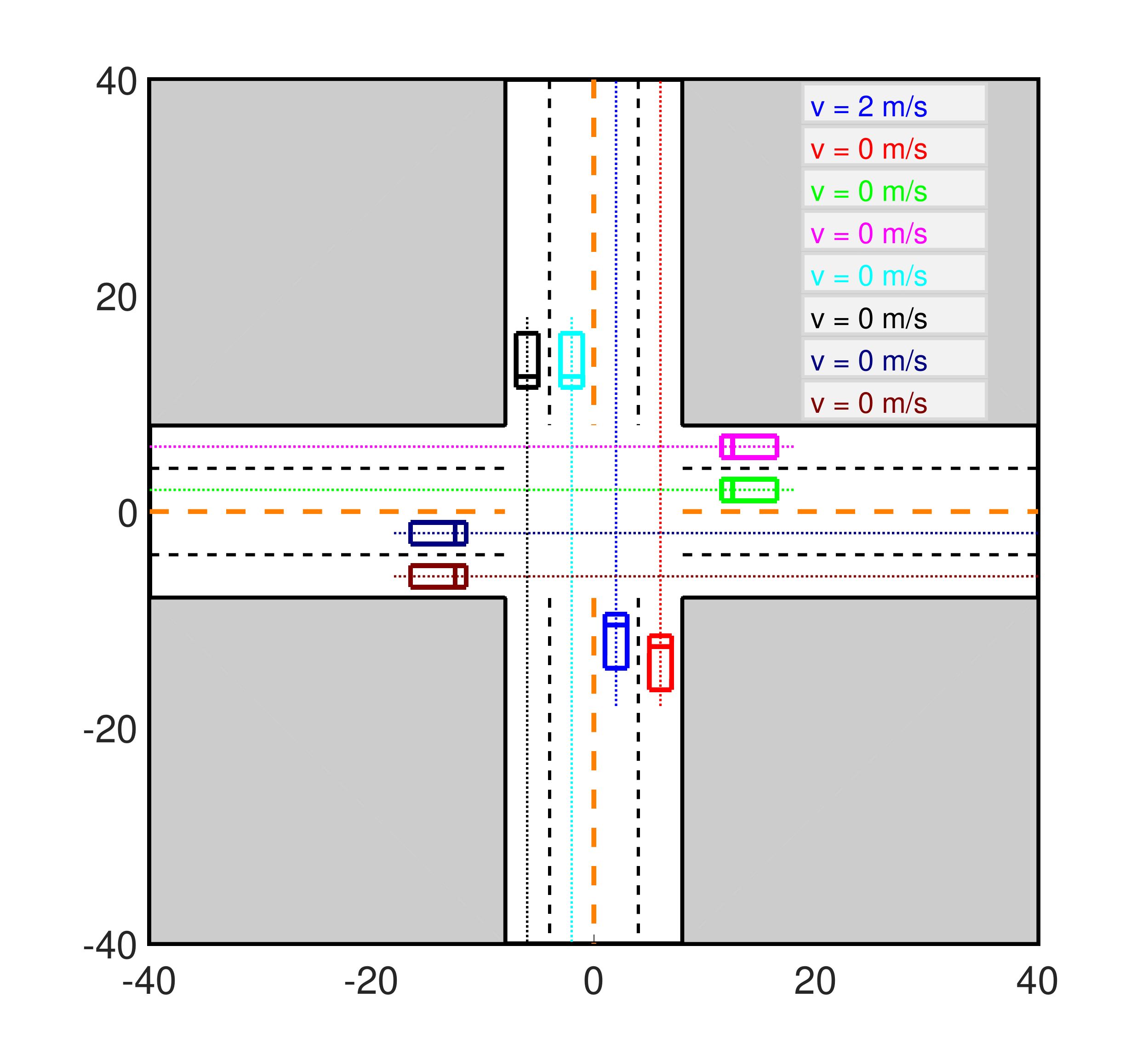,height=1.52in}}  
\put(  112,  100){\epsfig{file=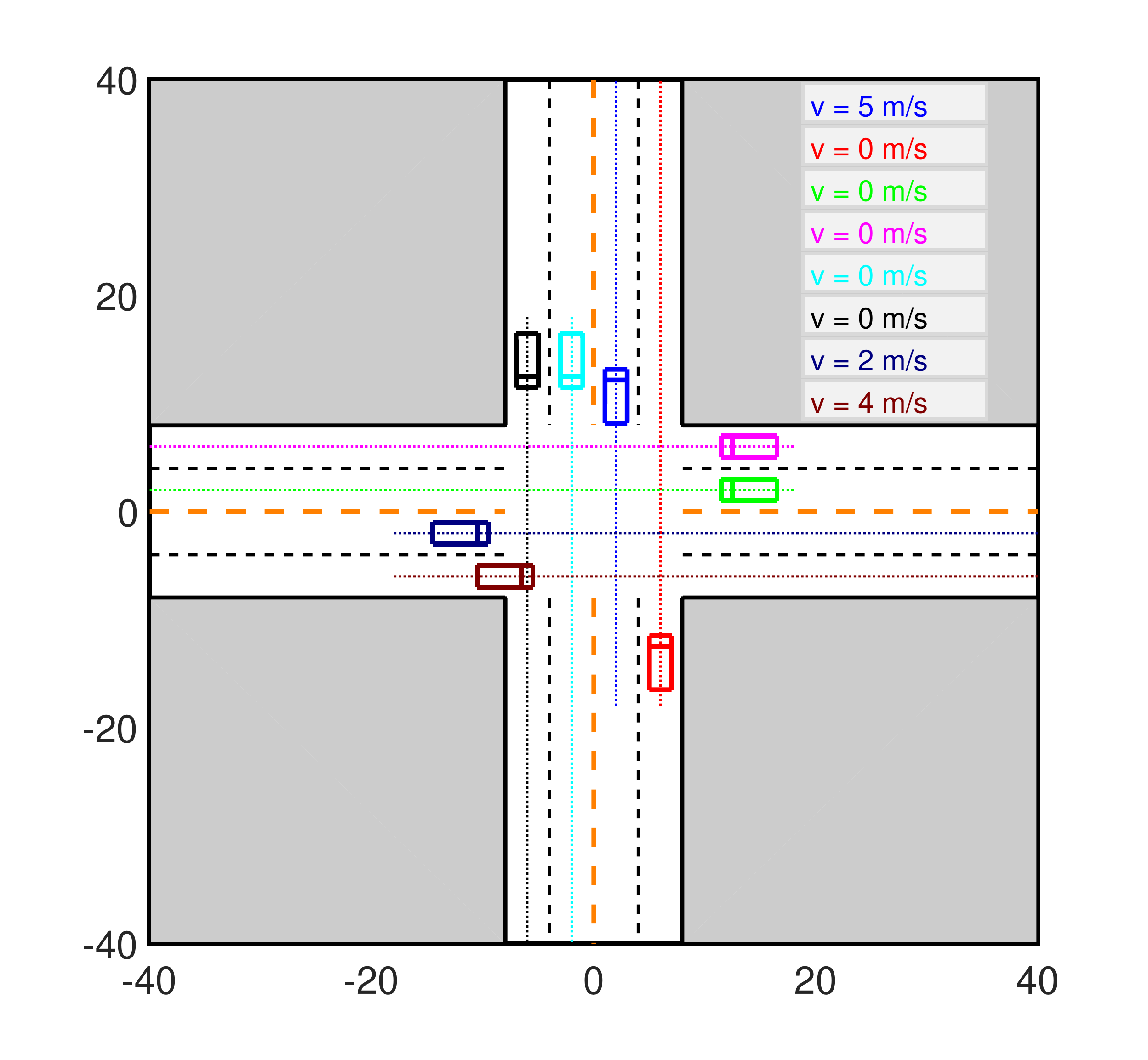,height=1.52in}}  
\put(  0,  -6){\epsfig{file=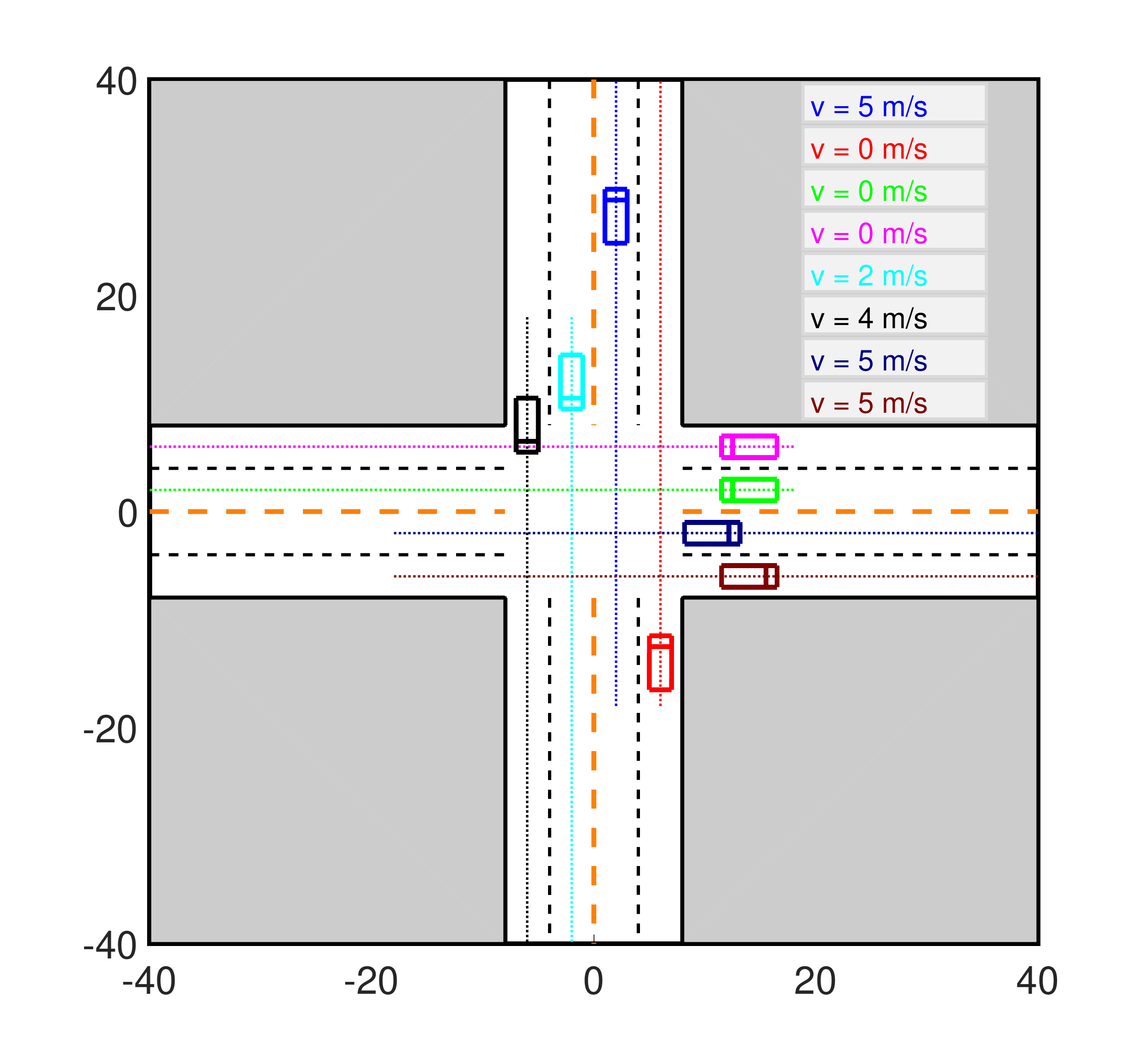,height=1.52in}}  
\put(  112,  -6){\epsfig{file=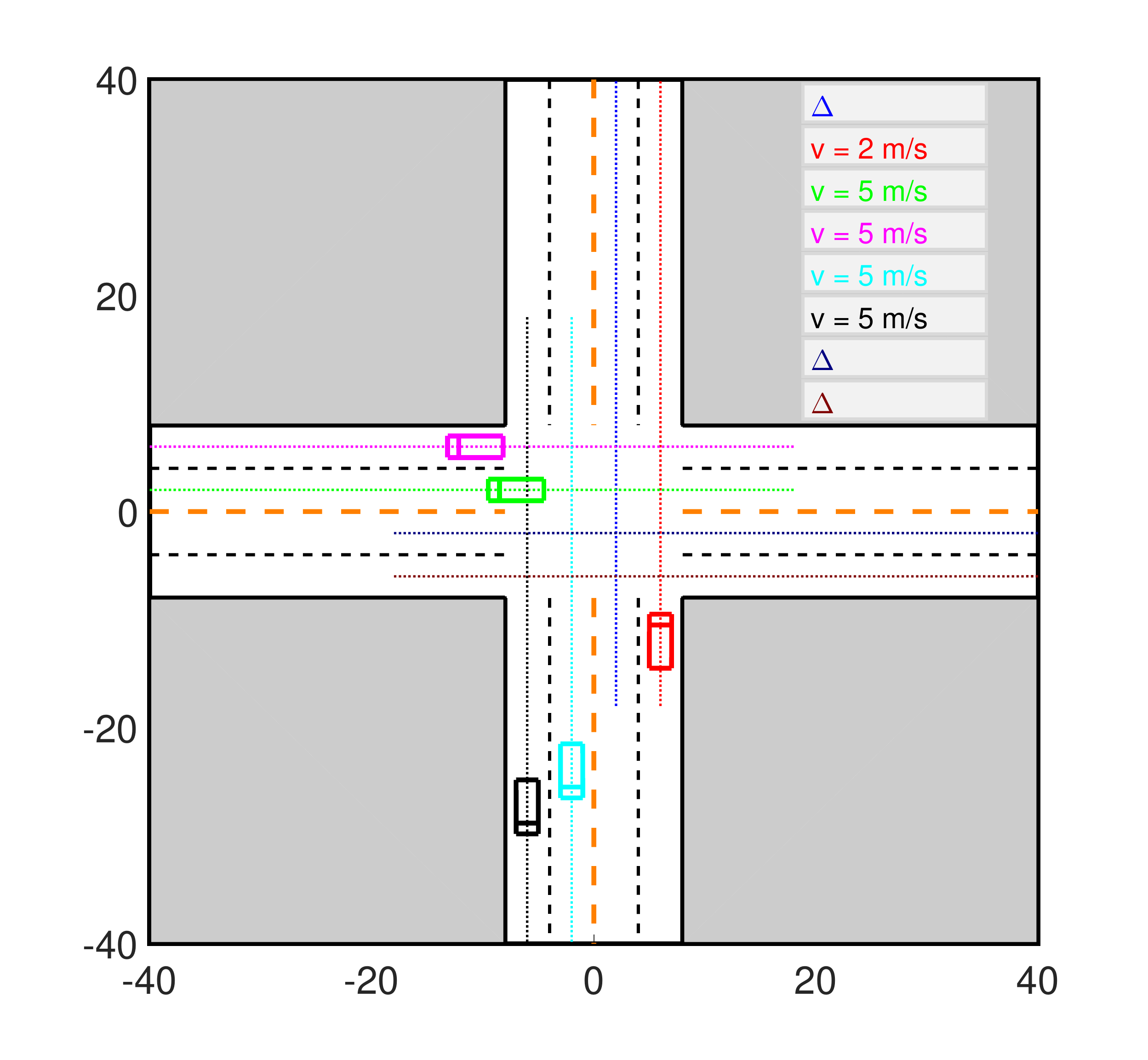,height=1.52in}}  
\small
\put(98, 310){(a)}
\put(210, 310){(b)}
\put(98, 204){(c)}
\put(210, 204){(d)}
\put(98, 98){(e)}
\put(210, 98){(f)}
\normalsize
\end{picture}
\end{center}
      \caption{Completely symmetric case~1. Figures~(a-f) show the simulation snapshots at a series of sequential steps.}
      \label{fig:completely_symmetric_1}
\end{figure}
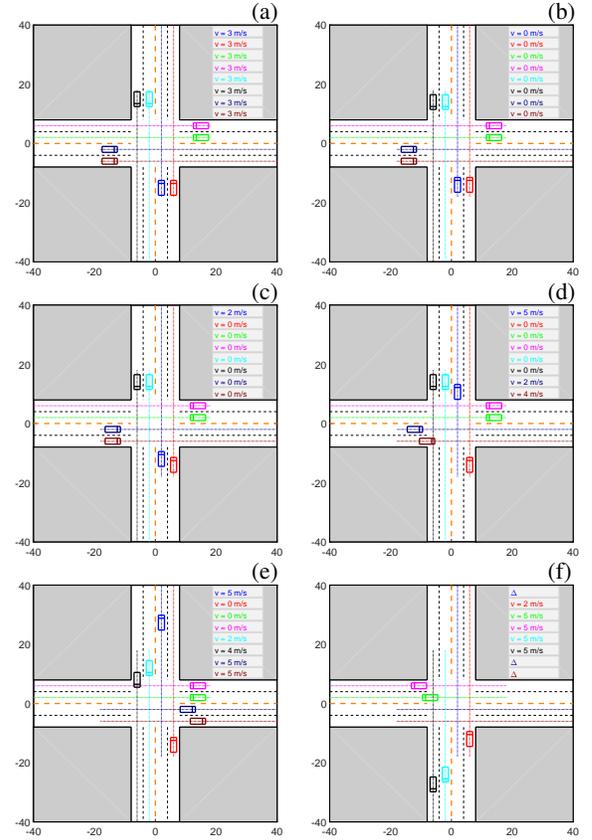

\begin{figure}[h!]
\begin{center}
\begin{picture}(234.0, 214.0)
\put(  0,  100){\epsfig{file=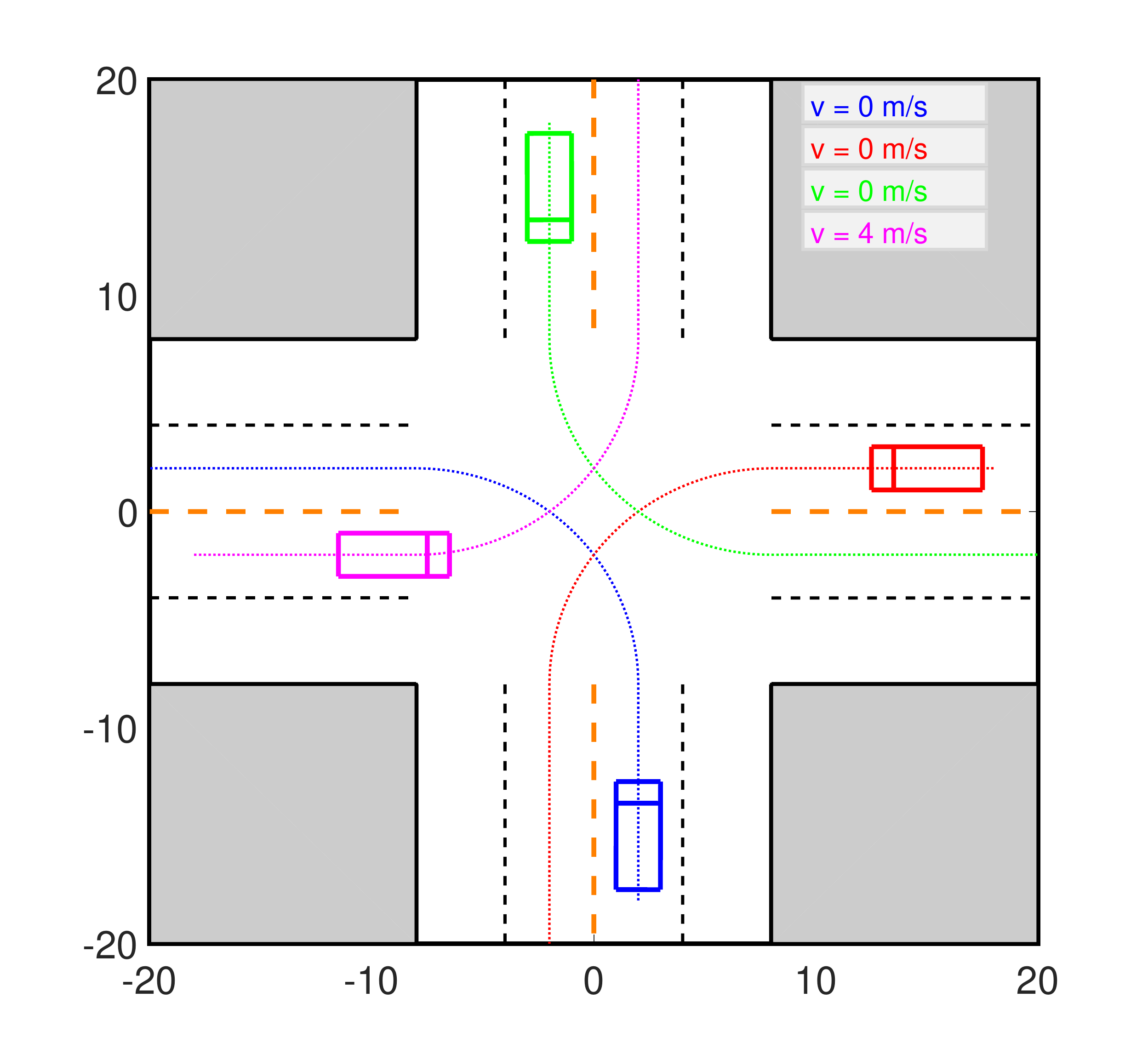,height=1.52in}}  
\put(  112,  100){\epsfig{file=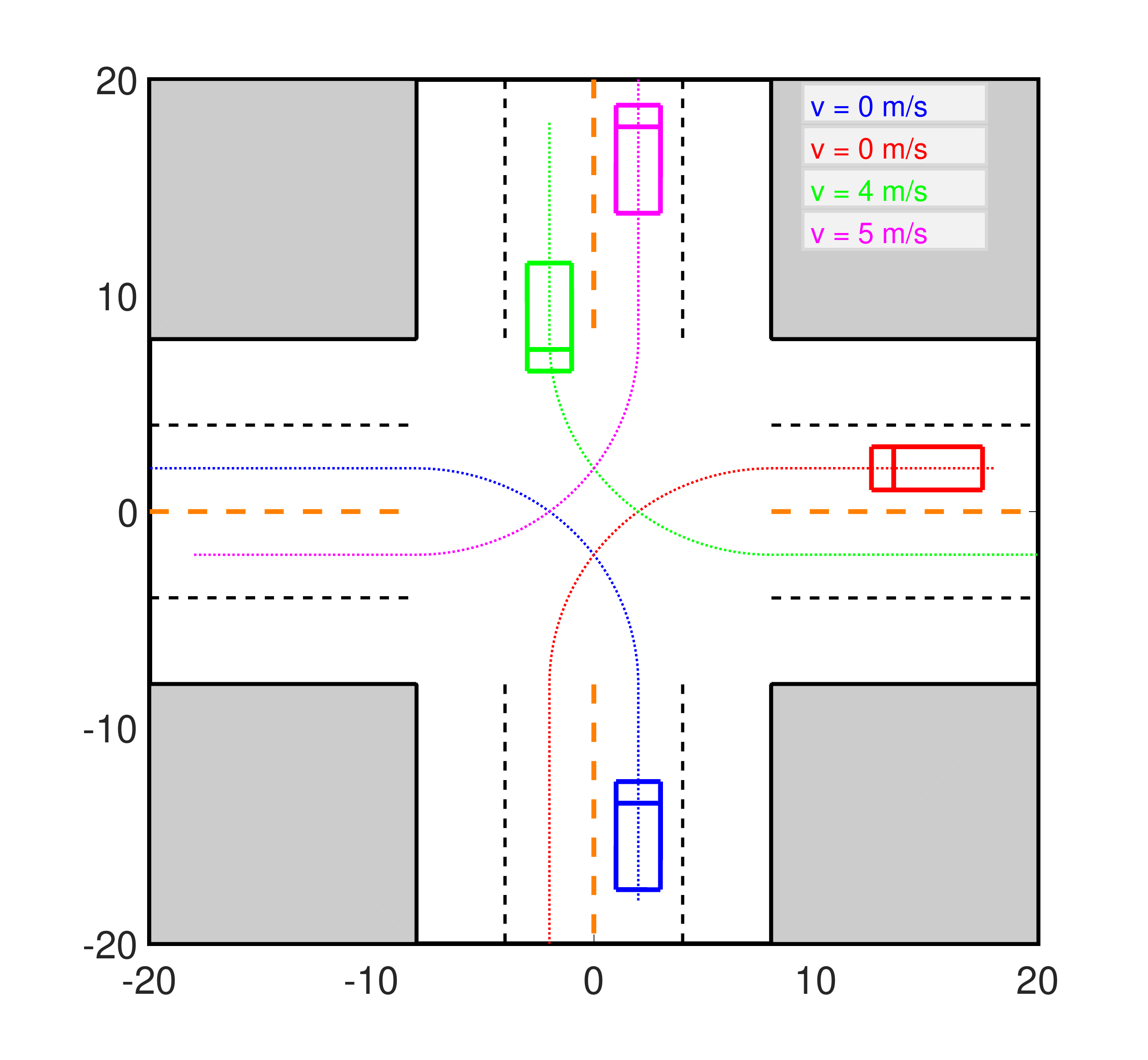,height=1.52in}}  
\put(  0,  -6){\epsfig{file=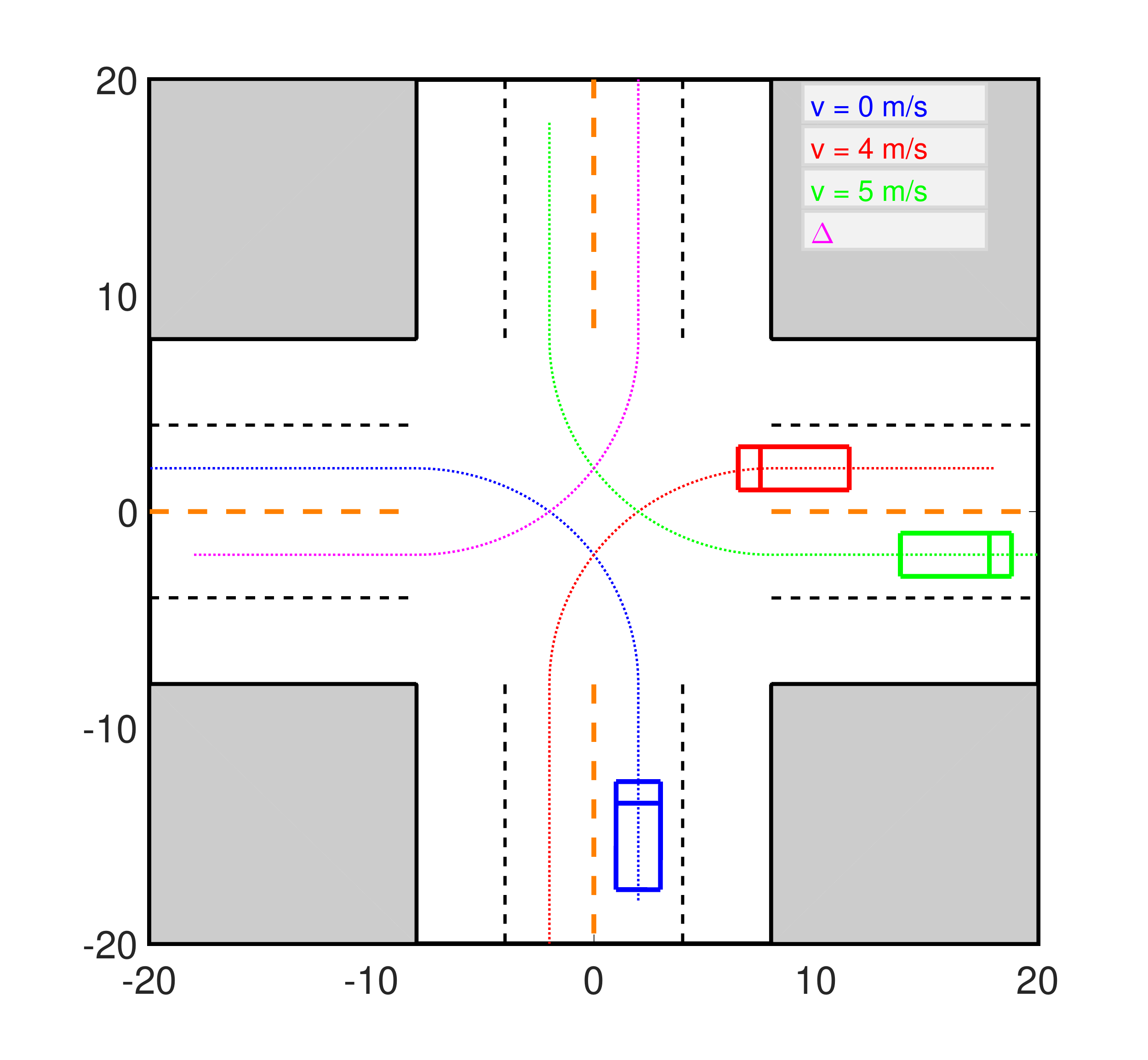,height=1.52in}}  
\put(  112,  -6){\epsfig{file=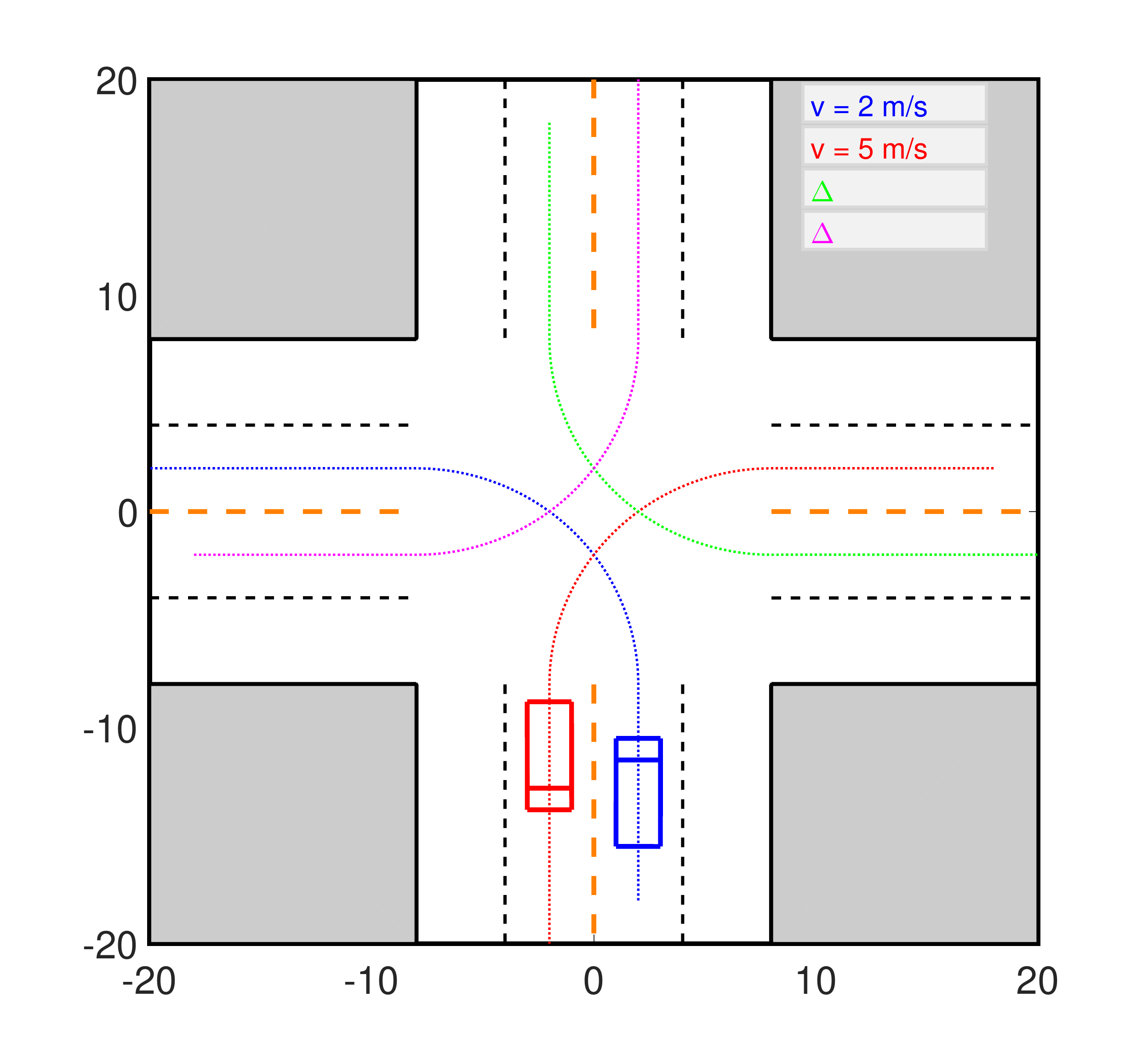,height=1.52in}}  
\small
\put(98, 204){(a)}
\put(210, 204){(b)}
\put(98, 98){(c)}
\put(210, 98){(d)}
\normalsize
\end{picture}
\end{center}
      \caption{Completely symmetric case~2. Figures~(a-d) show the simulation snapshots at a series of sequential steps.}
      \label{fig:completely_symmetric_2}
\end{figure}
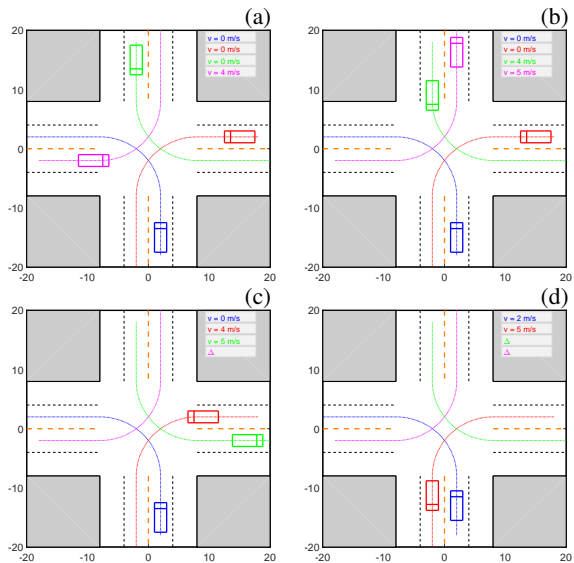

Both cases involve a geometrically symmetrical four-lane (two for each direction) four-way intersection. In the first case, 8 vehicles are approaching the entrances of the intersection from each of the eight forward lanes with the same initial distance to their corresponding entrance points $\Delta \rho^{\text{en}}(0)$ and the same initial speed $v(0)$. Their target lanes all correspond to maneuvers of going straight to cross the intersection. In the second case, 4 vehicles are approaching the entrances of the intersection from each of the four leftmost forward lanes of the road arms with the same $\Delta \rho^{\text{en}}(0)$ and $v(0)$. Their target lanes all correspond to maneuvers of making left turns. In both cases, all of the vehicles are using the decision-making model \eqref{equ:leader_follower_n_1} based on pairwise leader-follower games combined with Algorithm~\ref{alg:Break_D} to handle deadlocks. The simulation results are shown in Figs.~\ref{fig:completely_symmetric_1} and \ref{fig:completely_symmetric_2}.

In Fig.~\ref{fig:completely_symmetric_1}, the eight vehicles arrive at the entrances of the intersection at the same time (Fig.~\ref{fig:completely_symmetric_1}(a)). According to Algorithm~\ref{alg:Role}, no vehicle holds an overall leader role (being the leader in every pairwise interaction). As a result, all of the eight vehicles stop at the intersection entrances (Fig.~\ref{fig:completely_symmetric_1}(b)). According to Algorithm~\ref{alg:Break_D}, a deadlock is detected and the blue vehicle makes a probing acceleration (Fig.~\ref{fig:completely_symmetric_1}(c)). Then, the symmetry is broken: the blue vehicle becomes the overall leader and crosses the intersection first (Fig.~\ref{fig:completely_symmetric_1}(d)). The other vehicles cross the intersection in a clockwise order (Figs.~\ref{fig:completely_symmetric_1}(e)(f)).
In Fig.~\ref{fig:completely_symmetric_2}, after the four vehicles arrive at the entrances of the intersection at the same time, the left purple vehicle makes a probing acceleration (Fig.~\ref{fig:completely_symmetric_2}(a)) and gets to its target lane first (Fig.~\ref{fig:completely_symmetric_2}(b)). Similar to Fig.~\ref{fig:completely_symmetric_1}, the other vehicles then pass through the intersection in a clockwise order (Figs.~\ref{fig:completely_symmetric_2}(c)(d)).

From the above ``completely symmetric'' cases we can observe that our model exhibits reasonable behavior expected in traffic and has good capability in resolving traffic conflicts in challenging scenarios. We note that there is no centralized control or management in our model to guide the vehicles to resolve their conflicts -- the vehicles make their decisions independently. In particular, they take into account their interactions when making decisions. Such a process is consistent with drivers' decision-making in real-world traffic. Moreover, when a deadlock occurs, the vehicles try to resolve it through exploratory movements, which is also consistent with the way in which human drivers resolve deadlocks in real-world traffic.

\subsection{Case study 3: Leader-follower versus adaptive level-$\mathcal{K}$}

As discussed at the beginning of Section~\ref{sec:levelK}, a model representing driver interactive decision-making is supposed to have a reasonable capability of resolving traffic conflicts when interacting with other drivers whose driving styles are a priori unknown. To illustrate such capability of the decision-making model \eqref{equ:leader_follower_n_1} based on pairwise leader-follower games, we let vehicles that make decisions using \eqref{equ:leader_follower_n_1} interact with vehicles that make decisions using models different from \eqref{equ:leader_follower_n_1}, in particular, using the adaptive level-$\mathcal{K}$ decision-making model \eqref{equ:model_D}. Note that \eqref{equ:model_D} does not take into account the leader-follower relationships among vehicles, and thus, a vehicle using \eqref{equ:model_D} may not yield the right of way to another vehicle that is supposed to be the leader in their interactions.

\begin{figure}[h!]
\begin{center}
\begin{picture}(234.0, 522.0)
\put(  56,  412){\epsfig{file=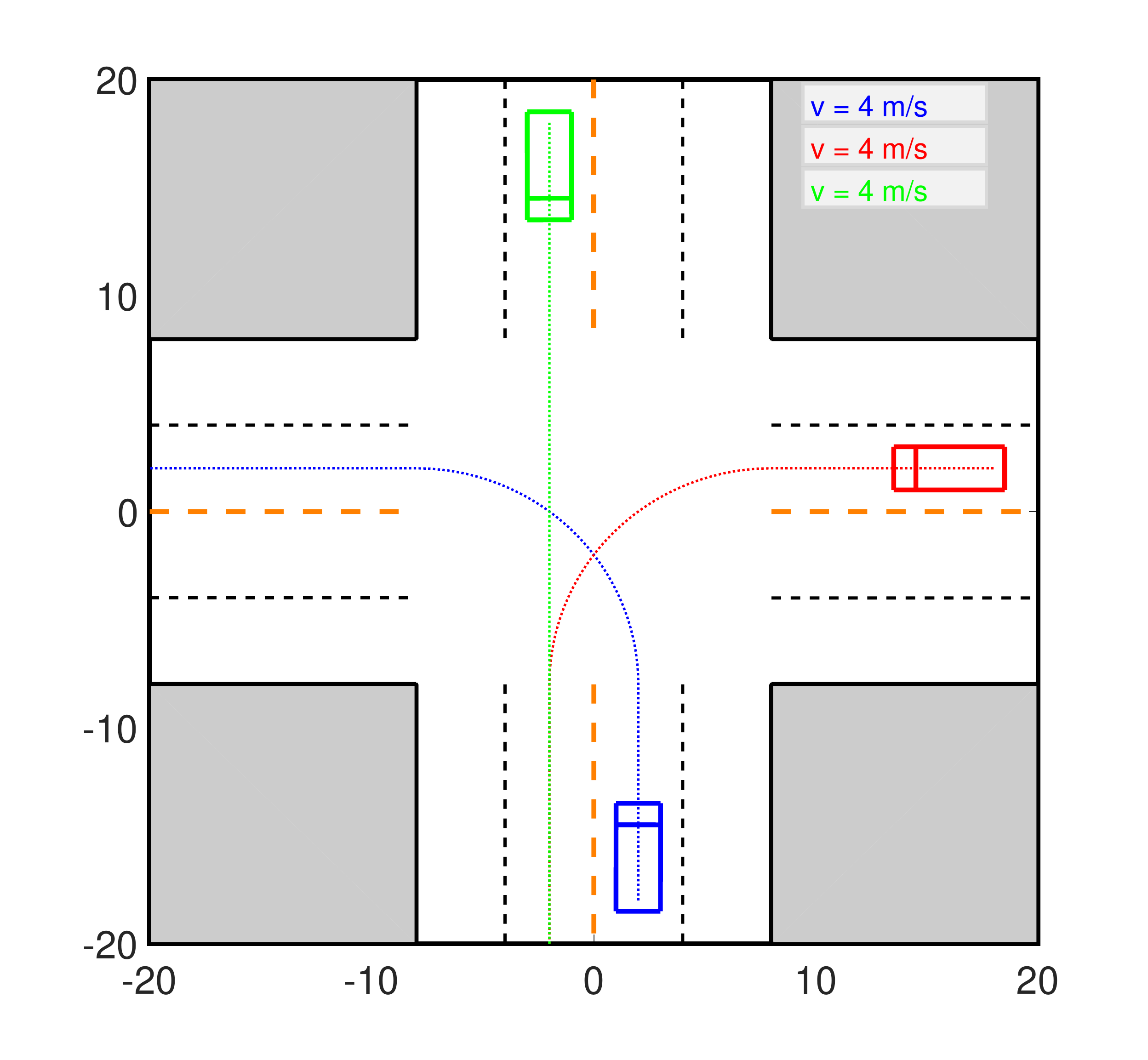,height=1.52in}}  
\put(  0,  307){\epsfig{file=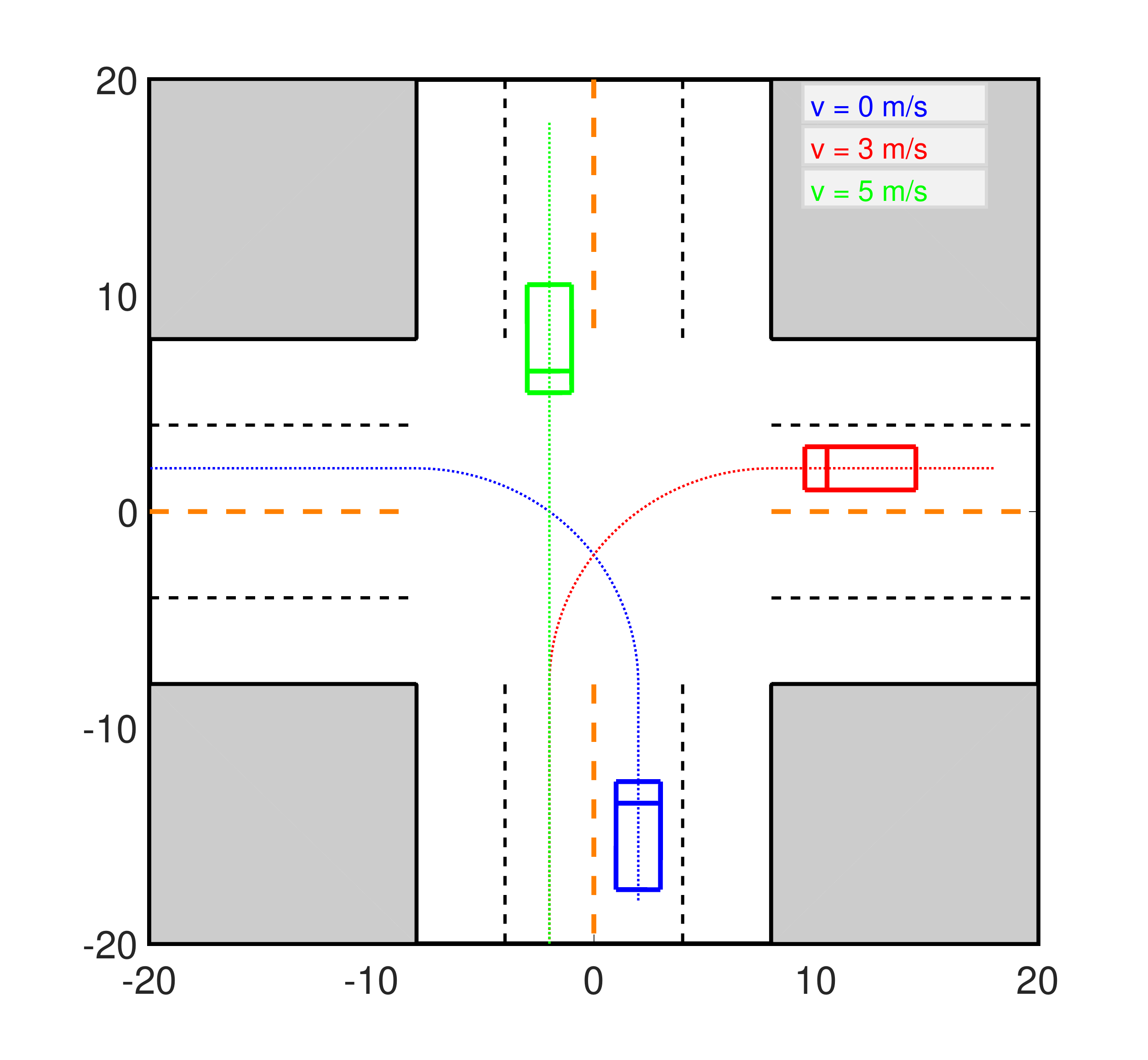,height=1.52in}}  
\put(  112,  307){\epsfig{file=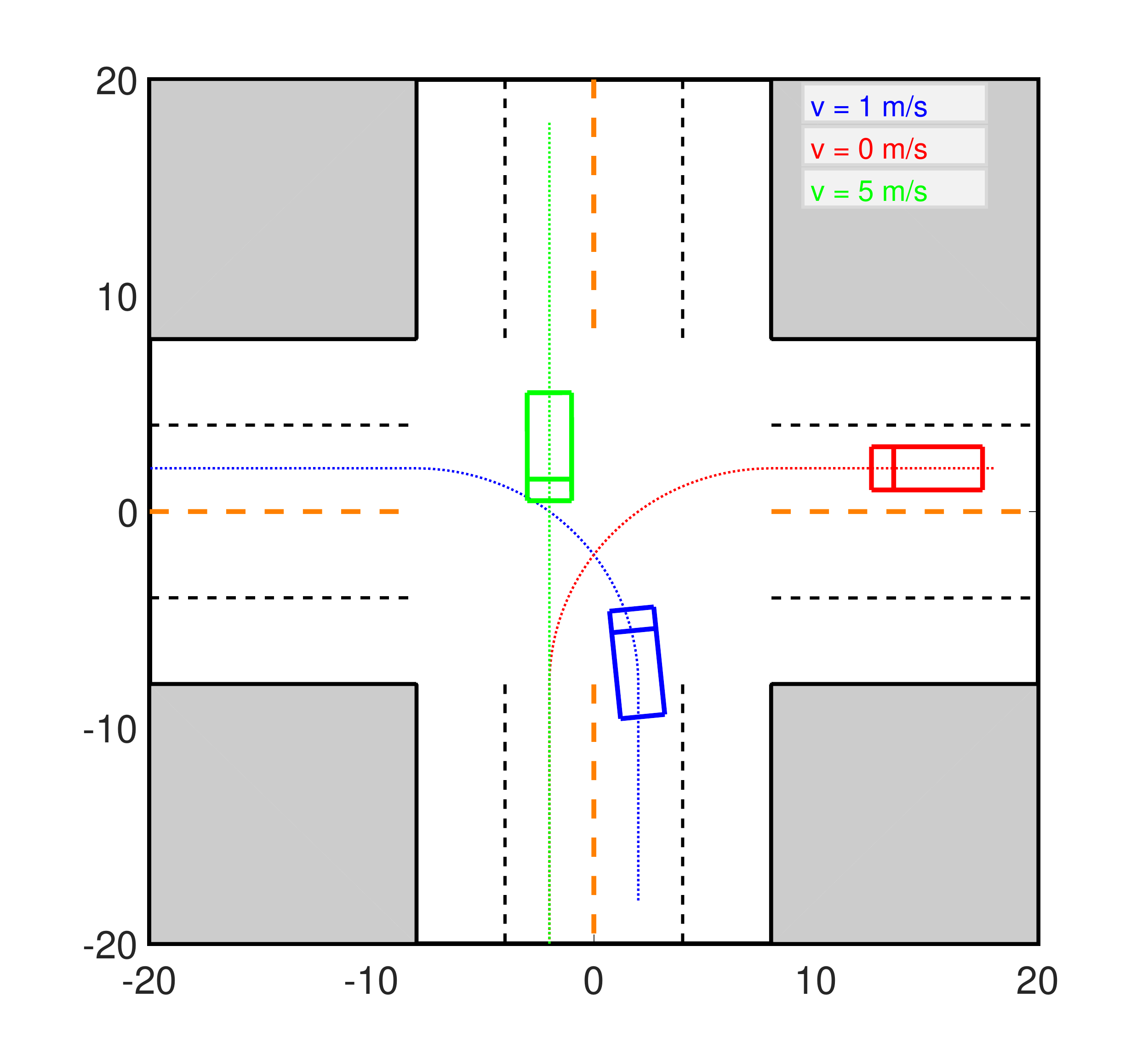,height=1.52in}}  
\put(  0,  202){\epsfig{file=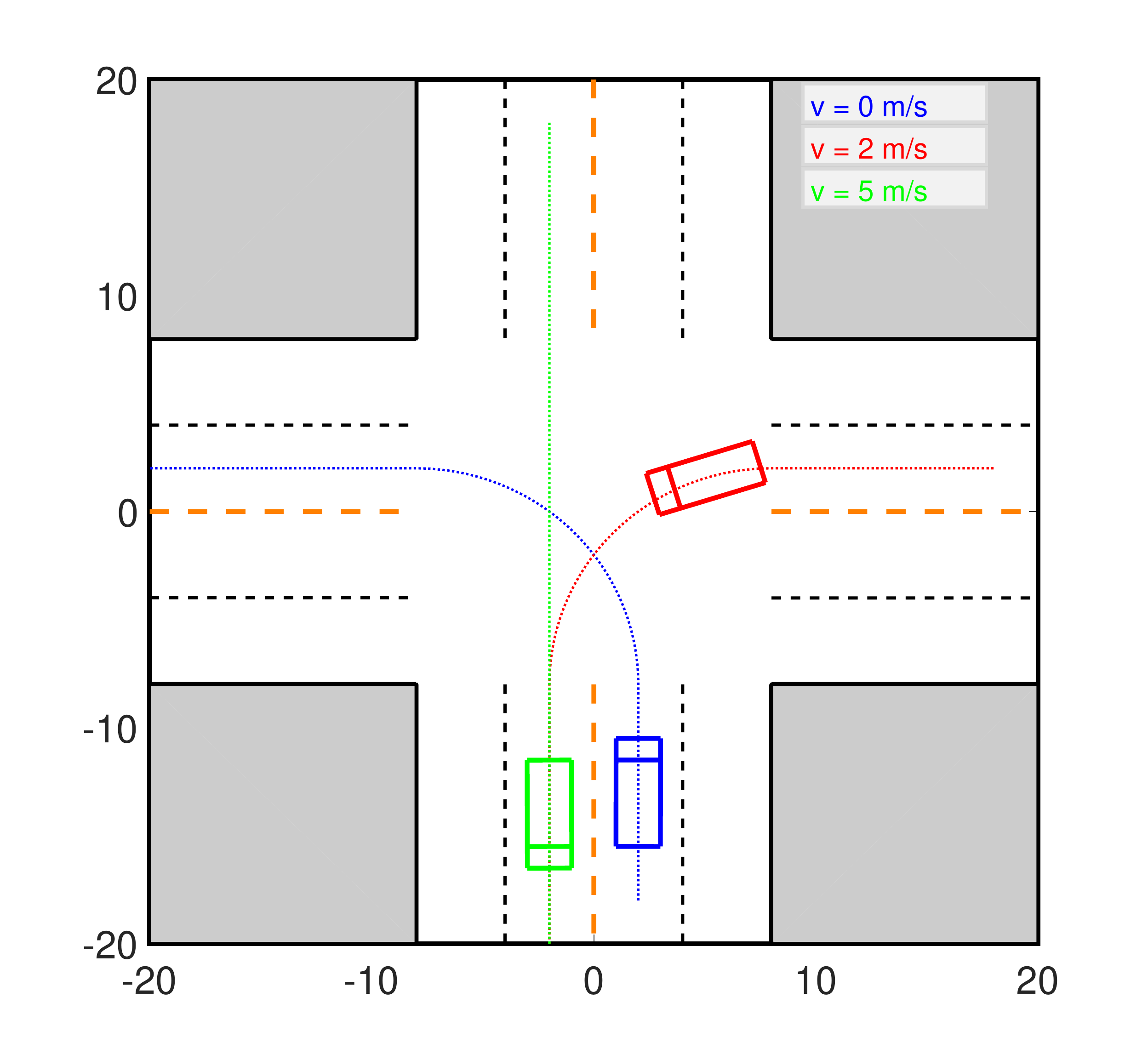,height=1.52in}}  
\put(  112,  202){\epsfig{file=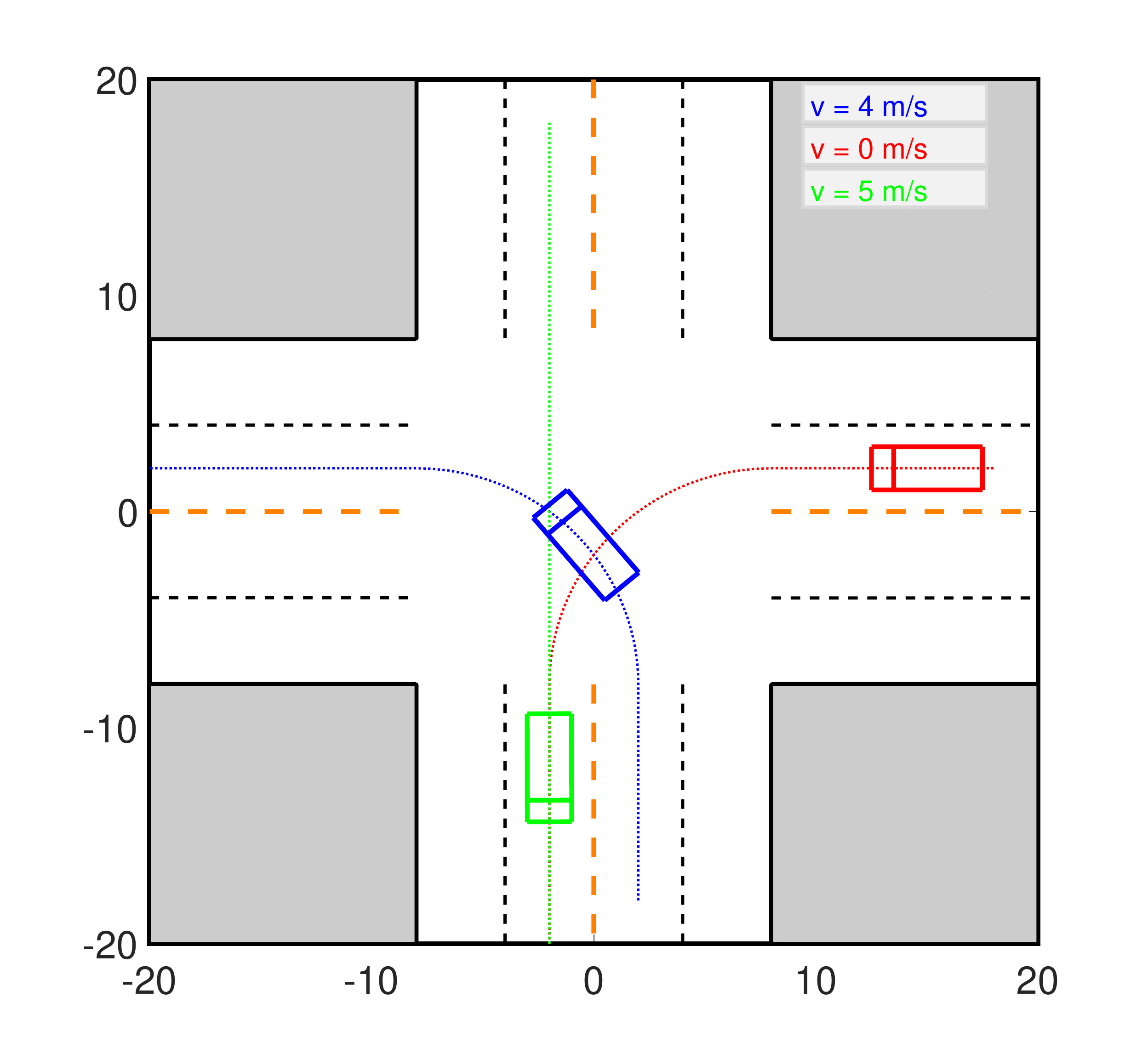,height=1.52in}}  
\put(  0,  97){\epsfig{file=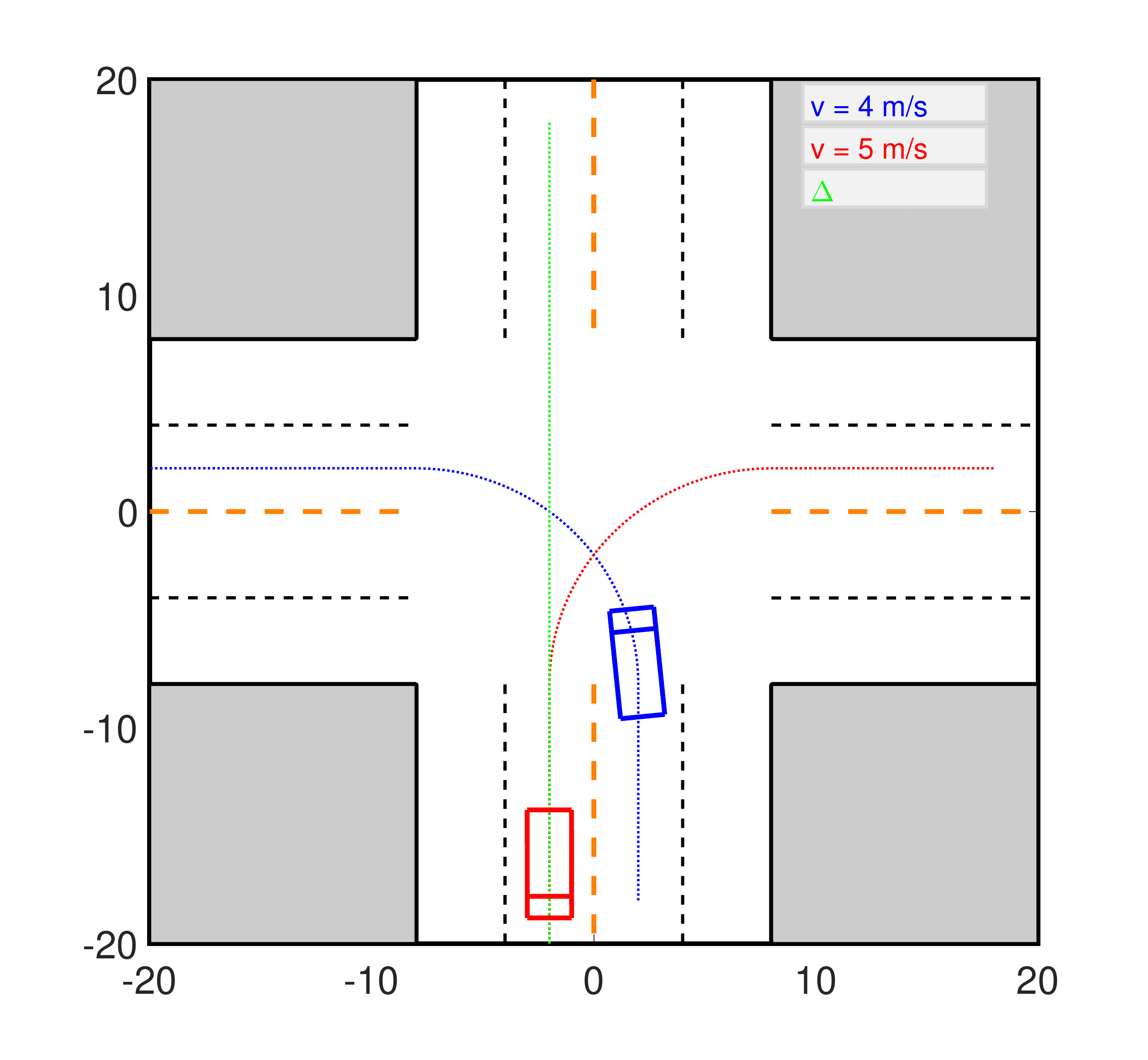,height=1.52in}}  
\put(  112,  97){\epsfig{file=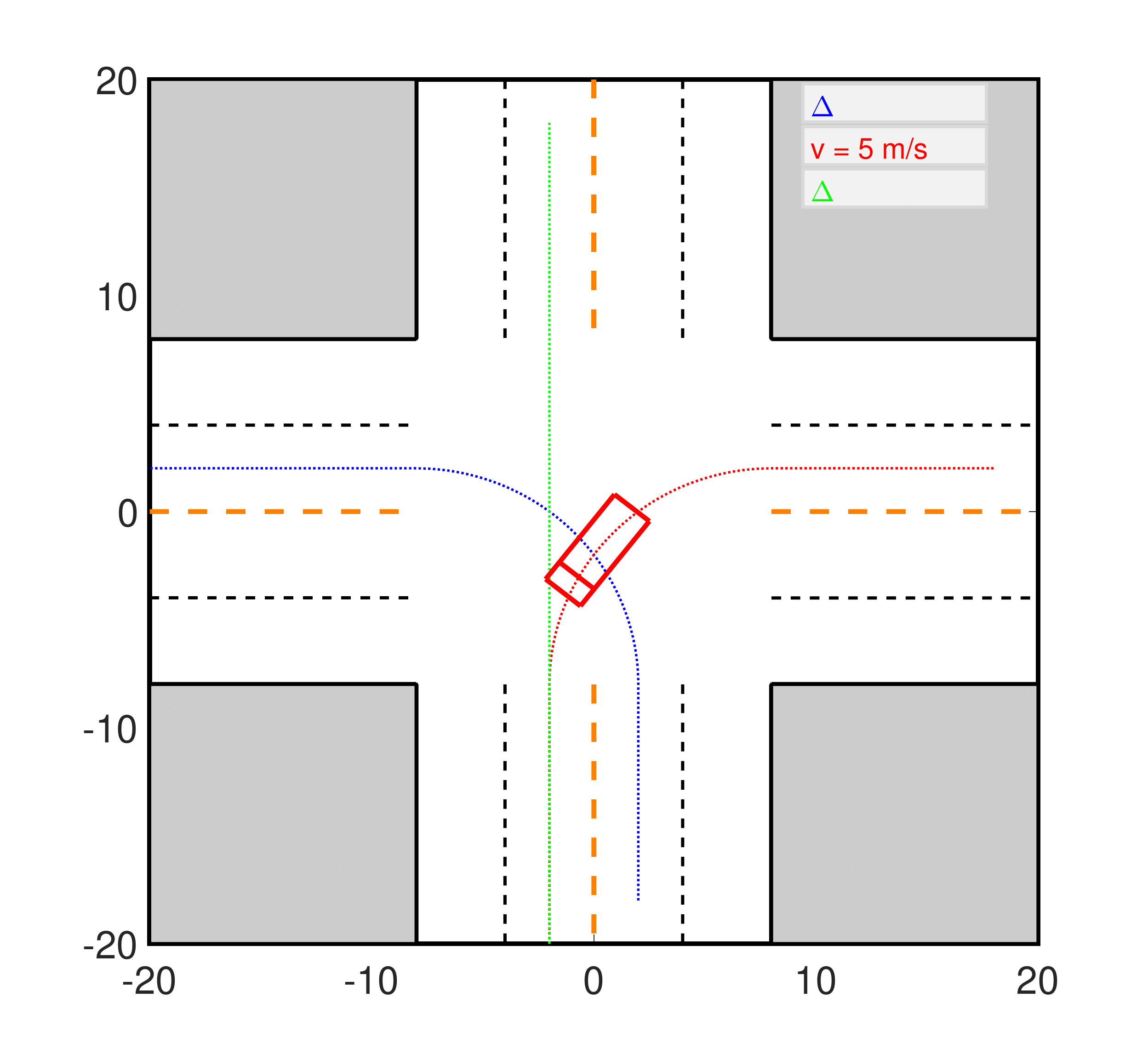,height=1.52in}}  
\put(  0,  -8){\epsfig{file=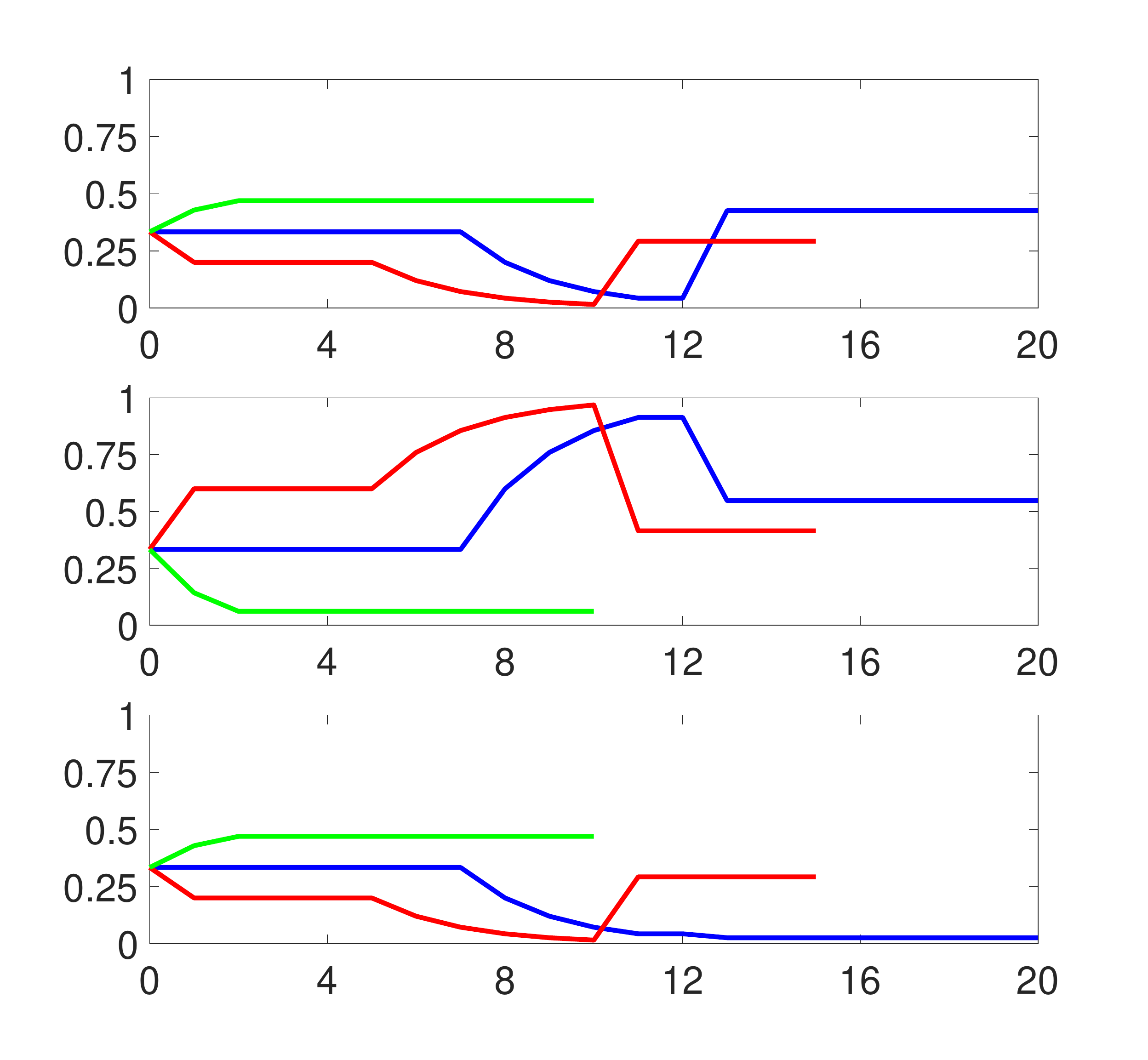,height=1.52in}}  
\put(  112,  -8){\epsfig{file=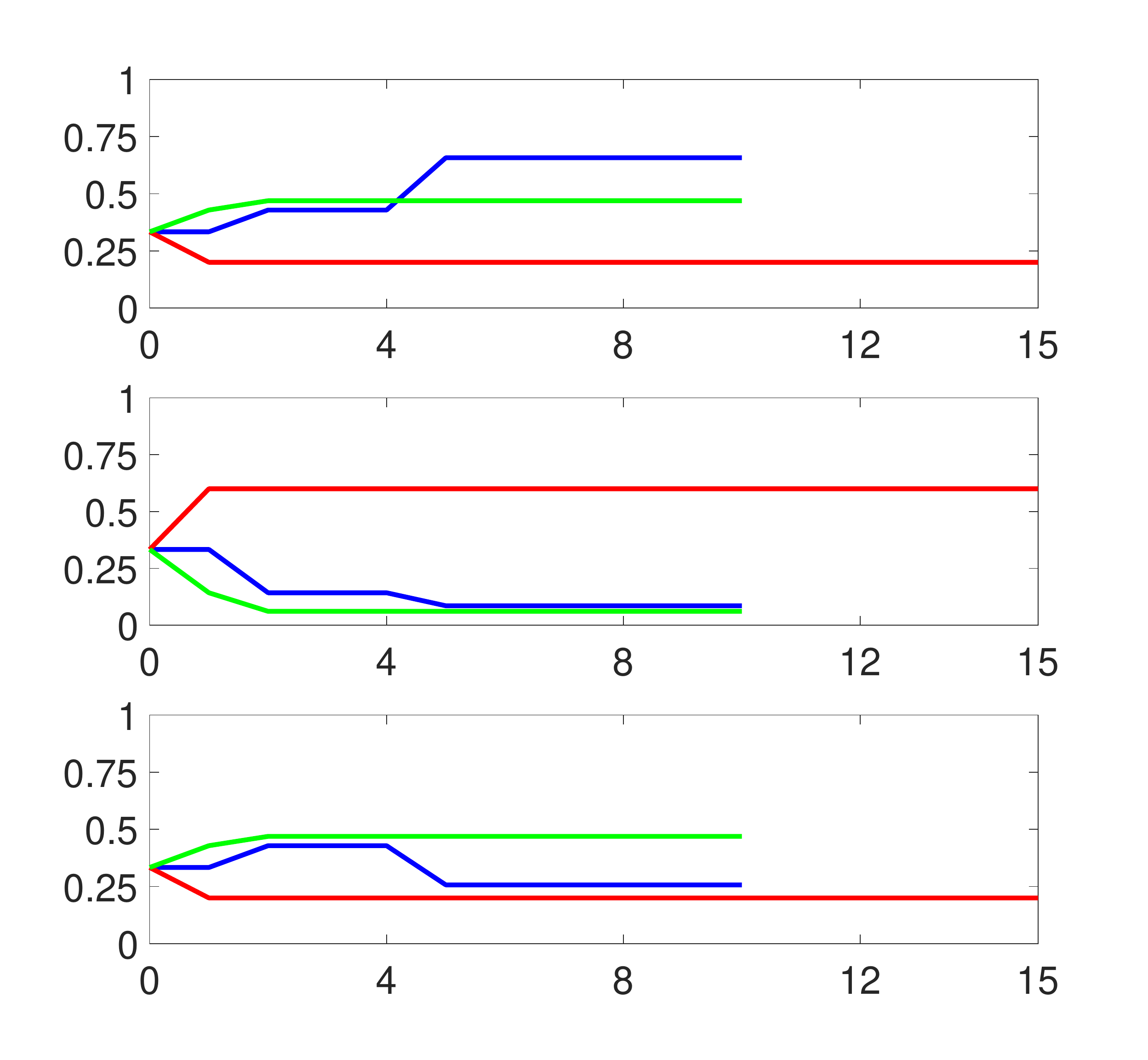,height=1.52in}}  
\small
\put(165, 506){(a)}
\put(98, 411){(b)}
\put(210, 411){(f)}
\put(98, 306){(c)}
\put(210, 306){(g)}
\put(98, 201){(d)}
\put(210, 201){(h)}
\put(98, 96){(e)}
\put(210, 96){(i)}

\footnotesize
\put(18, 87){level-0}
\put(18, 54){level-1}
\put(18, 21){level-2}
\put(130, 87){level-0}
\put(130, 54){level-1}
\put(130, 21){level-2}
\normalsize
\end{picture}
\end{center}
      \caption{Leader-follower versus adaptive level-$\mathcal{K}$ case~1 (left): the blue car is using the decision-making model \eqref{equ:leader_follower_n_1} and the other two cars are using \eqref{equ:model_D} and \eqref{equ:model_estimation}, and case~2 (right): the blue car is using \eqref{equ:model_D} and \eqref{equ:model_estimation} and the other two cars are using \eqref{equ:leader_follower_n_1}. Figures~(a-d) show the simulation snapshots at a series of sequential steps of case~1, and figure~(e) shows the corresponding model estimation time histories of the three vehicles (the blue curves correspond to the level-$0$, $1$, and $2$ belief histories of the blue car, etc); figures~(a) and (f-h) show the simulation snapshots at a series of sequential steps of case~2, and figure~(i) shows the corresponding model estimation time histories of the three vehicles.}
      \label{fig:lf_vs_k}
\end{figure}
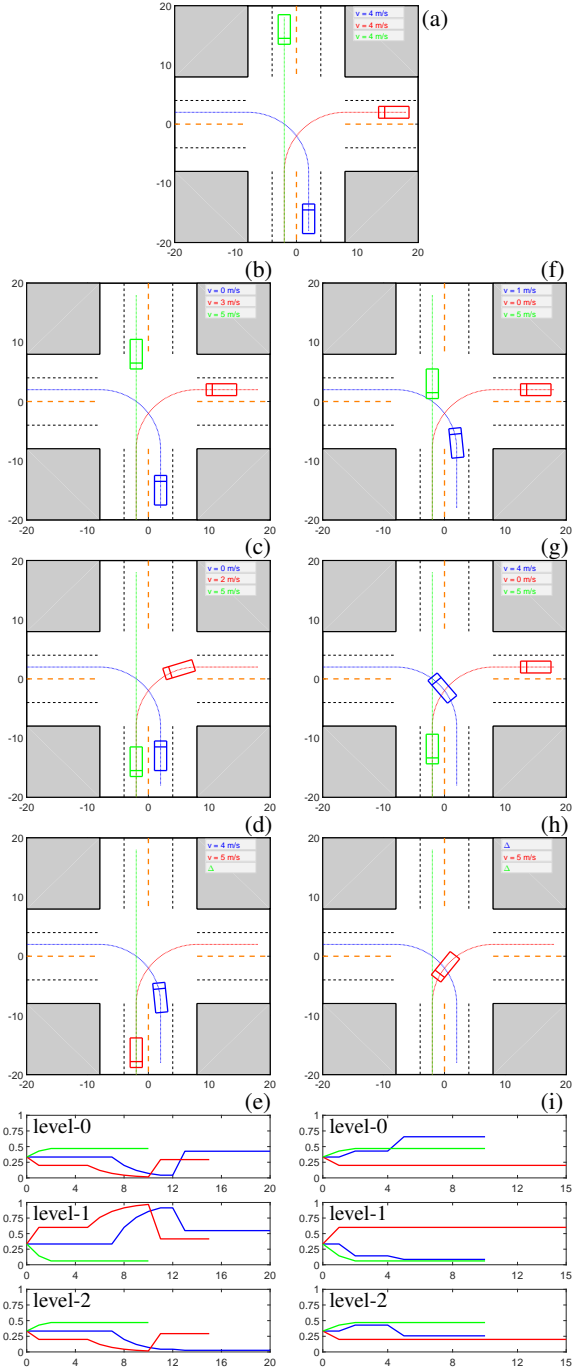

We initialize the traffic scenario as follows: Three vehicles are approaching the entrances of a geometrically symmetrical four-lane four-way intersection from three different road arms. Vehicle~1 (blue) is coming from the bottom arm and is making a left turn to the left arm; vehicle~2 (red) is coming from the right arm and is making a left turn to the bottom arm; and vehicle~3 (green) is coming from the top arm and is going straight to the bottom arm. The three vehicles are initialized with the same $\Delta \rho^{\text{en}}(0)$ and $v(0)$. At first, we let vehicle~1 using \eqref{equ:leader_follower_n_1} interact with vehicles~2 and 3 using \eqref{equ:model_D}, and the simulation results are shown in Fig.~\ref{fig:lf_vs_k} (left column). Then, we let vehicle~1 using \eqref{equ:model_D} interact with vehicles~2 and 3 using \eqref{equ:leader_follower_n_1}, and the simulation results are shown in Fig.~\ref{fig:lf_vs_k} (right column).

In Figs.~\ref{fig:lf_vs_k}(a-d) where vehicle~1 (blue) is using \eqref{equ:leader_follower_n_1}, according to Algorithm~\ref{alg:Role}, it is the follower of vehicles~2 (red) and 3 (green). As a result, it yields the right of way to both vehicles~2 and 3. Vehicles~2 and 3 pass through the intersection ahead of vehicle~1 based on \eqref{equ:model_D} and \eqref{equ:model_estimation}. Fig.~\ref{fig:lf_vs_k}(e) shows the model estimation time histories of the three vehicles: it can be observed that as vehicle~1 decides to yield the right of way, the belief that it can be modeled as level-$1$ (corresponding to the most conservative driver) increases over $7-12\,$[s] (when vehicles~2 and 3 are inside the intersection). In Figs.~\ref{fig:lf_vs_k}(a) and (f-i) where vehicles~2 (red) and 3 (green) are using \eqref{equ:leader_follower_n_1}, according to Algorithm~\ref{alg:Role}, vehicle~3 is the leader in every pairwise interaction. As a result, it decides to cross the intersection first. Such a result illustrates that our model \eqref{equ:leader_follower_n_1} is not conservative when it has the legitimate right of way. Different from Figs.~\ref{fig:lf_vs_k}(a-d), vehicle~1 decides to pass through the intersection ahead of vehicle~2 based on \eqref{equ:model_D} and \eqref{equ:model_estimation}. This is because when vehicle~2 yields the right of way to vehicle~3, vehicle~1 thinks that vehicle~2 is a conservative driver (which can be seen from the model estimation time histories Fig.~\ref{fig:lf_vs_k}(i), where the belief that vehicle~2 can be modeled as level-$1$ increases at the beginning). Thus, vehicle~1 decides to go ahead of this conservative driver. Note that the adaptive level-$\mathcal{K}$ decision-making model \eqref{equ:model_D} and \eqref{equ:model_estimation} does not account for the right-of-way traffic rules. Although vehicle~2 is supposed to have the right of way over vehicle~1 when they arrive at the entrances of the intersection, it decides to wait until vehicle~1 passes because of vehicle~1's aggressive preemption.

The above ``leader-follower versus adaptive level-$\mathcal{K}$'' cases illustrate that our model \eqref{equ:leader_follower_n_1} is capable of making effective use of its legitimate rights of way in resolving traffic conflicts. Nevertheless, when encountering unexpected disruptions, it can quickly adapt its strategy to avoid traffic incidents.

\subsection{Case study 4: Randomized traffic scenarios}

To show the capability of the proposed framework to model vehicle interactions at a wide range of uncontrolled intersection traffic scenarios and statistically evaluate its performance, we randomly generate the layout and geometry parameters $\{M_{\text{f}}^{(m)}\}_{m=1}^N$, $\{M_{\text{b}}^{(m)}\}_{m=1}^N$, and $\{\phi^{(m)}\}_{m=1}^N$ of the intersections for each fixed number of intersection arms $N \in \{3,4,5\}$, randomly generate the origin lanes, target lanes, initial distances to intersection entrances $\Delta \rho^{\text{en}}(0)$, and initial speeds $v(0)$ of the vehicles for each fixed number of vehicles $n \in \{2,4,6,8,10\}$, and have 100 simulation runs for each pair of $(N,n)$.

In particular, we sample $\{M_{\text{f}}^{(m)}\}_{m=1}^N$, $\{M_{\text{b}}^{(m)}\}_{m=1}^N$ based on categorical distributions and $\{\phi^{(m)}\}_{m=1}^N$ based on truncated normal distributions\footnote{with mean $\frac{2m\pi}{N}$ and standard deviation $\frac{\pi}{24}$ truncated to the range $\big[\frac{2m\pi}{N}-\frac{\pi}{8},\frac{2m\pi}{N}+\frac{\pi}{8}\big]$.} as follows:

\begin{align*} 
    M_{\xi}^{(m)} & \sim \text{Cat} \Big(\{1,2,3\}, \{0.15,0.7,0.15\}\Big), \quad \xi \in \{\text{f},\text{b}\}, \nonumber \\
    \phi^{(m)} & \sim \text{Normal} \Big(\frac{2m\pi}{N}, \frac{\pi}{24}, \big[\frac{2m\pi}{N}-\frac{\pi}{8},\frac{2m\pi}{N}+\frac{\pi}{8}\big]\Big),
\end{align*}
for each $m = 1,\cdots,N$. Once the intersection has been created, we assign each vehicle's origin road arm based on a uniform distribution over all road arms, and assign its origin lane based on a uniform distribution over all forward lanes of its origin road arm. After that, we assign its target road arm and target lane based on uniform distributions over, respectively, all acceptable road arms and all acceptable backward lanes of the assigned target road arms, where ``acceptable'' means satisfying the traffic rule constraints described at the beginning of Section~\ref{sec:path}. Then, each vehicle's $\Delta \rho^{\text{en}}(0)$ and $v(0)$ are initialized based on uniform distributions over the ranges $[10, 28]\,$[m] and $[2, 4]\,$[m/s]. Furthermore, we enforce a minimum initial separation, $\rho^{\text{sep}}$, between any two vehicles that are initialized on the same origin lane --  if $\Delta \rho^{\text{en}}_i(0)$ of vehicle $i$ is in the range of $[\Delta \rho^{\text{en}}_j(0) - \rho^{\text{sep}}, \Delta \rho^{\text{en}}_j(0) + \rho^{\text{sep}}]$ for any vehicle $j$ that has been initialized before vehicle $i$ and is on the same origin lane of vehicle $i$, then $\Delta \rho^{\text{en}}_i(0)$ is re-sampled as above.

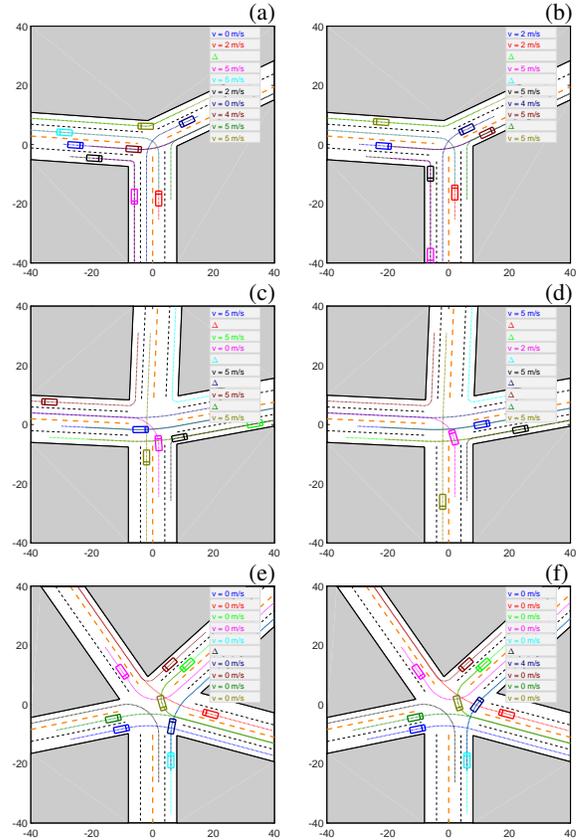
\begin{figure}[h!]
\begin{center}
\begin{picture}(230.0, 312.0)
\put(  0,  206){\epsfig{file=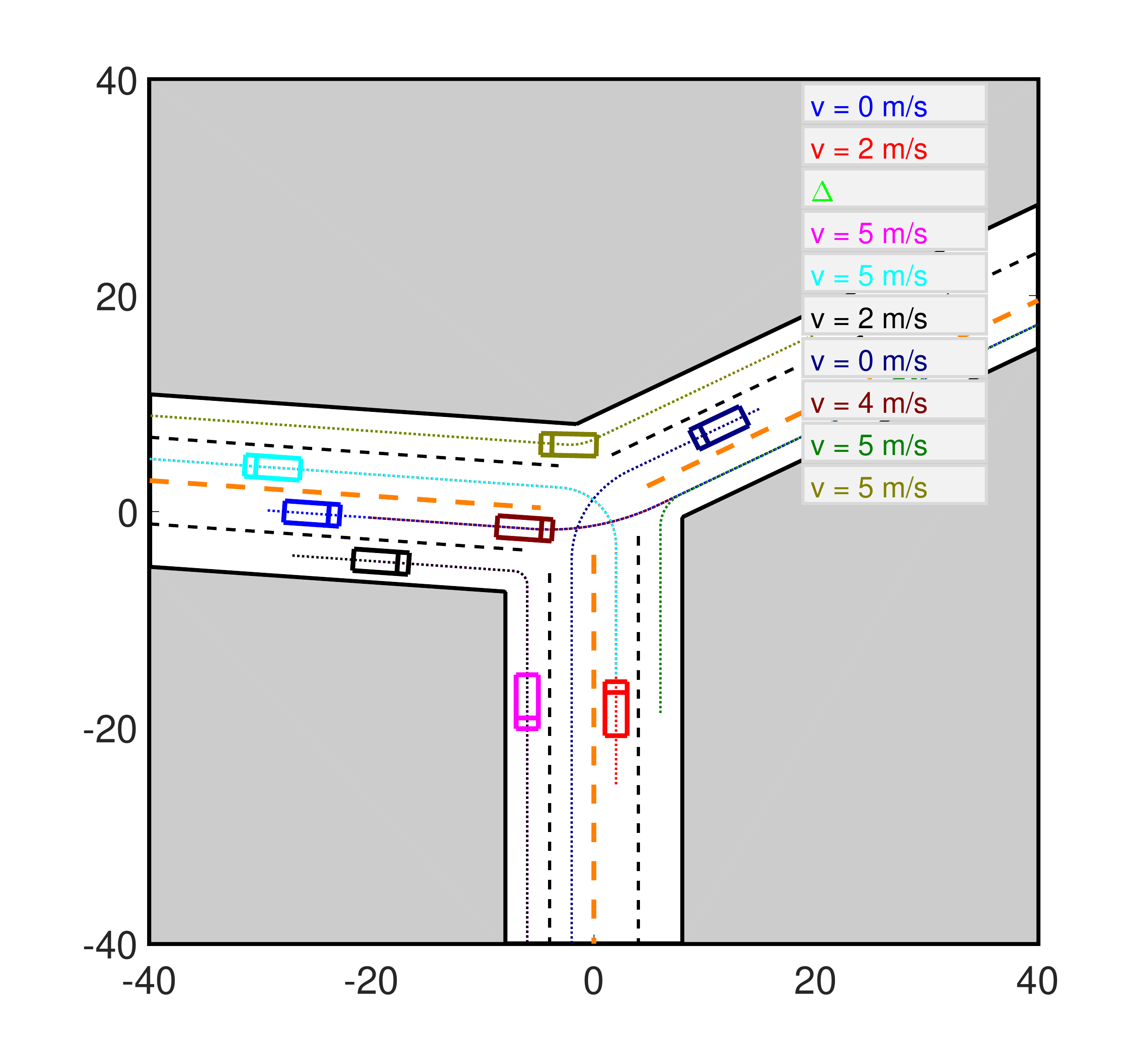,height=1.52in}}  
\put(  112,  206){\epsfig{file=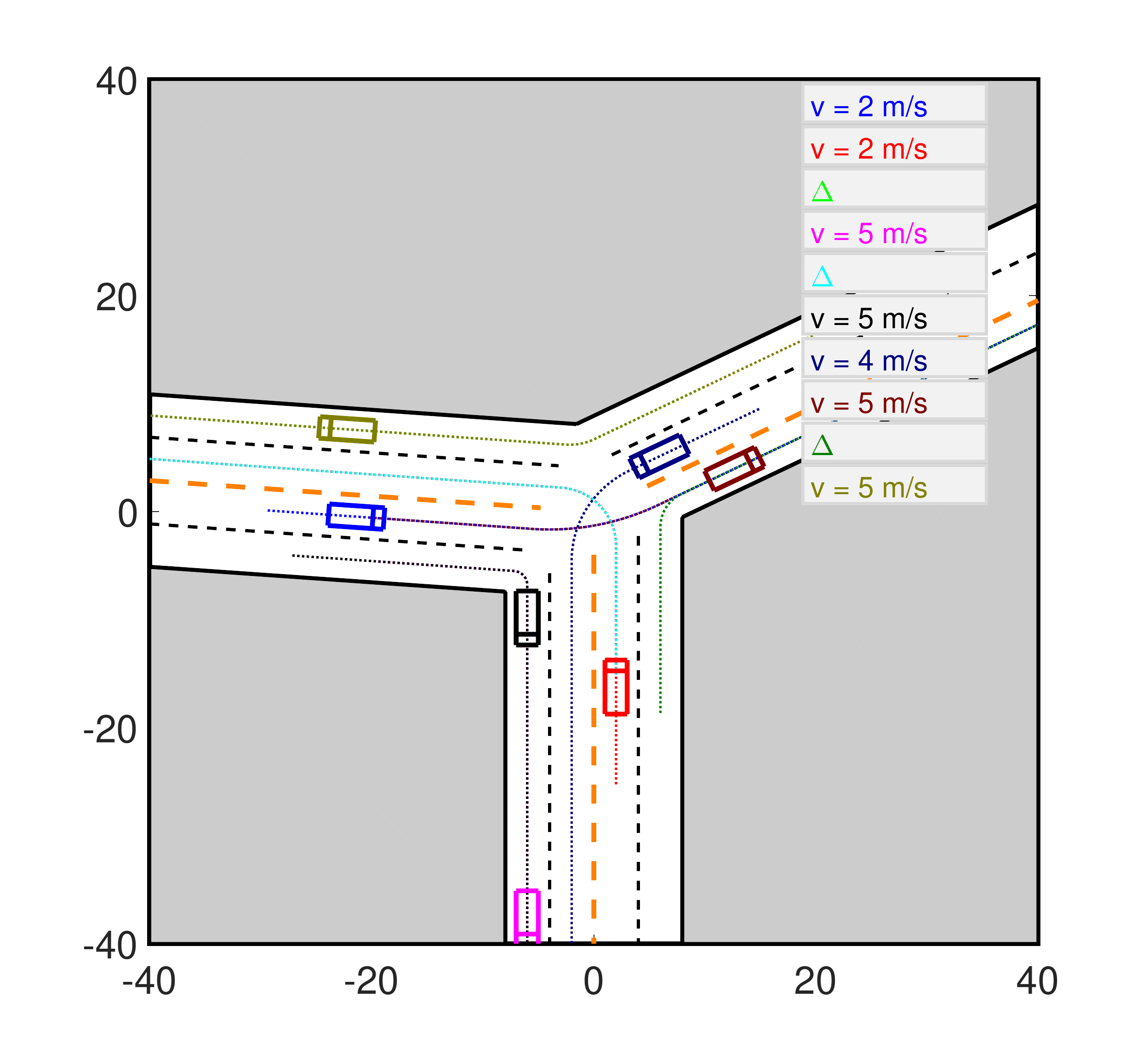,height=1.52in}}  
\put(  0,  100){\epsfig{file=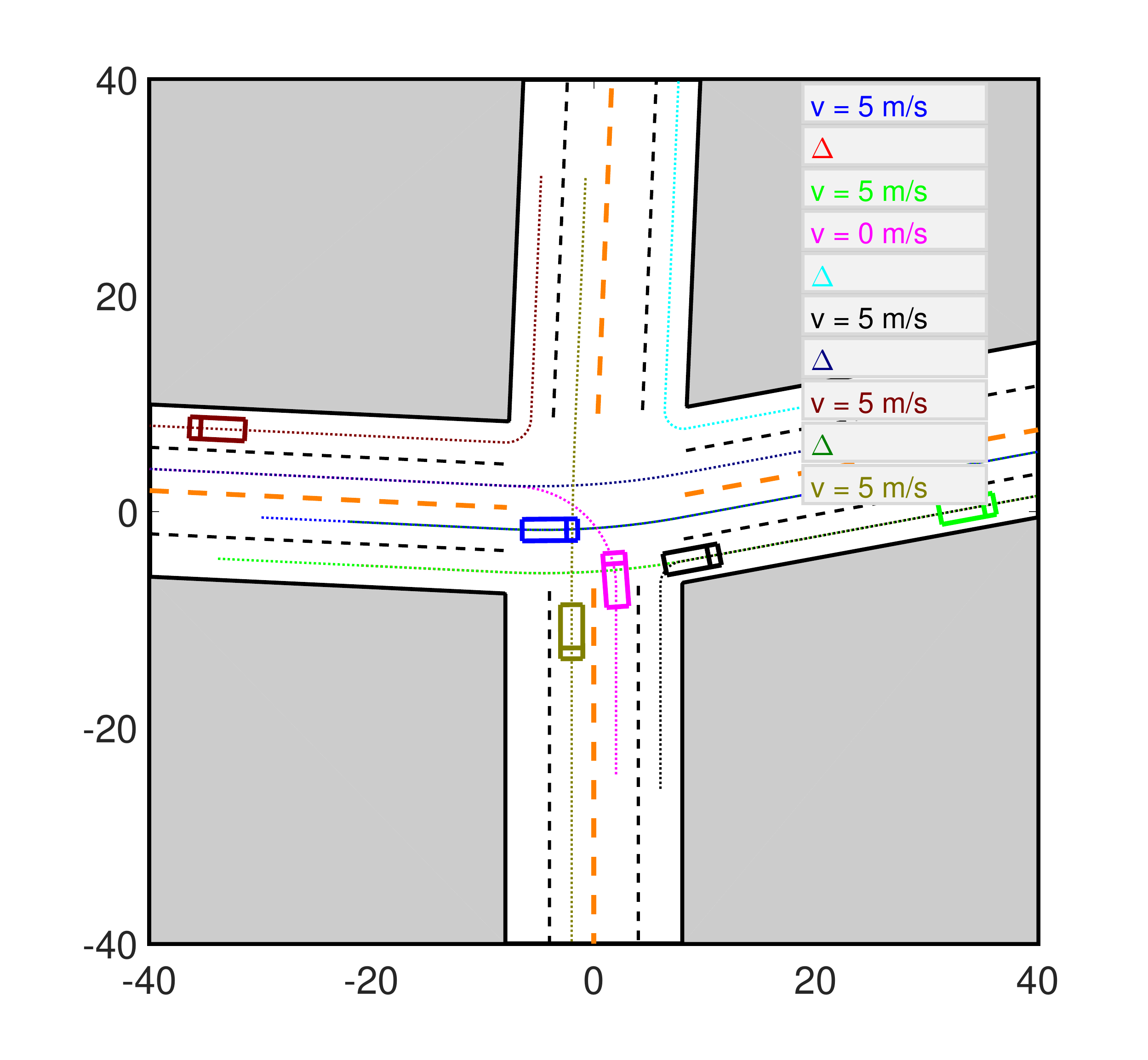,height=1.52in}}  
\put(  112,  100){\epsfig{file=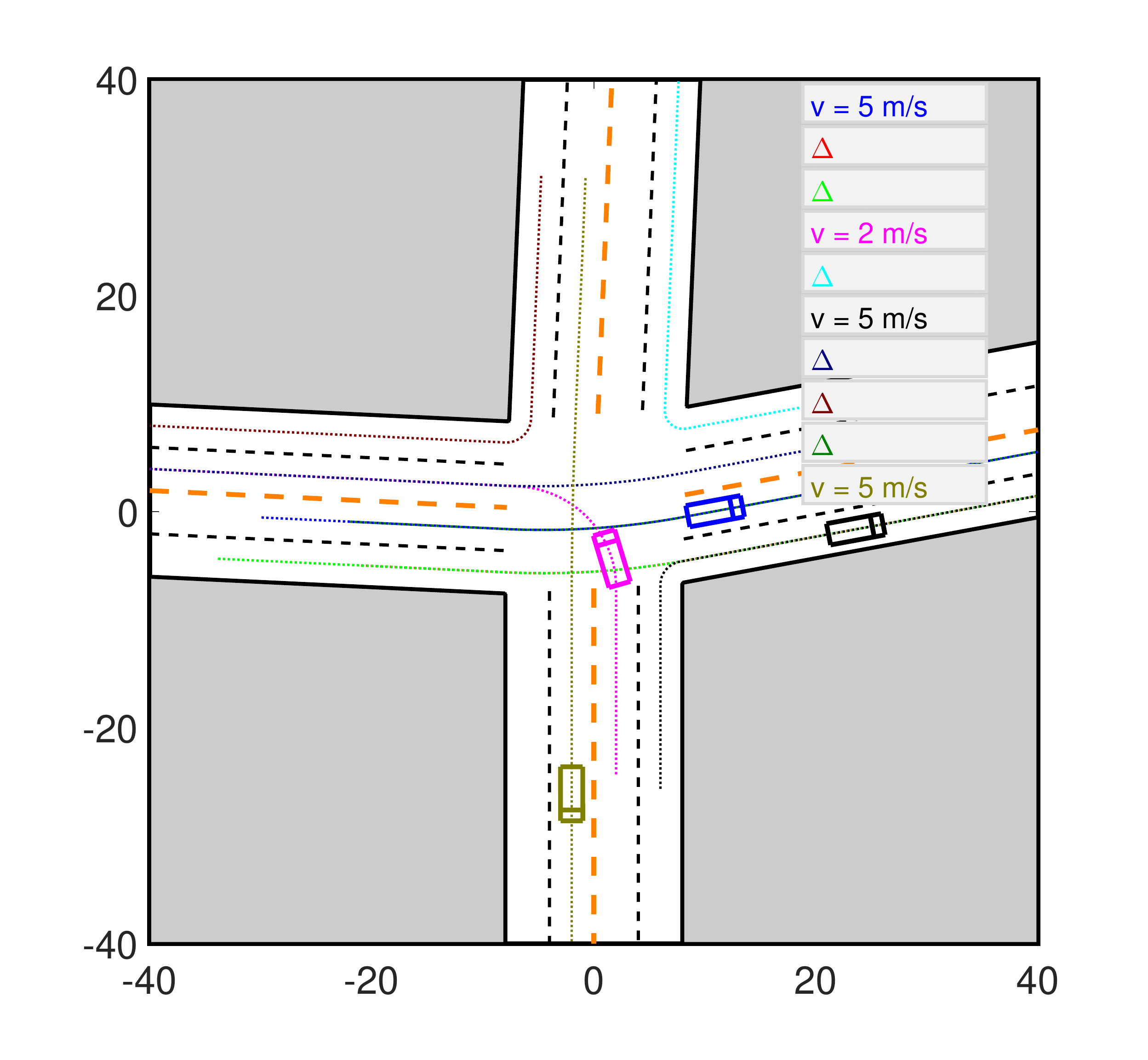,height=1.52in}}  
\put(  0,  -6){\epsfig{file=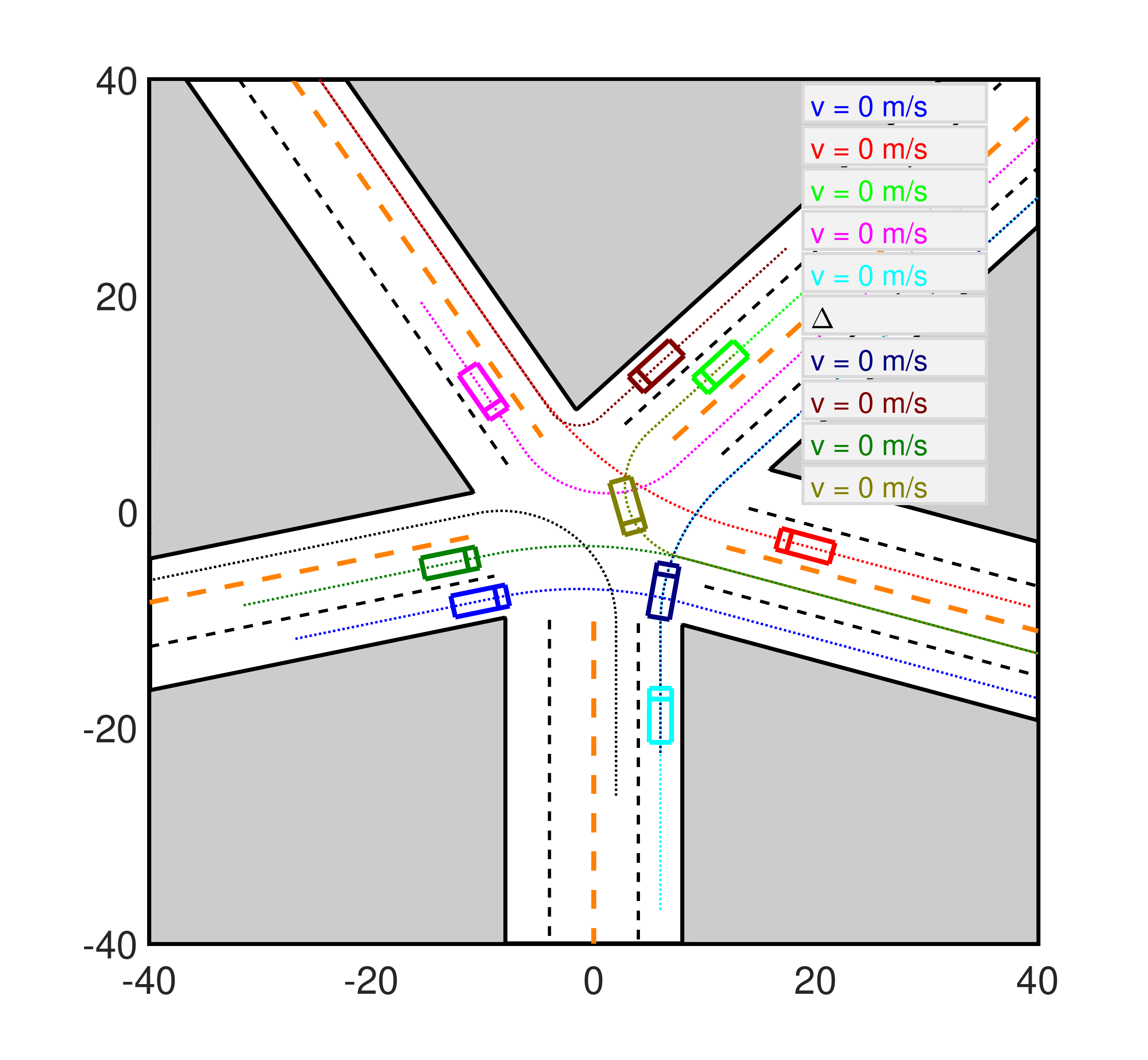,height=1.52in}}  
\put(  112,  -6){\epsfig{file=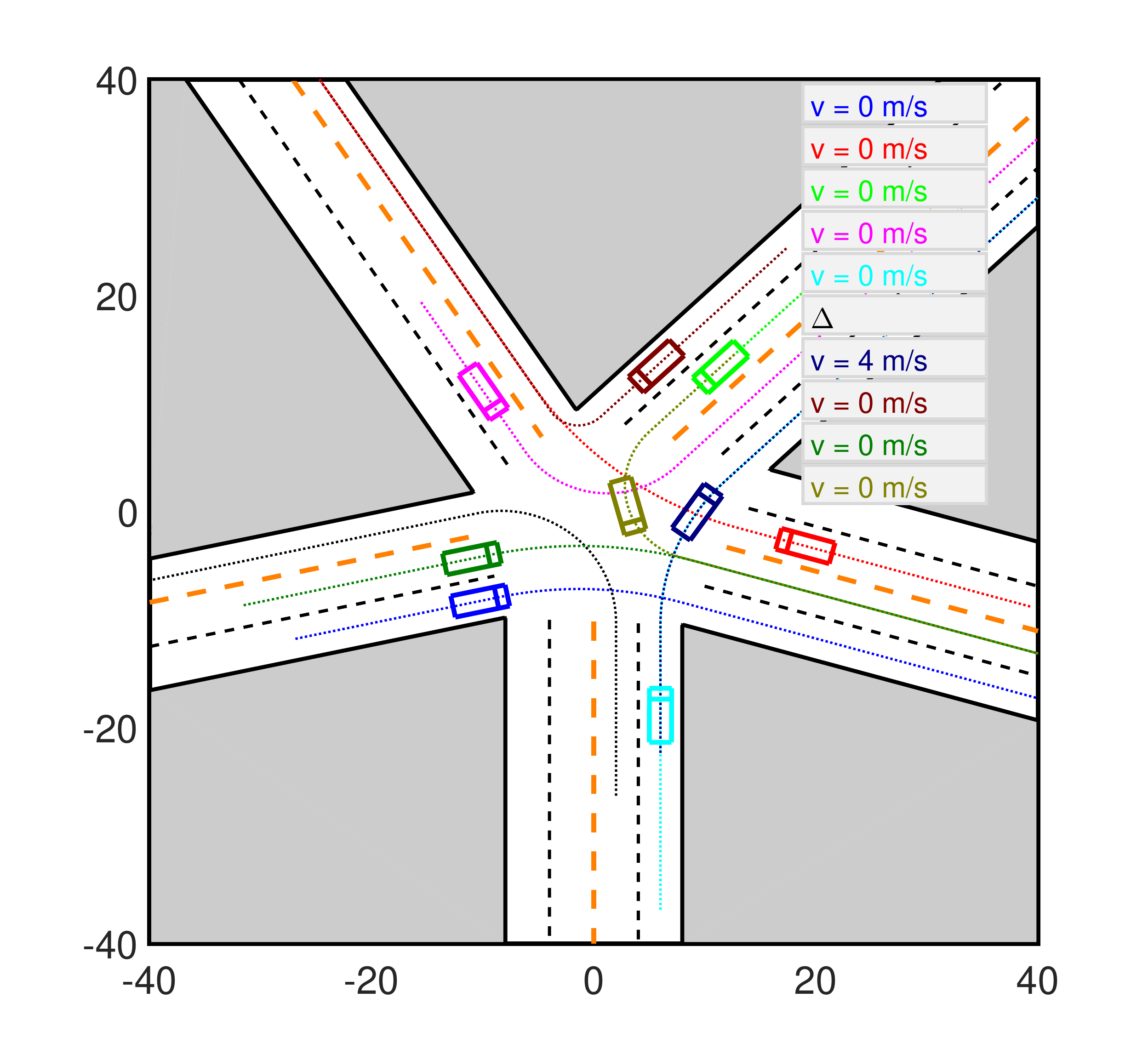,height=1.52in}}  
\small
\put(98, 310){(a)}
\put(210, 310){(b)}
\put(98, 204){(c)}
\put(210, 204){(d)}
\put(98, 98){(e)}
\put(210, 98){(f)}
\normalsize
\end{picture}
\end{center}
	\caption{Randomized traffic scenarios at randomized intersections. Figures~(a-b) show the snapshots of a three-way intersection simulation at two sequential steps, figures~(c-d) show the snapshots of a four-way intersection simulation at two sequential steps, and figures~(e-f) show the snapshots of a five-way intersection simulation at two sequential steps.}
	\label{fig:success_cases}
\end{figure}

Some of the simulated traffic scenarios are shown in Fig.~\ref{fig:success_cases}.

\subsubsection{Statistical evaluation}

We define several statistical metrics to evaluate the proposed framework to model vehicle interactions at uncontrolled intersections, including the rate of success (SR), the rate of collision (CR), and the rate of deadlock (DR). The rate of success is defined as the proportion of simulation runs where all the vehicles safely (without colliding with any other vehicles) reach their terminal points $(x^{\text{term}},y^{\text{term}})$ within $60\,$[s] of simulation time. The rate of collision is defined as the proportion of simulation runs where at least one vehicle collision occurs (once a vehicle collision occurs at a simulation step, the simulation run stops at that step). The rate of deadlock is defined as the proportion of simulation runs where no vehicle collision occurs but there is at least one vehicle that does not reach its terminal point $(x^{\text{term}},y^{\text{term}})$ within $60\,$[s] of simulation time. We note that based on their definitions, $\text{SR}+\text{CR} + \text{DR} = 1$.

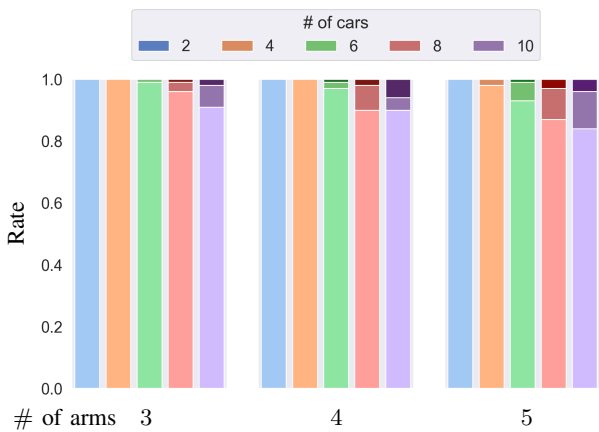
\begin{figure}[h!]
\begin{center}
\begin{picture}(210.0, 152.0)
\put(  0,  5){\epsfig{file=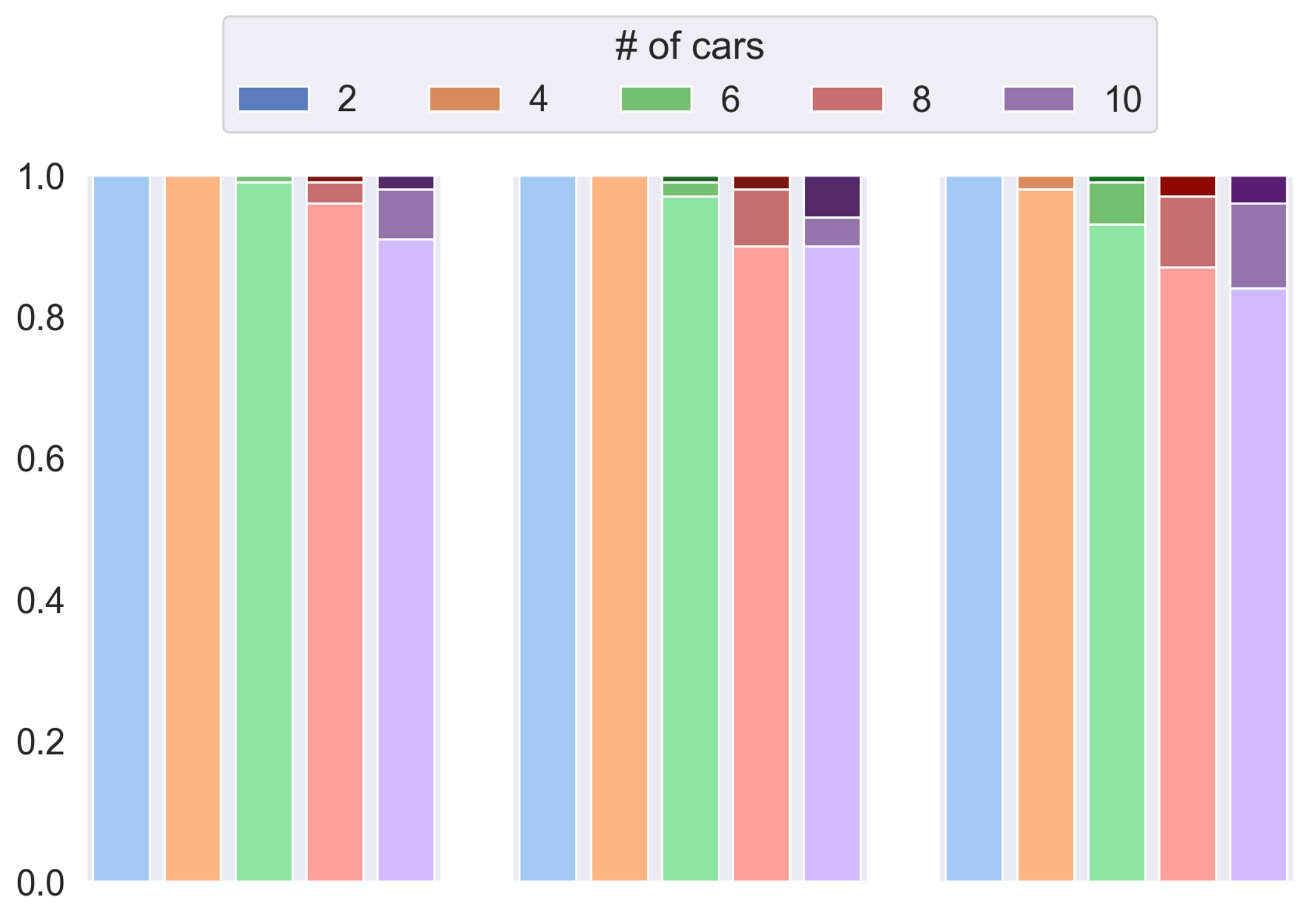,height=2.1in}}  
\small
\put(  -8,  -3){$\#$ of arms}
\put(  -10,  65){\rotatebox{90}{Rate}}
\put(  40,  -3){$3$}
\put(  112,  -3){$4$}
\put(  184,  -3){$5$}
\normalsize
\end{picture}
\end{center}
	\caption{Statistical evaluation of the vehicle interaction model. Light color: SR, medium color: DR, dark color: CR.}
	\label{fig:results}
\end{figure}

A model representing driver interactive decision-making is supposed to have reasonably high SR, and reasonably low CR and DR. The evaluation results of our model are shown in Fig.~\ref{fig:results}.

It can be observed that as the numbers of road arms and of vehicles increase, which correspond to traffic scenarios of increased complexity, the rates of collision and of deadlock also increase. In three-way and four-way intersection traffic scenarios with $2$ or $4$ vehicles, no collisions or deadlocks are observed. When up to $10$ vehicles are interacting at three-way or four-way intersections, the rates of success are higher than $0.9$. It can also be observed that five-way intersections are more challenging (with higher rates of collision and of deadlock compared to three-way and four-way intersections) -- the rate of success drops to $0.84$ for the case of $10$ interacting vehicles. This is because more vehicles may get to the entrances of the intersection or be inside the intersection at the same time for five-way intersections compared to three-way and four-way intersections, which may lead to higher chances of traffic conflicts. Indeed, five-way intersections are also more challenging to drivers compared to three-way and four-way intersections in real-world traffic scenarios.

\begin{figure}[h!]
\begin{center}
\begin{picture}(230.0, 98.0)
\put(  0,  0){\epsfig{file=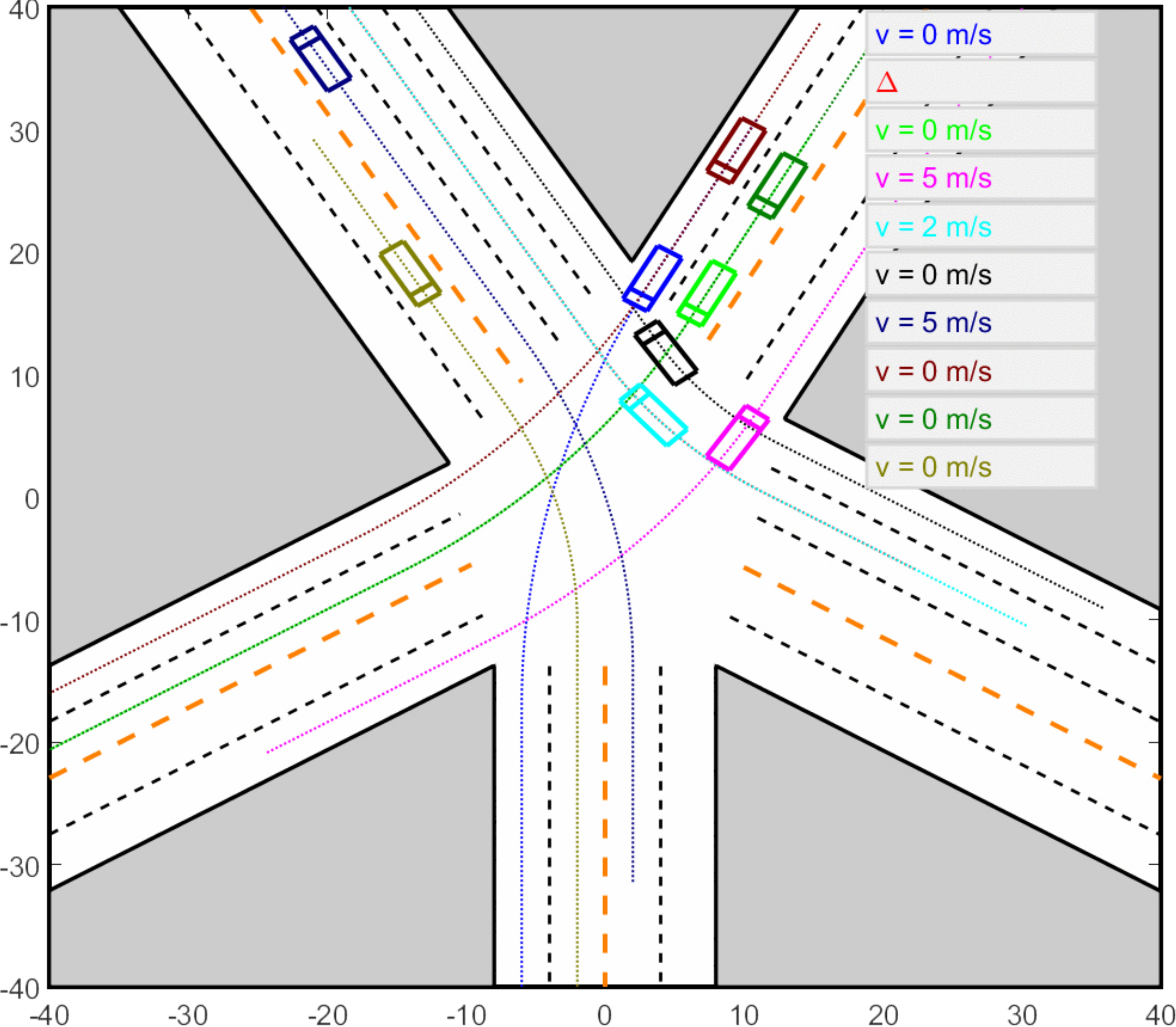,height=1.3in}}  
\put(  120,  0){\epsfig{file=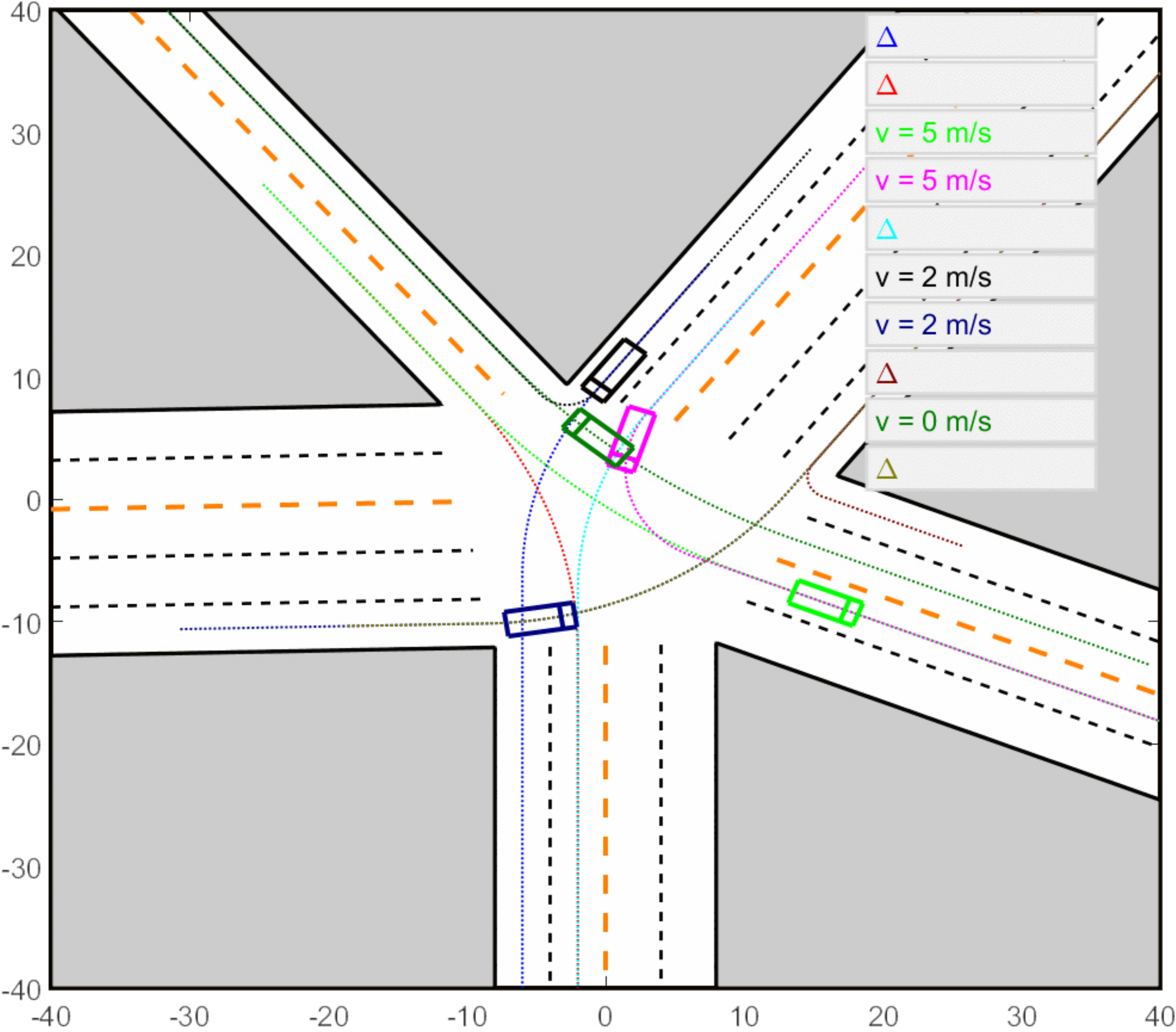,height=1.3in}}  
\small
\put(96, 96){(a)}
\put(215, 96){(b)}
\normalsize
\end{picture}
\end{center}
	\caption{Two failure cases. (a) A deadlock scenario. (b) A collision scenario. }
	\label{fig:failure_cases}
\end{figure}

Furthermore, by watching the animations of the simulation runs, we observe that most collisions are caused by simultaneous exploratory actions of two or more vehicles in deadlock scenarios. Note that in Algorithm~\ref{alg:Break_D}, a vehicle is not permitted to accelerate if its acceleration would lead to a collision when the other vehicles in conflict remain stopped. However, if two or more vehicles accelerate at the same time, it is possible that their simultaneous accelerations lead to a collision although each single acceleration would not. Two of the failure scenarios are shown in Fig.~\ref{fig:failure_cases}. The rates of failures ($\text{CR} + \text{DR}$) resulting from our framework are much lower than those resulting from the scheme in \cite{mandiau2008behaviour} ($3\%$ versus almost $50\%$ for the case of $6$ vehicles at four-way intersections). Note that communications and negotiations among human drivers in deadlock scenarios, such as through eye contacts or gestures, are not considered in our framework.

\begin{figure}[h!]
\begin{center}
\begin{picture}(240.0, 142.0)
\put(  10,  8){\epsfig{file=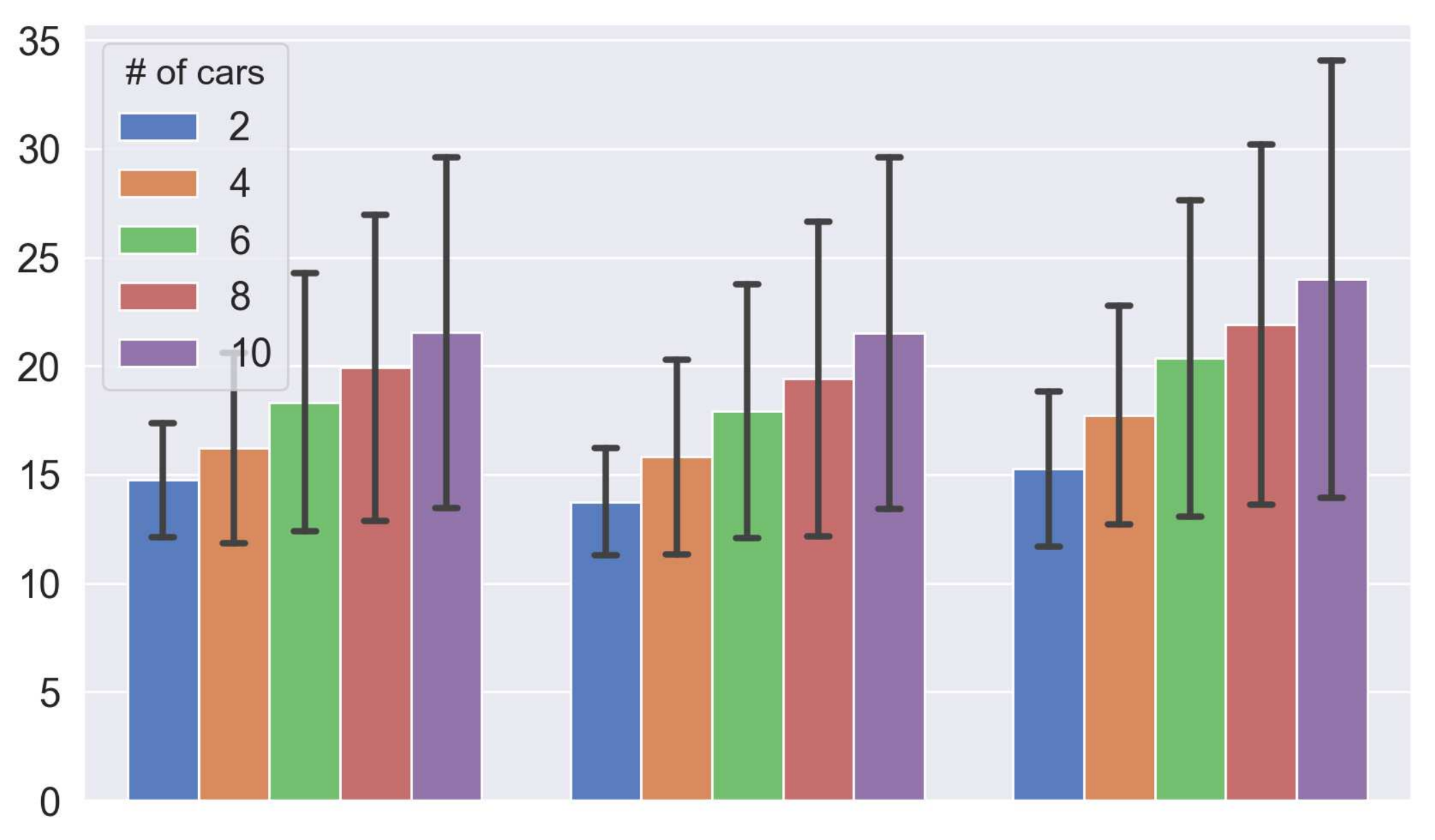,height=1.8in}}  
\small
\put(  2,  2){$\#$ of arms}
\put(  2,  60){\rotatebox{90}{ACT [s]}}
\put(  56,  2){$3$}
\put(  124,  2){$4$}
\put(  194,  2){$5$}
\normalsize
\end{picture}
\end{center}
	\caption{Average completion time (ACT). The black vertical bars represent the standard deviations.}
	\label{fig:ACT}
\end{figure}
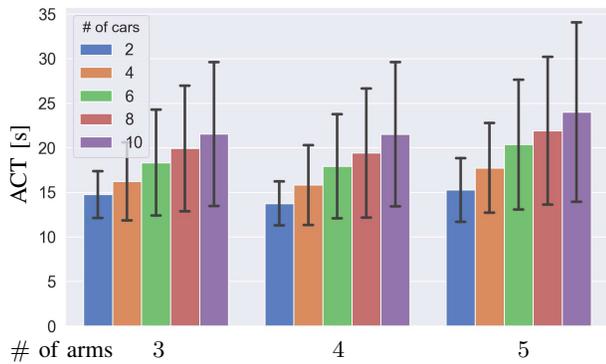

For the vehicles that safely reach their terminal points within $60\,$[s] of simulation time, we count their average completion time (CT), where a vehicle's CT is defined as the duration (in [s] of simulation time) from the simulation initialization to the time instant when the vehicle reaches its terminal point. The average CT can reflect how conservative the decision-making model is. The average CTs for different numbers of road arms and vehicles are shown in Fig.~\ref{fig:ACT}.

It can be observed that as the numbers of vehicles increase, the vehicles need more time to pass through the intersections. In particular, for the cases of $2$ and $4$ interacting vehicles, the average CTs exhibited by our model correspond to level-B in the real-world traffic quality rating system called the ``level-of-service'' (LOS) for unsignalized intersections defined based on the average control delay \cite{quiroga1999measuring}. LOS-B corresponds to traffic with a high degree of freedom and a small amount of interactions and is characterized by $10$-$15\,$[s] of average control delay \cite{HCM2000}. For the cases of $6$-$10$ interacting vehicles, the average CTs exhibited by our model correspond to LOS-C, which corresponds to traffic with restricted freedom due to significant interactions and is characterized by $15$-$25\,$[s] of average control delay. Furthermore, among three-way, four-way, and five-way intersections, the vehicles spend the least times to pass through four-way intersections --  this may be explained by the observation that the ``right-of-way'' rules (see Section~\ref{sec:leader_follower}) may function best for four-way intersections, which, as a matter of fact, are most common in real-world road networks.

\subsubsection{Computational complexity}

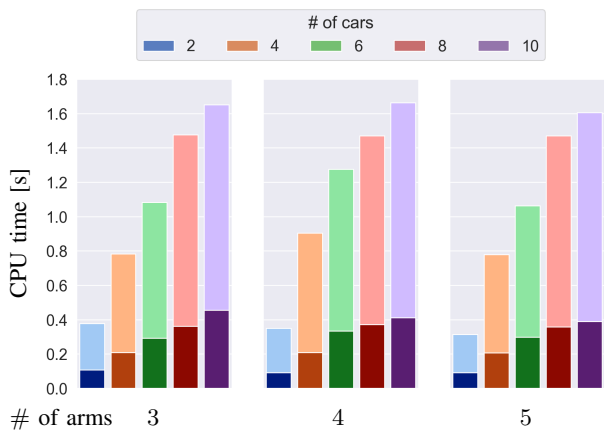
\begin{figure}[h!]
\begin{center}
\begin{picture}(240.0, 156.0)
\put(  11,  10){\epsfig{file=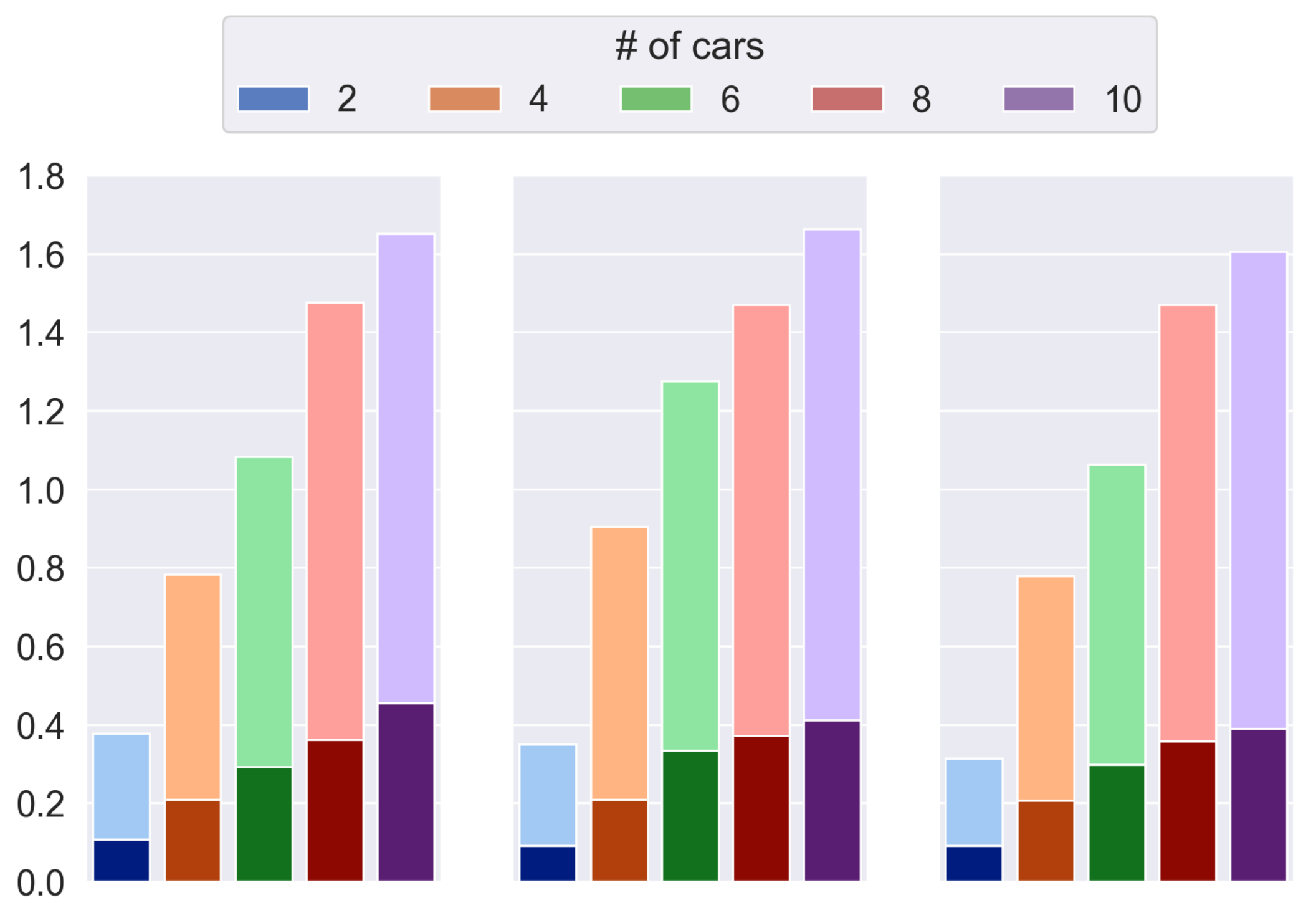,height=2.1in}}  
\small
\put(  0,  2){$\#$ of arms}
\put(  0,  50){\rotatebox{90}{CPU time [s]}}
\put(  52,  2){$3$}
\put(  122,  2){$4$}
\put(  193,  2){$5$}
\normalsize
\end{picture}
\end{center}
	\caption{Average (dark-colored bars) and worst (light-colored bars) computation times per vehicle per step.}
	\label{fig:CPUT}
\end{figure}

In addition to rate of success and average completion time, we also care about the computational effort of the proposed framework to model vehicle interactions at uncontrolled intersections, as it determines the framework's scalability to model traffic scenarios of increased complexity. As mentioned in Section~\ref{sec:game_n}, traditional generalizations of leader-follower game-theoretic decision-making models to $n$-player settings require exponentially increased computational efforts to solve for solutions as the number of players increases \cite{yoo2018predictive}. However, thanks to the pairwise decoupling of vehicle interactions, the computational complexity of our decision-making model \eqref{equ:leader_follower_n_1} (solved using a tree-search method) increases only linearly with the number of interacting vehicles increasing. We use the average and the worst computation times per vehicle per step (in [s] of real time) to represent our model's computational complexity, which are, respectively, the average and the worst CPU times for one vehicle to confirm its action choice over one step (including the time to compute the initial action choice using \eqref{equ:leader_follower_n_1} and the time to adjust the action choice using Algorithm~\ref{alg:Break_D} if a deadlock is detected). The results for different numbers of road arms and vehicles are shown in Fig.~\ref{fig:CPUT}. The simulations are performed on Matlab~R2016a platform using an Intel Core i7-4790 3.60~GHz PC with Windows~10 and 16.0~GB of RAM. The computation times are calculated using Matlab \textit{tic-toc} command.

It can be observed that: 1) As the number of vehicles increases, the computation time increases. 2) For four-way and five-way intersections, the increase in computation time slows down with the number of vehicles increasing. And 3) the increase in computation time slows down as the number of road arms increases from three to four and five. The explanations to 2) and 3) may be that as the numbers of vehicles and road arms increase, the vehicles are more likely to be outside each other's perception range, and as a result, the number of vehicles involved in the computation of model \eqref{equ:leader_follower_n_1} decreases (see Section~\ref{sec:ranges}). In general, the increase in computational effort for our decision-making model \eqref{equ:leader_follower_n_1} to solve for solutions is only linear in the increase in the number of interacting vehicles. And in turn, the total computation time for all vehicles to solve for decisions is quadratic in the number of vehicles. This makes our framework have reasonably good scalability to model traffic scenarios at uncontrolled intersections of increased complexity.

In contrast, the adaptive level-$\mathcal{K}$ decision-making model described in Section~\ref{sec:levelK} may not be as well-scalable as the decision-making model \eqref{equ:leader_follower_n_1} based on pairwise leader-follower games. This is because the adaptive level-$\mathcal{K}$ model needs to first compute the level-$k$ decision of each of the interacting vehicles for $k=0,1,\cdots,k_{\max}$ using \eqref{equ:levelK} in order to estimate their levels using \eqref{equ:model_estimation}, and then compute the optimal decision of the ego vehicle using \eqref{equ:model_D}, which itself involves combinatorially increased computational complexity in the number of interacting vehicles. 

\section{Summary}\label{sec:sum}

In this paper, we proposed a game-theoretic framework for modeling the interactive behavior of vehicles in multi-vehicle traffic scenarios at uncontrolled intersections. Our approach takes into account common traffic rules to designate a leader-follower relationship between each pair of interacting vehicles. A decision-making process based on pairwise leader-follower relationships is used to represent interactive decision-making of vehicles. Additional modeling considerations, representing courteous driving, limited perception ranges, and the capability of human drivers in resolving deadlock scenarios through probing, are also accounted for in the process.

In particular, uncontrolled intersections were modeled based on a parametrization scheme, to which the proposed vehicle interaction modeling approach was applied. This way, the interactive behavior of vehicles at a rich set of uncontrolled intersection traffic scenarios (with various numbers of interacting vehicles, intersection layouts and geometries, etc) could be modeled.

Simulation results were reported and showed that the vehicle interaction model exhibited reasonable behavior expected in traffic. The performance of the model was then evaluated based on several statistics, including the rate of success, the rate of collision, the rate of deadlock, the average completion time, as well as the average and the worst computation times. It was shown that the model had reasonably high rates of success in resolving traffic conflicts and average completion times matching the level-of-service criteria used for rating real-world traffic. Moreover, thanks to the pairwise decoupling of vehicle interactions, the computational complexity of the decision-making model increases linearly as the number of interacting vehicles increases, which improves the model's scalability.

Furthermore, another game-theoretic model to represent interactive decision-making of vehicles was considered. The approaches to modeling vehicle interactions based on the model proposed in this paper and this alternative model were discussed and compared in simulations.

The framework proposed in this paper for modeling multi-vehicle interactions at uncontrolled intersections can be used as simulation tool for calibration, validation and verification of autonomous driving systems \cite{li2018game_2,li2018game_3,li2019game}. In addition, it may also be used in high-level decision-making algorithms of autonomous vehicles \cite{li2018game,tian2018adaptive}, and to support intersection automation/autonomous intersection management \cite{chen2016cooperative}. Moreover, vehicle interactions in some other traffic scenarios, such as highway merging and driving in parking lots, may be modeled based on the proposed framework with modified road layouts and geometries. These are left as topics for our future research.

\bibliographystyle{IEEEtran}

\bibliography{ref}

\section*{Appendix~A}

The three segments of $\mathcal{P}$ are described using the following equations:
\small
\begin{align}\label{equ:appendix_a1}
	& a_1 x + b_1 y + c_1 = 0 \quad \text{if } \rho \leq \rho^{\text{en}}, \nonumber \\
	& (x-x_{\text{c}})^2 + (y-y_{\text{c}})^2 = r^2 \quad \text{if }\rho^{\text{en}} < \rho \leq \rho^{\text{ex}}, \nonumber \\
	& a_2 x + b_2 y + c_2 = 0 \quad \text{if } \rho > \rho^{\text{ex}},
\end{align}
\normalsize
in which $(a_1,b_1,c_1)$ and $(a_2,b_2,c_2)$ are the parameters of the line segments corresponding to the coefficients $(\sin(\phi^{(m)}),-\cos(\phi^{(m)}),\frac{k w_{\text{lane}}}{2})$ in \eqref{equ:lane_func}, where $m$ and $k$ are determined by the origin lane and target lane of the vehicle, and $(x_{\text{c}}, y_{\text{c}}, r)$ are the center coordinates and radius of the arc segment. The unknowns $(x_{\text{c}}, y_{\text{c}}, r)$ and $\rho^{\text{ex}}$ are computed by first solving the following function set:
\small
\begin{align}\label{equ:appendix_a2}
	& x_{\text{c}} = \frac{a_1 b_2 x(\rho^{\text{ex}}) - a_2 b_1 x(\rho^{\text{en}}) + a_1 a_2 y(\rho^{\text{en}}) - a_1 a_2 y(\rho^{\text{ex}})}{a_1 b_2 - a_2 b_1} \nonumber \\
	& y_{\text{c}} = \frac{b_1 b_2 x(\rho^{\text{ex}}) - b_1 b_2 x(\rho^{\text{en}}) + a_1 b_2 y(\rho^{\text{en}}) - a_2 b_1 y(\rho^{\text{ex}})}{a_1 b_2 - a_2 b_1} \nonumber \\
	& r = \frac{\sqrt{a_1^2+b_1^2} \big(a_2 y(\rho^{\text{en}}) - b_2 x(\rho^{\text{en}}) + b_2 x(\rho^{\text{ex}}) -a_2 y(\rho^{\text{ex}}) \big)}{a_1 b_2 - a_2 b_1} \nonumber \\
	& x(\rho^{\text{ex}}) = -\frac{-b_2^2 x_{\text{c}} + a_2 b_2 y_{\text{c}} + a_2 c_2}{a_2^2 + b_2^2} \nonumber \\
	& y(\rho^{\text{ex}}) = -\frac{-a_2^2 y_{\text{c}} + a_2 b_2 x_{\text{c}} + b_2 c_2}{a_2^2 + b_2^2},
\end{align}
\normalsize
where $(x(\rho^{\text{en}}), y(\rho^{\text{en}}))$ are given based on the intersection layout (see Section~\ref{sec:intersection}); and then computing $\rho^{\text{ex}}$ by:
\small
\begin{align}
	& \rho^{\text{ex}} = \rho^{\text{en}} + r\, \Delta\phi, \quad \Delta\phi =\arccos\Big(\frac{u^\top v}{\|u\|\|v\|}\Big), \\
	& u = \! \big[x(\rho^{\text{en}})- x_{\text{c}},  y(\rho^{\text{en}})-y_{\text{c}}\big]^{\!\top}, \quad v = \! \big[x(\rho^{\text{ex}})- x_{\text{c}},  y(\rho^{\text{ex}})-y_{\text{c}}\big]^{\!\top}. \nonumber
\end{align}
\normalsize

\begin{IEEEbiography}[{\includegraphics[width=1in,height=1.25in,clip,keepaspectratio]{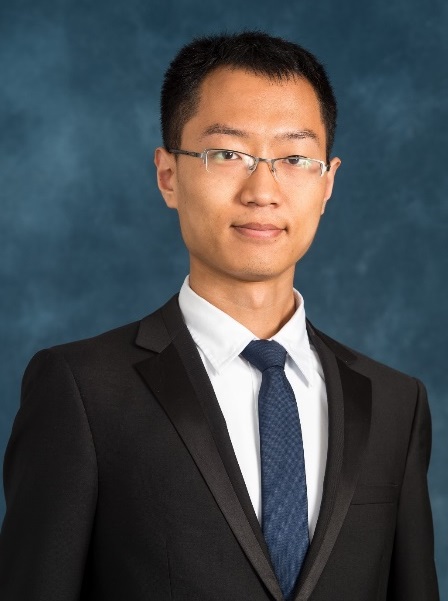}}]
{\textbf{Nan Li}} received the B.S. degree in automotive engineering from Tongji University, Shanghai, China, in 2014, and the M.S. degree in mechanical engineering from the University of Michigan, Ann Arbor, MI, USA, in 2016, where he is currently pursuing the Ph.D. degree in aerospace
engineering. His current research interests include predictive control, stochastic control, and application of game theory to multi-agent systems.
\end{IEEEbiography}

\newpage
\section*{Appendix~B}

\begin{table}[ht!]
    \scriptsize
	\centering
	\caption{Simulation parameter values.}
	\begin{tabular}{cccc}
		\toprule[1pt]
		Variable(s) & Value(s) & Unit & Remarks \\
		\midrule
		$[v_{\min},v_{\max}]$ & $[0,5]$ & m/s & speed range \\
		\midrule
		$\Delta t$ & $1$ & s & sampling period \\
		\midrule
		$A$ & $\{-4,-2,0,2\}$ & m/s$^2$ &\makecell{$\{$hard brake, decelerate, \\ maintain, accelerate$\}$} \\
		\midrule
		$\delta$ & $0.5$ & m & \makecell{threshold for  \\ differentiating distances }\\
		\midrule
		$w_{1,2,3}$, $\hat{w}$ & $\{100,5,1\}$, $\frac{1}{4}$ & & reward function weights \\
		\midrule
		$(l_c,w_c)$ & $(6,2.4)$ & m & $c$-zone size \\
		\midrule
		$(l_{s,\text{f}}^l,l_{s,\text{r}}^l,w_s^l)$ &$(5,4,2.8)$ & m & \makecell{$s$-zone size \\ for leader} \\
		\midrule
		$(l_{s,\text{f}}^f,l_{s,\text{r}}^f,w_s^f)$ & $(14,4,2.8)$& m & \makecell{$s$-zone size \\ for follower} \\
		\midrule
		$\mathcal{N}$ & 2 & & prediction horizon \\
		\midrule
		$\lambda$ & 0.6 & & discount factor \\
		\midrule
		\makecell{$\omega_i,$\\$\forall\, i \in \{1,\cdots,n\}$ } & 30 & m & \makecell{maximum perception \\ distance } \\
		\midrule
		\makecell{$p_i,$\\$\forall\, i \in \{1,\cdots,n\}$} & 0.25 & & \makecell{probing movement \\probability } \\
		\midrule	
		$(l_{s,\text{f}}^\mathcal{K},l_{s,\text{r}}^\mathcal{K},w_s^\mathcal{K})$ & $(9.5,4,2.8)$ & m & \makecell{$s$-zone size for level-$\mathcal{K}$ \\ models, $\mathcal{K}=0,\cdots,$ \\ $k_{\max}$, and for model $\mathfrak{K}$} \\
		\midrule
		$k_{\max}$ & 2 & & highest level-$\mathcal{K}$ model \\
		\midrule
		$\Delta \mathbb{P}$ & $\frac{2}{3}$ & & \makecell{model $\mathfrak{K}$ belief \\ update parameter} \\
		\bottomrule[1pt]
	\end{tabular}\label{tab:para}
\end{table}

\begin{IEEEbiography}[{\includegraphics[width=1in,height=1.25in,clip,keepaspectratio]{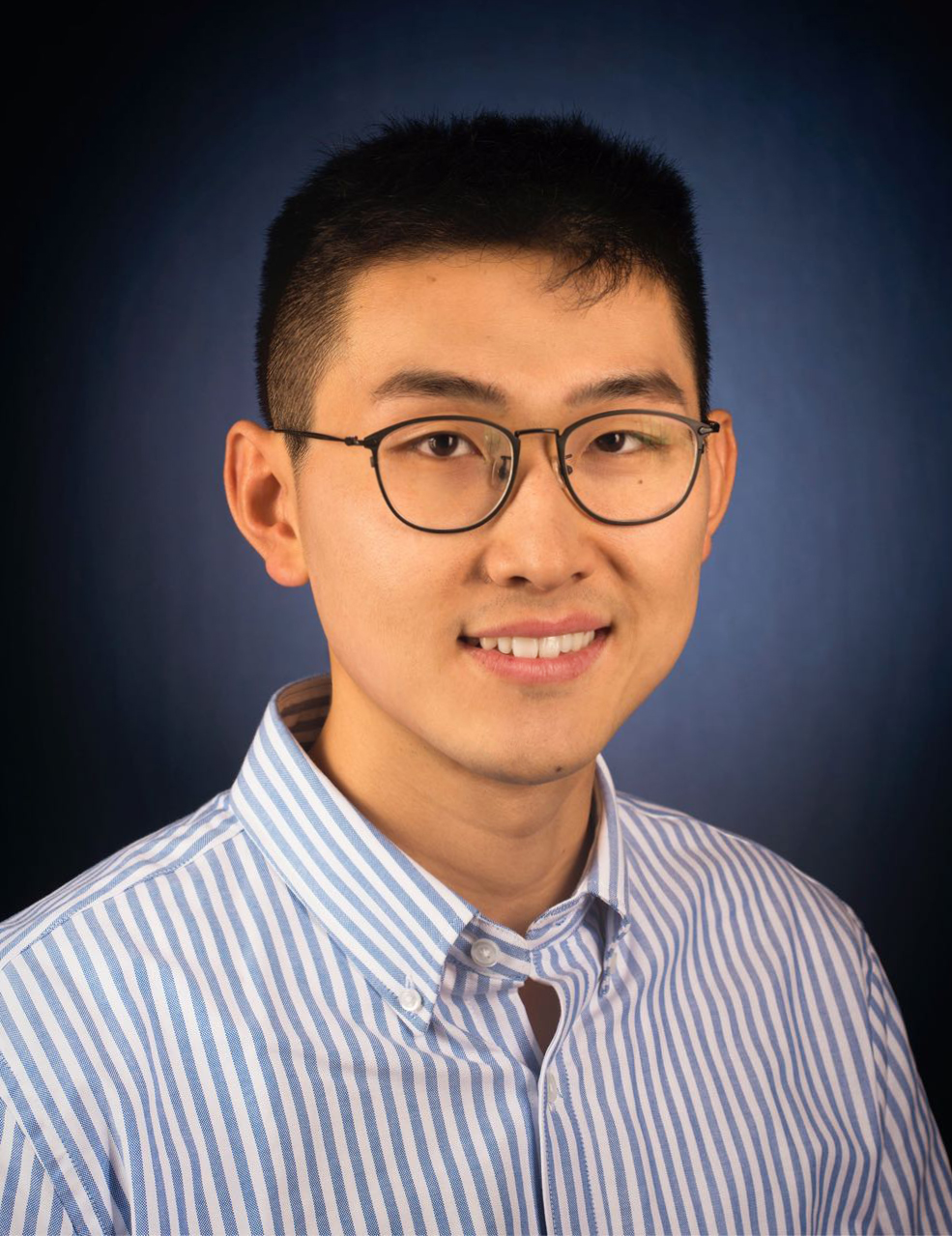}}]
{\textbf{Yu Yao}} received the B.Eng. degree in Aerospace Engineering from Beijing Institute of Technology in 2015 and M.S. degree in Robotics from the University of Michigan in 2017. He is currently a Ph.D. candidate at the University of Michigan Robotics Institute. His research interests include anomaly detection, action recognition/prediction, scene understanding and their applications to autonomous vehicles and intelligent transportation systems.
\end{IEEEbiography}

\begin{IEEEbiography}[{\includegraphics[width=1in,height=1.25in,clip,keepaspectratio]{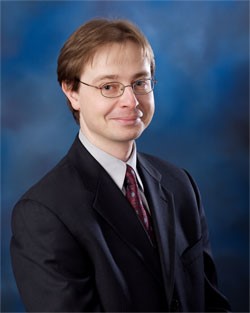}}]
{\textbf{Ilya Kolmanovsky}} received the M.S. and Ph.D. degrees in aerospace engineering and the M.A. degree in mathematics from the University of Michigan, Ann Arbor, MI, USA, in 1993, 1995, and
1995, respectively. He is currently a Full Professor with the Department of Aerospace Engineering, University of Michigan. He is named as an inventor on 100 U.S. patents. His current research interests include control theory for systems with state and control constraints and control applications to aerospace and automotive systems. Dr. Kolmanovsky was a recipient of the Donald P. Eckman Award of
American Automatic Control Council and of two IEEE Transactions on Control Systems Technology outstanding paper awards.
\end{IEEEbiography}

\begin{IEEEbiography}[{\includegraphics[width=1in,height=1.25in,clip,keepaspectratio]{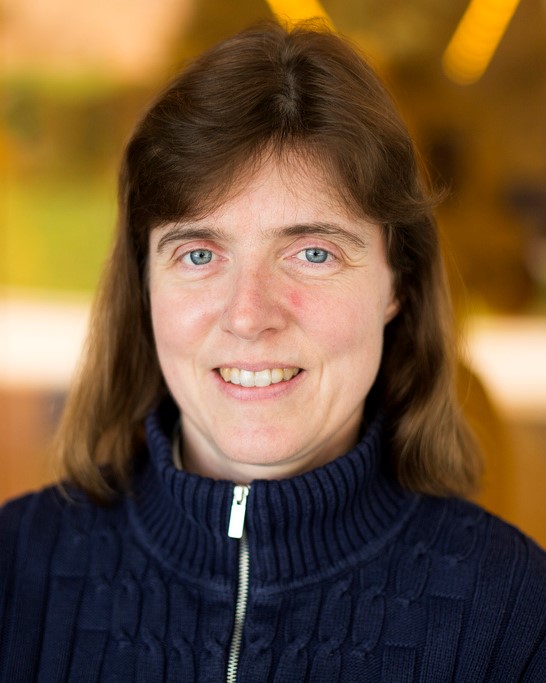}}]
{\textbf{Ella Atkins}} is a Professor of Aerospace Engineering at the University of Michigan, where she directs the Autonomous Aerospace Systems Lab and is Associate Director of Graduate Programs for the Robotics Institute.  Dr. Atkins holds B.S. and M.S. degrees in Aeronautics and Astronautics from MIT and M.S. and Ph.D. degrees in Computer Science and Engineering from the University of Michigan.  She is Editor-in-Chief of the AIAA Journal of Aerospace Information Systems (JAIS) and past-chair of the AIAA Intelligent Systems Technical Committee. She has served on the National Academy's Aeronautics and Space Engineering Board, and pursues research in Aerospace and robotic system contingency planning, autonomy, and safety.
\end{IEEEbiography}

\begin{IEEEbiography}[{\includegraphics[width=1in,height=1.25in,clip,keepaspectratio]{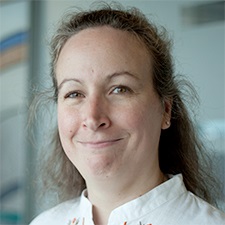}}]
{\textbf{Anouck R. Girard}} received the Ph.D. degree in ocean engineering from the University of
California, Berkeley, CA, USA, in 2002. She has been with the University of Michigan, Ann Arbor, MI, USA, since 2006, where she is currently an Associate Professor of Aerospace Engineering. She has co-authored the book Fundamentals of Aerospace Navigation and Guidance (Cambridge University Press, 2014). Her current research interests include flight dynamics and control systems. Dr. Girard was a recipient of the Silver Shaft Teaching Award from the University of Michigan and a Best Student Paper
Award from the American Society of Mechanical Engineers.
\end{IEEEbiography}

\end{document}